\documentclass[prx,superscriptaddress,longbibliography,twocolumn,floatfix,showpacs,preprintnumbers,amsmath,amssymb,amsfonts,lengthcheck]{revtex4-1}

\usepackage[dvips]{graphics}

\DeclareMathAlphabet{\mathsfit}{\encodingdefault}{\sfdefault}{m}{sl}
\SetMathAlphabet{\mathsfit}{bold}{\encodingdefault}{\sfdefault}{bx}{sl}

\usepackage[dvips]{graphics}
\usepackage{geometry}
 \usepackage{ctable}
 \usepackage[percent]{overpic}

\usepackage{textgreek}
\usepackage{bm}
\usepackage{contour}
\usepackage{braket}
\usepackage{soul}

\usepackage{blindtext}
\usepackage[percent]{overpic}
\usepackage{mathtools}
\usepackage{xcolor}
\usepackage{multirow}
\usepackage{array}

\usepackage[percent]{overpic}

\usepackage{mathtools}
\renewcommand{\vec}[1]{\mathbf{#1}}

\renewcommand{\vec}[1]{\mathbf{#1}}

\newcommand{\tens}[1]{\mbox{\textsf{\textbf{#1}}}}

\newcommand{\dif}{\!\! \mathrm{d}}
\newcommand{\mi}{\textrm{i}} 
\newcommand{\me}{\mathrm{e}}

\newcommand{\Eclo}{\mathcal{E}^{(1)}}
\newcommand{\Eclt}{\mathcal{E}^{(2)}}
\newcommand{\Ecli}{\mathcal{E}^{(i)}}
\newcommand{\Eclj}{\mathcal{E}^{(j)}}
\newcommand{\Ei}{\hat{E}^{(i)}}
\newcommand{\Ej}{\hat{E}^{(j)}}
\newcommand{\Eo}{\hat{E}^{(1)}}

\begin{document}

\title{Entanglement Harvesting from Electromagnetic Quantum Fields}

\author{Frieder Lindel}
\affiliation{\freiburg}
\author{Alexa Herter}
\affiliation{\ethZ}
\author{Valentin Gebhart}
\affiliation{\Vale}
\author{J\'{e}r\^{o}me Faist}
\affiliation{\ethZ}
\author{Stefan Y. Buhmann}
\affiliation{\kassel}

\newcommand{\freiburg}{Physikalisches Institut, Albert-Ludwigs-Universit\"{a}t Freiburg, Hermann-Herder-Stra{\ss}e 3, D-79104, Freiburg, Germany}
\newcommand{\Vale}{QSTAR, INO-CNR and LENS, Largo Enrico Fermi 2, 50125 Firenze, Italy}
\newcommand{\kassel}{Institut f\"{u}r Physik, Universit\"{a}t Kassel, Heinrich-Plett-Stra{\ss}e 40, 34132 Kassel, Germany}
\newcommand{\ethZ}{ETH Zurich, Institute of Quantum Electronics, Auguste-Piccard-Hof 1, 8093 Zurich, Switzerland}

\date{\today}

\begin{abstract}
In many states of the quantum electromagnetic field, including the vacuum state, entanglement exists between different space-time regions---even space-like separated ones. These correlations can be harvested and, thereby, detected by quantum systems which locally interact with the field. Here, we propose an experimental implementation of such an entanglement-harvesting scheme which is based on electro-optic sampling (EOS). We demonstrate that state-of-the-art EOS experiments enable one to harvest entanglement from the vacuum field and to study quantum correlations within general THz fields. We further show how Bell nonlocality present in the vacuum field can be probed. Finally, we introduce a novel approach to mitigate shot noise in single-beam EOS configurations. These findings pave the way for experimental inquiries into foundational properties of relativistic quantum field theory, and empower EOS as a diagnostic tool in THz quantum optics.
\end{abstract}

\maketitle

\section{Introduction}

Entanglement is considered as a key element concerning the foundations of quantum theory \cite{einstein1935can,bell1964einstein,horodecki2009quantum}, and it is thought of as being a quantum resource \cite{horodecki2013quantumness,chitambar2019quantum} bringing along an advantage of quantum technologies over their classical counterparts. Usually, for applications such as quantum computing \cite{nielsen2010quantum} or cryptography \cite{gisin2002quantum}, it is considered in low dimensional, few-particle settings, e.g., for a couple of qubits. Entanglement has also been investigated in relativistic quantum field theory (QFT) \cite{summers_vacuum_1985,summers_bells_1987,calabrese2004entanglement,witten2018aps,nishioka2018entanglement}, with the finding that it is a rather generic feature in infinite-dimensional quantum fields. For example, the vacuum state of, e.g., a massless scalar field sustains entanglement between space-like separated regions \cite{summers_vacuum_1985,summers_bells_1987}. Furthermore, the ground state is even capable of violating a Bell inequality, revealing the presence of nonlocality in the vacuum field \cite{summers_vacuum_1985,summers_bells_1987}. Growing interest in the entanglement structure of relativistic quantum fields was further driven by its connection to fundamental questions in quantum gravity. For instance, it was shown to be at the heart of the black-hole information paradox \cite{preskill1992black,hawking2005information,susskind1993stretched}, and has been proposed as a key feature to explain the emergence of classical space-times from quantum field theories \cite{van2010building,lashkari2014gravitational,cao2018bulk}.

\begin{figure}
\includegraphics[width=1.\columnwidth]{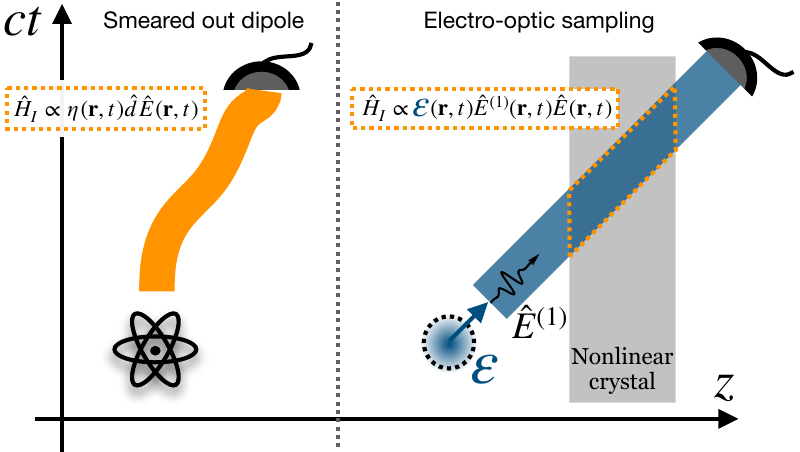}
\caption{\textit{Local probes of quantum fields.} To probe relativistic quantum fields $\hat{E}$ locally in space and time, usually two-level Unruh--DeWitt detectors (here illustrated as atom) are considered (left hand side). While propagating through space and time, they locally interact via their dipole moment $\hat{d}$ with the quantum field $\hat{E}$ within the support of their smearing and switching function $\eta(\vec{r}, t)$ (orange shaded region). Analogously, in electro-optic sampling (right hand side), a focused, ultra-short laser pulse $\mathcal{E}$ (blue), propagating through a nonlinear crystal, induces a local interaction between the quantum field $\hat{E}$ and a co-propagating near-infrared probe mode $\hat{E}^{(1)}$ (black wavy arrow, the domain of the interaction is highlighted by the orange dashed line). After the interaction, the final state of the atom or the probe mode contains information about the quantum field localized to the space-time volume of the atoms or of the laser pulse inside the crystal, respectively.  }
\label{fig:Scheme} 
\end{figure}

The entanglement present in quantum fields can be swapped to local quantum probes (Unruh-DeWitt detectors, see Fig.~\ref{fig:Scheme}) in so-called entanglement-harvesting protocols \cite{valentini_non-local_1991,reznik_violating_2005,franson_generation_2008,salton2015acceleration,pozas-kerstjens_harvesting_2015,simidzija_harvesting_2018,henderson_harvesting_2018,tjoa_when_2021,gooding2023vacuum}: Two initially uncorrelated quantum systems (local probes), localized to different regions in space and time, can become entangled after locally interacting with a quantum field for a finite time. It was further shown that the entanglement harvested by the probes is distillable, i.e., the local probes can violate a Bell inequality after suitable local quantum operations and classical communication \cite{reznik_violating_2005,matsumura_violation_2020}. Two different processes can be distinguished which lead to correlations or entanglement between the local probes \cite{tjoa_when_2021,lindel2023separately}: the two systems can exchange source radiation (i.e., communicate with each other) or harvest entanglement from the quantum field. We will refer to these processes as communication-based or genuine entanglement harvesting, respectively. When both processes are present, we will speak of communication-assisted entanglement harvesting. If the two local probes remain space-like separated throughout the process, communication-based entanglement harvesting would require faster-than-light-signalling, making genuine entanglement harvesting the only possible source of the correlations \cite{tjoa_when_2021,lindel2023separately}. In this case, the generated entanglement between the local probes can only be due to entanglement pre-existing in the quantum field which is swapped to the two probes \cite{de2023entanglement}. Entanglement-harvesting protocols thus allow one to probe and thereby witness the space-time structure of quantum properties of general states of a quantum field \cite{de2023entanglement}, as well as providing a route to extracting these resources for possible use in quantum technologies \cite{martin-martinez_sustainable_2013}. This makes it one of the main workhorses in the field of relativistic quantum information \cite{mann2012relativistic}, which studies the interplay between relativity and quantum information theory. 

In electro-optic sampling (EOS) \cite{wu1995free,wu1996ultrafast}, tightly focused, coherent laser pulses propagate through a nonlinear crystal, in which they effectively induce an interaction between a localized probe-field mode (acting as an Unruh-DeWitt detector \cite{unruh1976notes,unruh1984happens,onoe2022realizing}) and the THz quantum field inside the crystal, see Fig.~\ref{fig:Scheme}. The interaction is switched on and off by the laser pulses entering and leaving the crystal. EOS was originally introduced to detect the field amplitude of classical THz fields on sub-cycle time-scales \cite{wu1995free,wu1996ultrafast,leitenstorfer1999detectors}. More recently, novel applications to THz quantum optics, such as, e.g., the detection of fluctuations in the (squeezed) vacuum \cite{riek_direct_2015,riek2017subcycle} and correlation measurements on the vacuum and thermal states \cite{benea-chelmus_electric_2019,settembrini2022detection} have been reported; chip-based implementations have been introduced \cite{benea2020electro,salamin2019compact}; and the possibility for full quantum-state tomography of the THz field state was discussed \cite{kizmann_quantum_2022,hubenschmid2022complete,hubenschmid2023optical,onoe2023direct,yang2023}. In Refs.~\cite{settembrini2022detection,onoe2022realizing,lindel2023separately}, EOS was identified as an experimental realization of local quantum probes, which interact with the electromagnetic THz field on subcycle time-scales similar to the Unruh--DeWitt detectors considered in generic entanglement-harvesting protocol, see Fig.~\ref{fig:Scheme}. It has been shown that using two probe beams, both genuine and communication-based two-point correlation harvesting from the polaritonic vacuum inside the nonlinear crystal is possible \cite{lindel2023separately}. The former has already been implemented experimentally, revealing the presence of correlations in the vacuum field between space-like separated regions \cite{settembrini2022detection}. So far, all considerations were limited to probing two-point correlation functions of the probe modes, and the question of whether \textit{quantum} correlations, i.e. entanglement, can be harvested remained open.

Here, we fill this gap by studying the harvesting of \textit{quantum} correlations in EOS experiments. By treating EOS in the interaction picture, we find the reduced density matrix of the two local probes after the interaction with a general quantum field inside the nonlinear crystal in Section~\ref{sec:InteractionPicture}. In Section \ref{sec:Correlations}, we connect this result to standard two-beam EOS measurements recovering previous results on two-point correlation harvesting from quantum-vacuum fluctuations. We further show that no additional two-point correlations can be harvested from displacing the vacuum state into a coherent state, and how thermal fluctuations alter the harvesting protocol. In the main Section of the manuscript, Section \ref{sec:Entanglement}, entanglement harvesting in EOS experiments is discussed. We evaluate the negativity as an entanglement measure, to find a sufficient condition for the existence of entanglement between the two probes in Section \ref{sec:EntanglementMeasure}. In Section \ref{sec:EntanglementWitness}, we construct an entanglement witness which is based on existing EOS detection schemes only, and which thus makes it possible to experimentally reveal the presence of the harvested entanglement. In Sections \ref{sec:EntanglementVacuum} and \ref{sec:EntanglementThermal}, we analyze entanglement harvesting in EOS for different quantum fields and find that communication-based as well as genuine entanglement harvesting from the vacuum can be achieved with state-of-the-art EOS experiments, while no entanglement can be harvested from thermal fluctuations. In Section \ref{sec:Bell}, we examine a Bell inequality which is based only on standard EOS correlation measurements, and show how distillable Bell nonlocality can be harvested from the vacuum in EOS. Also, in Appendix \ref{appsec:ShotNoiseRemoved}, we introduce a shot-noise removed method to probe fluctuations locally in a quantum field with a single-beam EOS setup. Here, this scheme is used to construct the entanglement witness, but it may find applications beyond that in the future.

\section{EOS in the interaction picture} \label{sec:InteractionPicture}

EOS with quantized fields is usually treated in the Heisenberg picture \cite{moskalenko_paraxial_2015,lindel2020theory,lindel2021macroscopic,lindel2022probing}. In this section, in order to discuss entanglement harvesting, an interaction-picture description of EOS is developed, which closely connects EOS to previous work on entanglement harvesting \cite{pozas-kerstjens_entanglement_2016,simidzija_harvesting_2018,tjoa_when_2021}.

\subsection{EOS Setup} \label{sec:EOSSetup}

\begin{figure}
\includegraphics[width=1.\columnwidth]{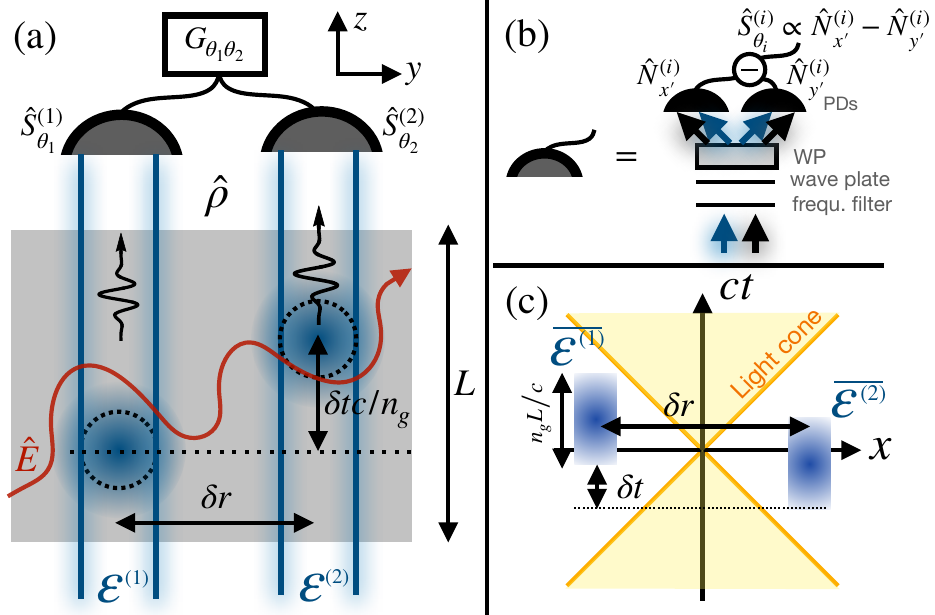}
\caption{(a) \textit{Two-beam EOS setup}. Two coherent laser pulses $\Ecli$ separated by $\delta r$ and $\delta t$ in space and time, respectively, propagate through a nonlinear crystal (gray rectangle) of length $L$. Via the nonlinear coupling they mix with the THz field (red) inside the crystal, to generate the probe field (black wave line) given by the density matrix $\hat{\rho}$. (b) Ellipsometry detection scheme: Each of the probe-field modes emerging from the crystal passes through a frequency filter, a wave plate inducing a relative phase $\theta_i$ between $\Ecli$ and the probe field, and a Wollaston prism (WP). The latter separates two orthogonal field-polarization directions $\hat{E}_{x^\prime}$ and $\hat{E}_{y^\prime}$, which are rotated with respect to $\hat{E}_{x}$ and $\hat{E}_{y}$, and which are then individually detected by photodetectors (PD). Eventually, the difference frequency current proportional to $\hat{N}_{x^\prime}^{(i)}- \hat{N}_{y^\prime}^{(i)}$ is recorded, see Appendix \ref{app:EOSSetup} for details. (c) Sketch of the two space-like separated laser pulses localized in space and spending a finite time $n_g L/c$ inside the crystal ($n_g$ is the group refractive index in the NIR).  }
\label{fig:Setup}
\end{figure}

The two-beam EOS setup under consideration here is illustrated in Fig.~\ref{fig:Setup} (a), and was experimentally realized in Refs.~\cite{benea-chelmus_electric_2019,settembrini2022detection}. It consists of two $y$-polarized, ultra-short, near-infrared (NIR) coherent pump laser pulses $\Eclo$ and $\Eclt$, which propagate through a nonlinear crystal. Inside the crystal, they mix via the nonlinear coupling with the $x$-polarized THz field $\hat{E} $ to generate excitations in the two $x$-polarized NIR probe fields $ \hat{E}^{(1)}$ and $ \hat{E}^{(2)}$. In contrast to the THz field $\hat{E} $, the near-infrared fields $ \hat{E}^{(1)}$ and $ \hat{E}^{(2)}$ can be treated (as the laser pulses) in the paraxial approximation \cite{lindel2020theory} and are thus co-propagating with the laser pulses $\Eclo$ and $\Eclt$, respectively. The effective interaction between the different fields is governed by a Hamiltonian \cite{onoe2022realizing,lindel2023separately}
\begin{align} \label{eq:HiEOS}
\hat{H}_{I}(t) = 2 \chi^{(2)} \sum_{i = 1,2} \int_{V_C} \dif ^3 r  \mathcal{E}^{(i)}(\vec{r}, t) \hat{E} (\vec{r}, t)  \hat{E}^{(i)} (\vec{r}, t),
\end{align}
where $ \chi^{(2)} $ is the nonlinear susceptibility of the crystal with volume $V_C$. We see in Eq.~\eqref{eq:HiEOS} that the laser pulses induce an effective interaction between $\hat{E} $ and the two field modes $\hat{E}^{(i)}$, which only takes place inside the two space-time volumes of the laser pulses inside the nonlinear crystal. The laser pulses are focused to a space-time volume much smaller than a THz wave-length. Thus, the two NIR field modes $\hat{E}^{(1)}$ and $ \hat{E}^{(2)}$, which are initially in their ground states, can be seen as local quantum probes that probe the THz field on subcycle time-scales within the space-time volume of the two laser pulses inside the crystal, respectively, see Fig.~\ref{fig:Setup}(c). $\hat{E}^{(1)}$ and $ \hat{E}^{(2)}$ thus play the role of the Unruh-DeWitt detectors considered in generic entanglement-harvesting protocols \cite{pozas-kerstjens_harvesting_2015}, see Fig.~\ref{fig:Scheme}.

We assume that the spatial parts of the probe fields and of the laser pulses are given by the lowest-order Laguerre-Gauß mode propagating in $z$ direction \cite{benea-chelmus_electric_2019,settembrini2022detection}. This was shown to be a suitable approximation for current experimental setups \cite{lindel2020theory}. The coherent amplitudes of the laser pulses are given by (see Appendix \ref{app:EOSSetup})
\begin{multline} \label{eq:laserPulse}
\Ecli(\vec{r}, t) = \sqrt{\frac{\hbar \omega_c N_p}{\eta}}\int \dif \omega \, g_{i}(\vec{r}_\parallel) \\
\times \mathcal{E}(\omega) \me^{\mi \omega \left(\frac{n(\omega) z}{c} -t - \delta t_i\right)},
\end{multline}
where $N_p$ is the total number of photons in each pulse, and $\eta = 4 \pi \epsilon_0 c n_c$, with $\epsilon_0$ the vacuum permittivity, $c$ the speed of light in vacuum, and $n_c = n(\omega_c) $ the refractive index [$n(\omega)$] at the central frequency of the laser pulses $\omega_c$. Both pulses are assumed to have the same Gaussian spectrum $\mathcal{E}(\omega)$ with width $\sigma_\omega$ leading to pulses of duration $\tau_\sigma = 2/\sigma_\omega$, and $g_{1}(\vec{r}_\parallel)  = g_{ 2}(\vec{r}_\parallel- \delta \vec{r}_\parallel)  = \sqrt{2/\pi w^2}  \me^{- r_\parallel^2/w^2}$ are their transversal Gaussian profiles with beam waist $w$. Furthermore, $\delta t_i $ and $ \delta \vec{r}_\parallel = \delta r \vec{e}_y$ are the time delay and spatial separation between the two laser pulses, see Fig.~\ref{fig:Setup}(a). The two probe modes co-propagating with the two laser pulses, respectively, read
\begin{multline} \label{eq:SignalFieldModeExp}
\hat{E}^{(i)}(\vec{r}, t) =  \mi  \int_{|\omega| > \Lambda}\hspace{-.5cm} \dif \omega \mathrm{sgn}[\omega] \sqrt{\frac{\hbar |\omega|}{4\pi \epsilon_0 n(\omega) c}} \\
\times g_{i}(\vec{r}_\parallel) \me^{\mi \omega \left(\frac{n(\omega) z}{c} -t - \delta t_i \right)} \hat{a}_i(\omega) ,
\end{multline}
where $\hat{a}_i(\omega)$ is the bosonic annihilation operator of the $x$-polarized lowest-order Laguerre-Gauss mode, and $\Lambda$ separates the NIR field from the THz frequency range. In the THz frequency range, dispersive and absorptive effects cannot be neglected in general, so we use macroscopic QED \cite{scheel2009macroscopic,buhmann2013dispersion} to find the quantized THz field $\hat{E}$ inside the nonlinear crystal, see App.~\ref{appsec:Polaritonic}.

Each probe field is detected by an ellispsometry detection scheme, compare Fig.~\ref{fig:Setup}(b). It relies on overlapping the $x$-polarized probe field with the $y$-polarized laser pulse, which acts as broadband local oscillator, to measure the probe-field \textit{amplitude}. It is thus similar to balanced homodyne detection \cite{loudon2000quantum}. For each mode, the resulting EOS signal reads \cite{kizmann_quantum_2022} (see Appendix \ref{app:EOSSetup} for details) 
\begin{align} \label{eq:SignalRed}
\hat{S}^{(i)}_{\Theta_i} = \sqrt{N_d} \left[ \mi P(\Theta_i) \hat{a}_i  -\mi P^{\ast}(\Theta_i) \hat{a}_i^\dagger \right],
\end{align}
where we defined new bosonic creation and annihilation operators of the probe fields 
\begin{align} \label{eq:creation}
\hat{a}_i = \sqrt{\frac{\eta}{N_d}} \int\limits_{\omega_\mathrm{min}}^{\omega_\mathrm{max}} \dif \omega \, \frac{\mathcal{E}(\omega)}{\sqrt{\hbar \omega}}  \hat{a}_{ i}(\omega) .
\end{align}
Here, $N_d$ is the number of photons in each of the two laser pulses within the frequency range $[\omega_\mathrm{min}, \omega_\mathrm{max}]$ that is not filtered out by the frequency filter. Furthermore, $\Theta_i = \{ \theta_i, \mathrm{sgn}[\alpha_i]\}$ where $\theta_i \in [\pi/2, 3\pi/2]$ is the phase shift induced by the wave-plate, $\mathrm{sgn}[\alpha_i]$ is the sign of the angle $\alpha_i$ of the fast axis of the waveplate against the $y$-axis, see Appendix \ref{app:EOSSetup}, and $P(\Theta_i) = \mathrm{sgn}[\alpha_i] [\sqrt{-\mathrm{cos}(\theta_i)} + \mi \sqrt{2} \mathrm{cos}(\theta_i/2)] $. Often we will set $\mathrm{sgn}[\alpha_i] = 1$ and thus define $P(\theta_i) \equiv P(\theta_i,\mathrm{sgn}[ \alpha_i] = 1) $.

\subsection{Reduced Density Matrix}

In this section, we obtain the state $\hat{\rho}$ of the two NIR probe fields emerging from the crystal. Before the laser pulses have entered the crystal, the probe and THz fields are in their vacuum states $ \hat{\rho}_\mathrm{vac}$ and $ \hat{\rho}_\mathrm{vac}^{(\mathrm{THz})}$, respectively. The mixing of the different fields inside the nonlinear crystal is given by the interaction Hamiltonian in Eq.~\eqref{eq:HiEOS}. The probe field emerging from the crystal is thus given by 
\begin{align} \label{eq:rhotGen}
\hat{\rho} = \mathrm{tr}_{\mathrm{THz}} \left\{ \hat{U} \hat{\rho}_\mathrm{vac} \otimes \hat{\rho}_\mathrm{vac}^{(\mathrm{THz})} \hat{U}^\dagger  \right\} .
\end{align}
with the time-evolution operator given by a time-ordered exponential
\begin{align} \label{eq:TimeEvo}
\hat{U} = \mathcal{T} \me^{-\mi \int_{-\infty}^{\infty} \dif t^\prime \hat{H}_I(t^\prime) /\hbar} .
\end{align}
In Eq.~\eqref{eq:rhotGen}, we traced out the THz field. Expanding the time-evolution operator in Eq.~\eqref{eq:TimeEvo} up to second order in $H_I$, we can evaluate Eq.~\eqref{eq:rhotGen} to obtain the multimode state of the NIR field emerging from the crystal, see Appendix \ref{app:RedDenstiy} for details. We represent the resulting state in the measurement basis of the EOS detection scheme $B_\text{EOS} =\{\ket{0,0} , \ket{1,0} , \ket{0,1} ,\ket{1,1}  ,\ket{2,0}, \ket{0,2}  \}$ with $\ket{n,m}$ being Fock states with respect to the bosonic operators $\hat{a}_1 $ and $\hat{a}_2 $ defined in Eq.~\eqref{eq:creation}. The state $\hat{\rho}$ expressed in the EOS basis contains all information necessary to describe EOS experiments and reads
\begin{align} \label{eq:rhoDensitym}
\hat{\rho} =   \left( \begin{array}{cccccc}
1- \! \! \! \!\sum\limits_{i=1,2} \! \! L_{ii}  - X & L_1^\ast & L_2^\ast & M^\ast & K_{11}^\ast  & K_{22}^\ast\\
L_1 & L_{11} & L_{12}^\ast & 0 & 0 & 0  \\
L_2 & L_{21}& L_{22}  & 0 & 0 & 0 \\
M & 0 & 0 & X &0 &0 \\
K_{11}& 0 & 0 &0 &0 &0 \\
K_{22} & 0 & 0 &0 &0 &0
\end{array} \right).
\end{align}
Here, we have defined
\begin{subequations} \label{eq:RhoEntries}
\begin{align} \label{eq:L01w} 
L_{i} & =  -   \int_{\vec{r},t}  F_i(\vec{r},t) \braket{\hat{E}(\vec{r},t)},\\ \label{eq:L12w}
L_{ij} & = \int_{\vec{r},\vec{r}^\prime, t,t^\prime} \hspace{-0.8cm}      F_i(\vec{r}, t)  F^\ast_j(\vec{r}^\prime, t^\prime) [  \ \mathcal{C}(\boldsymbol{\rho},\tau) - \mi \hbar  \mathcal{R}^{\prime\prime }(\boldsymbol{\rho},\tau) ], \\ \label{eq:Mw}
M  & =  \int_{\vec{r},\vec{r}^\prime, t, t^\prime} \hspace{-0.8cm}       F_1(\vec{r},t)  F_2(\vec{r}^\prime, t^\prime)  [  \ \mathcal{C}(\boldsymbol{\rho},\tau) - \mi \hbar  \mathcal{R}^\prime(\boldsymbol{\rho},\tau) ],  \\ 
K_{ii} & = \frac{1}{\sqrt{2}}\int_{\vec{r},\vec{r}^\prime, t,t^\prime} \hspace{-0.8cm}       F_i(\vec{r}, t)  F_i(\vec{r}^\prime, t^\prime) [\mathcal{C}(\boldsymbol{\rho},\tau)  - \mi \hbar  \mathcal{R}^\prime(\boldsymbol{\rho},\tau)  ], \\
X & = \mathcal{O}(\chi^{(2) 4}),
\end{align}
\end{subequations}
with the shorthand notation $\boldsymbol{\rho} = \vec{r}-\vec{r}^\prime$, $\tau = t-t^\prime$, $\int_{\vec{r}, t} = \int_{-\infty}^\infty \dif t \int_{V_C} \dif^3 r$, and we have introduced
\begin{multline} \label{eq:Fi}
F_i(\vec{r}, t)  =  2 \chi^{(2)}   \sqrt{\frac{2\pi L \omega_c }{\hbar \eta} } \overline{\Ecli}(\vec{r}, t)  g_i (\vec{r}_\parallel) \\
\times  \int_{\omega_\mathrm{min}}^{\omega_\mathrm{max}} \dif \omega  \mathcal{E}(\omega) \me^{\mi (\omega-\omega_c)(t + \delta t_i -n_g z/c)} ,
\end{multline} 
where $\overline{\Ecli}$ is the normalized pulse envelope, compare Eq.~\eqref{eq:PulseEnvelopes}, and  $n_g$ is the group refractive index in the NIR, see Appendix \ref{app:EOSSetup}. In case no frequency filtering is applied, we can set $\omega_\mathrm{min} = 0$ and $\omega_\mathrm{max} \to \infty $ to find the simplified expression
\begin{align} \label{eq:FiNofilter}
F_i(\vec{r}, t) = \frac{1}{2} \sqrt{C N_d} \;\overline{\Ecli} ^2 (\vec{r},t ) ,
\end{align}
with the detection efficiency 
\begin{align}
    \sqrt{C} = \frac{2 L \chi^{(2)}  \omega_c }{\epsilon_0 c n_c}.
\end{align}
Furthermore, in Eq.~\eqref{eq:RhoEntries}, we introduced the correlation $\mathcal{C}$ and response function $\mathcal{R}$ of the THz field 
\begin{align} \label{eq:CorrelationFunction}
\mathcal{C}(\boldsymbol{\rho}, \tau) & = \frac{1}{2}\langle \{ \hat{E}(\vec{r}, t ), \hat{E}(\vec{r}^\prime, t^\prime) \} \rangle, \\ \label{eq:ResponseFunction}
\mathcal{R}(\boldsymbol{\rho} ,\tau)  & = \frac{\mi }{\hbar}  \theta(\tau) \braket{ [\hat{E}(\vec{r}, t) , \hat{E}(\vec{r}^\prime, t^\prime) ] },
\end{align}
as well as the symmetric (reactive) and antisymmetric (dissipative) part of the response function $\mathcal{R}^\prime(\boldsymbol{\rho}, \tau) = [\mathcal{R}(\boldsymbol{\rho}, \tau) +\mathcal{R}(\boldsymbol{\rho}, -\tau)]/2$, and $\mathcal{R}^{\prime\prime}(\boldsymbol{\rho}, \tau) = [\mathcal{R}(\boldsymbol{\rho}, \tau) -\mathcal{R}(\boldsymbol{\rho}, -\tau)]/2$. While the correlation function in Eq.~\eqref{eq:CorrelationFunction} depends on the state of the THz field, the response function $\mathcal{R}$ is independent of it. In case of a crystal with constant refractive index in the THz $n$, it reads 
\begin{align}
 \mathcal{R}(\boldsymbol{\rho}, \tau) & = \frac{ \mu_0}{4 \pi   } \square_n \frac{1}{\rho}  \delta\left(\frac{\rho}{c_n}-\tau \right) , 
\end{align}
with $\square_n =\frac{\partial^2}{\partial t \partial t^\prime} - c_n^2 \frac{\partial^2}{\partial x \partial x^\prime}$, $\mu_0$ the vacuum permeability, and $c_n \equiv c/n$. We see that it vanishes outside the light-cone, i.e., for $\rho > c_n t$. As was shown in Ref.~\cite{lindel2023separately} for EOS and discussed in Ref.~\cite{tjoa_when_2021} for generic entanglement-harvesting protocols, its appearance in Eq.~\eqref{eq:RhoEntries} describes the exchange of source radiation between the space-time points $\vec{r}, t$ and $\vec{r}^\prime, t^\prime$. It thus governs communication-based processes, which are independent of the state of the THz field. All contributions proportional to the correlation function $\mathcal{C}$, on the other hand, can be attributed to genuine harvesting from the THz field. 
 
 Note that we assumed here that the THz field is stationary and homogeneous, such that $\mathcal{C}$ only depends on the space and time distances $\boldsymbol{\rho}$ and $\tau$. Extending our approach to inhomegeneous or non-stationary THz fields is straightforward, see Appendix \ref{app:RedDenstiy}. Also, in Eq.~\eqref{eq:rhoDensitym}, we included the density matrix element $X$. As $X$ is a term of at least fourth order in $\chi^{(2)}$, it can be neglected throughout most parts of this manuscript, except for Section \ref{sec:Bell} where it will be further discussed.

Equation \eqref{eq:rhoDensitym} is the reduced state of the probe modes emerging from the crystal. What can we learn from it about the THz quantum field? $L_i$ is proportional to the \textit{amplitude} of the THz field and is measured in standard EOS experiments \cite{wu1995free,wu1996ultrafast}: Without applying any frequency filtering ($\omega_\mathrm{min} = 0 $, $\omega_\mathrm{max} \to \infty$) and using a quarter-wave plate ($\theta = \pi/2$), we evaluate the EOS signal in Eq.~\eqref{eq:SignalRed} with the reduced density matrix in Eq.~\eqref{eq:rhoDensitym} and obtain
\begin{align}
    \braket{\hat{S}^{(i)}_{\frac{\pi}{2} } } = 2 \mathrm{Re}[L_i] =  \sqrt{C} N_d \int_{\vec{r},t}  \overline{\Ecli} ^2 (\vec{r},t ) \braket{\hat{E}(\vec{r},t)} .
\end{align}
Such a measurement can thus be used to locally probe the field \textit{amplitude} $\braket{\hat{E}(\vec{r},t)}$ of the THz field inside the space-time volume of the laser pulses $\overline{\Ecli} ^2 (\vec{r},t ) $. Furthermore, we can use the reduced density matrix in Eq.~\eqref{eq:rhoDensitym} to analyze whether the two probe modes are correlated, entangled or can violate a Bell inequality after emerging from the crystal, to see if EOS can be used for correlation, entanglement or nonlocality harvesting from the THz field. These three scenarios are further discussed in Sections \ref{sec:Correlations}, \ref{sec:Entanglement}, and \ref{sec:Bell}, respectively.

\section{Two-Point Correlations} \label{sec:Correlations}

We here analyze how two-point correlations are genuinely harvested and probed in two-beam EOS experiments. This has been discussed already in previous works \cite{benea-chelmus_electric_2019,lindel2021macroscopic,settembrini2022detection,lindel2023separately} by considering the correlations between the two EOS signals \footnote{Note that we do not normalize the signal by the detection efficiency as in Refs.~\cite{benea-chelmus_electric_2019,lindel2021macroscopic,settembrini2022detection,lindel2023separately}. Also, we included the term $\propto -\langle \hat{S}^{(1)}(\theta_1) \rangle \langle \hat{S}^{(2)}(\theta_2)   \rangle$ as in Ref.~\cite{lindel2023separately}, which vanishes in second order in $\chi^{(2)}$ in case the THz field is in its vacuum or thermal state, as considered in Refs.~\cite{benea-chelmus_electric_2019,lindel2021macroscopic,settembrini2022detection}. }
\begin{align} \label{eq:SignalG}
 G_{\theta_1 \theta_2} = \frac{1}{N_d^2}\left(\langle \hat{S}^{(1)}_{\theta_1}  
 \hat{S}^{(2)}_{\theta_2}   \rangle
  -\langle \hat{S}^{(1)}_{\theta_1} \rangle \langle \hat{S}^{(2)}_{\theta_2}   \rangle \right).
\end{align}
We are only interested in genuine correlation harvesting probing correlations present in the THz field, and not in the THz-field independent, communication-based harvesting. It was shown previously that this is achieved by using two quarter wave-plates in the detection setup, i.e., $\theta_{1,2} = \pi/2$ \cite{lindel2023separately}. We thus evaluate the EOS correlation signal in Eq.~\eqref{eq:SignalG} with $\theta_{1,2} = \pi/2$ using the density matrix in Eq.~\eqref{eq:rhoDensitym}. We neglect any frequency filtering of the detected field modes by setting $\omega_\mathrm{min} = 0$ and $\omega_\mathrm{max}  \to \infty$ such that we can use Eq.~\eqref{eq:FiNofilter}, and find
\begin{align} \label{eq:GenuineCorrHarv}
G_{\frac{\pi}{2}\frac{\pi}{2}}   &  =   \frac{2}{N_d} ( \mathrm{Re}[ M + L_{12}  ] -2 L_{1}L_{2})   \\ \label{eq:EOSCorrelationsField}
& = C\int_{\vec{r},\vec{r}^\prime, t, t^\prime} \hspace{-0.8cm}     \overline{\Eclo}^2(\vec{r}, t)  \overline{\Eclt}^2(\vec{r}^\prime, t^\prime) \overline{\mathcal{C}}(\vec{r}, \vec{r}^\prime, \tau).
\end{align}
Here, $\overline{\mathcal{C}}(\vec{r}, \vec{r}^\prime, \tau) =\mathcal{C}(\vec{r}, \vec{r}^\prime, \tau)   -  \braket{ \hat{E}(\vec{r}^\prime, t^\prime)} \braket{ \hat{E}(\vec{r}, t) } $. Equation \eqref{eq:GenuineCorrHarv} shows that $G_{\frac{\pi}{2}\frac{\pi}{2}} $ is given by the correlation function of the THz field $\overline{\mathcal{C}}$ averaged over the two space-time envelopes of the laser pulses. This generalizes the results of Ref.~\cite{lindel2023separately} to probing quantum fields that do not vanish on average. In the limit $\braket{\hat{E}(\vec{r},t)} = 0$ our result in Eq.~\eqref{eq:GenuineCorrHarv} reduces to the one obtained in Ref.~\cite{lindel2023separately}, showing the consistency of our interaction picture calculation with the Heisenberg picture approach employed in Refs.~\cite{moskalenko_paraxial_2015,lindel2020theory,lindel2023separately}. In the following, $\overline{\mathcal{C}}$ and $G_{\frac{\pi}{2}\frac{\pi}{2}}  $ are evaluated for three exemplary THz states.

\subsection{Vacuum State}

We briefly recap correlation harvesting from the vacuum inside the nonlinear crystal as previously discussed in Ref.~\cite{benea-chelmus_electric_2019,settembrini2022detection,lindel2021macroscopic,lindel2023separately}. For the THz field in the vacuum state, one finds that $\braket{ \hat{E}(\vec{r},t)} = 0$ such that $\overline{\mathcal{C}}_\mathrm{vac} = \mathcal{C}_\mathrm{vac}$ with \cite{lindel2023separately}
\begin{align} \label{eq:CorrelationVacuumClean}
\mathcal{C}_\mathrm{vac}(\vec{r}, \vec{r}^\prime, \tau) 
& = \frac{\mu_0 \hbar}{8 \pi^2   } \square_n \frac{1}{\rho} \left(\frac{\mathcal{P}}{\frac{\rho}{c}- \tau} + \frac{\mathcal{P}}{\frac{\rho}{c}+ \tau}\right),
\end{align}
where $\mathcal{P}$ denotes the principal value. To obtain Eq.~\eqref{eq:CorrelationVacuumClean}, we have neglected dispersion and absorption effects in the nonlinear crystal by setting $n(\Omega) = n \in \mathbb{R}$. For the more general case of a complex, frequency-dependent refractive index $n(\Omega)$, we use macroscopic quantum electrodynamics \cite{scheel2009macroscopic,buhmann2013dispersion} to find the two-point correlation function of the polaritonic vacuum inside the nonlinear crystal (see Appendix \ref{appsec:Polaritonic}).

\begin{figure}
\includegraphics[width=1.\columnwidth]{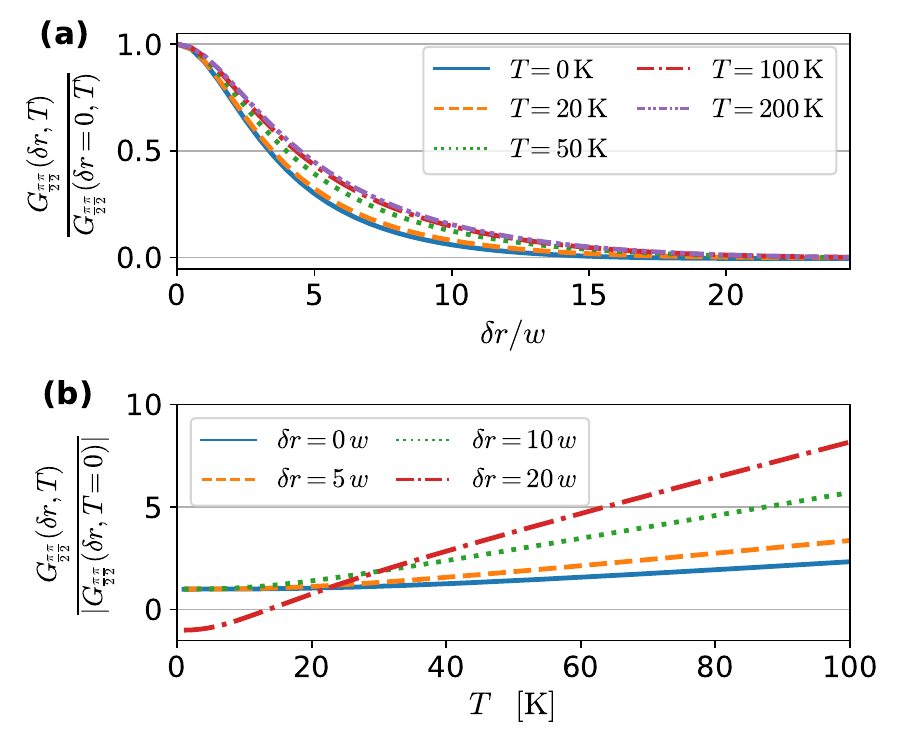}
\caption{\textit{Two-point correlation harvesting from thermal fields.} We plot the two-beam EOS correlation signal $G_{\frac{\pi}{2}\frac{\pi}{2}}$ as a function of the beam separation $\delta r$ [(a)] and temperature $T$ [(b)] in case the THz field is initially in a thermal state. We fix all parameters to those realized experimentally in Ref.~\cite{settembrini2022detection}, i.e., $w = 10\,\mu$m, $\tau_\sigma = 195\,$fs, $L = 1\,$mm, $n_g = 3.18$. We further used a Drude-Lorentz model for the refractive index in the THz frequency range $n(\Omega)$ with the parameters measured in Ref.~\cite{leitenstorfer1999detectors}, with the only exception that we use $\epsilon_\infty = 7.38 $ to match the experimental data of the real part of the refractive given in Ref.~\cite{settembrini2022detection}. For simplicity, we neglect the temperature dependence of the refractive index. Note that $T=0$ indicates that the THz field is in its vacuum state and that $G_{\frac{\pi}{2}\frac{\pi}{2}}(\delta r, T)$ can become negative for large beam separations.   }
\label{fig:Thermal}
\end{figure}

We see from Eq.~\eqref{eq:CorrelationVacuumClean} that there are correlations between space-like separated space-time regions, between which no communication via source radiation is possible. Thus, also for space-like separated laser pulses and the THz field in the vacuum state, two-point correlations are induced between $\hat{E}^{(1)}$ and $\hat{E}^{(2)}$. This has been experimentally observed in Ref.~\cite{settembrini2022detection}, see also Fig. \ref{fig:Thermal}.

\subsection{Coherent State}

We define a general multimode coherent state via its classical amplitude $\boldsymbol{\alpha}(\vec{r}, \Omega)$:
\begin{align} \label{eq:CoherentState}
\ket{\{\boldsymbol{\alpha}(\vec{r}, \Omega)\}} \equiv  \me^{\int_{-\infty}^\infty \dif \Omega \int \dif^3 r \boldsymbol{\alpha}(\vec{r}, \Omega) \cdot \hat{\vec{f}}(\vec{r}, \Omega) - \mathrm{h.c.}} \ket{\{0\}} .
\end{align}
This is an eigenstate of the polaritonic annihilation operator $\hat{\vec{f}}(\vec{r}, \Omega)$ \cite{scheel2009macroscopic,buhmann2013dispersion,lindel2021macroscopic} (see Appendix \ref{appsec:Polaritonic})
such that 
\begin{align}
\braket{\{\boldsymbol{\alpha}(\vec{r}, \Omega)\} |\hat{E}(\vec{r}, t) | \{\boldsymbol{\alpha}(\vec{r}, \Omega)\} } = E(\vec{r}, t).
\end{align}
Here, $E(\vec{r}, t) $ is the $x$-component of the classical electric field amplitude with all polaritonic creation and annihilation operators replaced by $\boldsymbol{\alpha}(\vec{r}, \Omega)$ and $\boldsymbol{\alpha}^\ast(\vec{r}, \Omega) $, respectively. For the coherent state in Eq.~\eqref{eq:CoherentState}, one finds
\begin{align} \label{eq:CorrCoh}
\mathcal{C}_\mathrm{coh}(\vec{r}, \vec{r}^\prime, \tau) = \mathcal{C}_\mathrm{vac}(\vec{r}, \vec{r}^\prime, \tau)  +   E(\vec{r}, t) E(\vec{r}^\prime, t^\prime) ,
\end{align}
such that 
\begin{align} \label{eq:CRpCoh}
\overline{\mathcal{C}}_\mathrm{coh}(\vec{r}, \vec{r}^\prime, \tau) = \overline{\mathcal{C}}_\mathrm{vac}(\vec{r}, \vec{r}^\prime, \tau).
\end{align}
It immediately follows from Eq.~\eqref{eq:EOSCorrelationsField} that
\begin{align}
G_{\theta_1 \theta_2}^{(\mathrm{coh})} & = G_{\theta_1 \theta_2}^{(\mathrm{vac})} .
\end{align}
As the vacuum state itself is a coherent state with $\boldsymbol{\alpha}(\vec{r}, \Omega) = 0$, this shows that the same amount of correlations can be harvested from all coherent states. In other words, coherent states \textit{resolved in space and time} consist of completely uncorrelated photons, except for their vacuum contribution.

\subsection{Thermal State}

A thermal state is defined as 
\begin{align} \label{eq:ThermalState}
\hat{\rho}_T = \frac{\me^{-\hat{H}_F/(k_B T )}}{\mathrm{tr}[\me^{-\hat{H}_F/(k_B T )}]},
\end{align}
where $\hat{H}_F$ is the free-field Hamiltonian, $k_B$ the Boltzmann constant and $T$ the temperature. For a thermal state $\hat{\rho}_T$ inside the nonlinear crystal, accounting for dispersion and absorption effects via the permittivity $\epsilon(\Omega)$, one finds using macroscopic quantum electrodynamics \cite{scheel2009macroscopic,buhmann2013dispersion}
\begin{multline} \label{eq:CorrelationTemp}
\overline{\mathcal{C}}(\vec{r}, \vec{r}^\prime, \tau) =  \frac{\mu_0 \hbar}{\pi} \int_{0}^\infty \dif \Omega \Omega^2 [2 n_T(\Omega) + 1] \\
\times  \mathrm{Im}[\mathsf{D}(\vec{r}, \vec{r}^\prime, \Omega)] \mathrm{cos}[\Omega \tau].
\end{multline}
Here, $\mathsf{D}$ is the $xx$-component of the Green tensor of the vector Helmholtz equation, see Appendix \ref{appsec:Polaritonic} for details, and $n_T(\Omega)$ is given by
\begin{align} \label{eq:ThermalDistr}
 n_T(\Omega) = \frac{1}{\me^{\hbar \Omega/k_B T}- 1}.
\end{align}
As $n_T(\Omega) $ increases monotonically with $T$, we expect that also $\overline{\mathcal{C}}$ increases with temperature. Inserting Eq.~\eqref{eq:CorrelationTemp} into Eq.~\eqref{eq:EOSCorrelationsField}, we obtain the EOS signal in case of a thermal THz field, which is further evaluated in Appendix \ref{appsec:EvaluateThermal}. In Fig.~\ref{fig:Thermal} (a), we plot $G_{\frac{\pi}{2}\frac{\pi}{2}}$ as a function of the beam separation $\delta r$ for different temperatures, and in Fig.~\ref{fig:Thermal} (b) as a function of temperature for different beam separations. Note that we recover two-point correlation harvesting from the vacuum for $T= 0$. In Fig.~\ref{fig:Thermal} (a) and (b) we see that for $T=0$ the correlation signal decreases with the beam separation. It even becomes negative for large beam separation $\delta r $ [hardly visible in Fig.~\ref{fig:Thermal}~(a)]. In general, the presence of thermal fluctuations does not change this qualitative behavior but leads to an increase in the amount of harvested correlations, showing that two-point correlations can be harvested from thermal fluctuations. 
Remarkably, there is a regime, e.g., for $\delta r = 20w$, in which the vacuum fluctuations lead to anti-correlations which get canceled by the positive correlations harvested from thermal fluctuations, such that overall no correlations are harvested and we find $G_{\frac{\pi}{2}\frac{\pi}{2}}(T = 14\, \mathrm{K}) = 0$.

\section{Entanglement} \label{sec:Entanglement}

In this section, we introduce EOS as an experimentally realizable platform for entanglement harvesting. First, we examine the negativity as an \textit{entanglement measure}, which offers a sufficient condition to show that entanglement was harvested by the probe modes. We then construct an entanglement witness, i.e., an experimentally accessible observable capable of detecting the entanglement present in the probe field emerging from the crystal. In Sec.~\ref{sec:EntanglementVacuum} and \ref{sec:EntanglementThermal}, we show that entanglement can be harvested from the vacuum in state-of-the-art EOS experiments via genuine or communication-based harvesting, respectively, while no entanglement harvesting is possible from thermal fluctuations.

\subsection{Entanglement Measure} \label{sec:EntanglementMeasure}

The reduced density matrix of the probe modes in Eq.~\eqref{eq:rhoDensitym} has a very similar structure to the state of the local probes found in previous works on entanglement harvesting \cite{tjoa_harvesting_2020}. We thus closely follow Ref.~\cite{tjoa_harvesting_2020} and use the negativity $\mathcal{N}$ as an entanglement measure for the mixed state of the two field modes emerging from the crystal in Eq.~\eqref{eq:rhoDensitym}. It is defined by 
\begin{align}
\mathcal{N} = \frac{|| \hat{\rho}^{ T_2} ||_1 -1}{2},
\end{align}
where $||\cdot ||_1$ is the trace norm, and $\rho^{T_2}$ is the partial transpose of $\hat{\rho}$. As an entanglement measure, $\mathcal{N}$ quantifies the amount of entanglement contained in the bipartite quantum state $\rho$. $\mathcal{N}$ vanishes for all separable states, such that $\mathcal{N}> 0 $ is a sufficient condition that the underlying state is entangled. For the state in Eq.~\eqref{eq:rhoDensitym}, we find in second-order in the nonlinear coupling $\chi^{(2)}$ \cite{tjoa_when_2021}
\begin{subequations} \label{eq:Negativity} 
\begin{align} 
\mathcal{N} & = \mathrm{max}[0,  E_1], \\ 
E_1 & =  |\bar{M}| -\bar{L}_{11},
\end{align}
\end{subequations}
with
\begin{align} \label{eq:Mbar}
\bar{M} = M - L_{1} L_{1} , \\ \label{eq:Liibar}
\bar{L}_{11} = L_{11}  - |L_{1}|^2. 
\end{align}
Here, we used that, for homogeneous, stationary THz fields and equivalent laser pulses up to spatial and temporal relative shifts, we have $L_{11} = L_{22}$ and $L_1 = L_2$. 

We can immediately conclude from the structure of $\bar{M}$ and $\bar{L}_{11}$ in Eqs.~\eqref{eq:Mbar} and \eqref{eq:Liibar}, respectively, that the same amount of entanglement can be harvested from all coherent states---including the vacuum state. This shows that no additional entanglement can be harvested if the vacuum field is coherently displaced. To see this, we evaluate the negativity for a coherent state \eqref{eq:CoherentState} by inserting the expressions for $L_{i}$, $L_{ii}$ and $M$ in Eqs.~\eqref{eq:L01w}, \eqref{eq:L12w}, and \eqref{eq:Mw}, respectively, into Eq.~\eqref{eq:Mbar} and \eqref{eq:Liibar}. Using Eq.~\eqref{eq:CorrCoh}, we find that $\bar{M}$ and $\bar{L}_{11}$ and thus also the negativity $\mathcal{N}$ are independent of the coherent amplitude $\boldsymbol{\alpha}(\vec{r}, \Omega)$. 

For the thermal and vacuum states, which we will consider exclusively in the following, we find $L_{1} = L_2  = 0$ such that $E_1 =   L_{11} - |M|^2$. Thus, whether the state is entangled or not depends on the interplay between the local generation of a photon ($ L_{11} $) and the coherences between $\ket{00}$ and $\ket{11}$ ($|M|$). This is again in close analogy to the case of generic entanglement-harvesting protocols, see for example Refs.~\cite{reznik_violating_2005,pozas-kerstjens_harvesting_2015}. We can further distinguish communication-based and genuine entanglement harvesting following Refs.~\cite{tjoa_when_2021,lindel2023separately}. The former (latter) is given by all terms proportional to the THz state independent response function $\mathcal{R}$ (THz state dependent correlation function $\mathcal{C}$) in the expression of $M$ in Eq.~\eqref{eq:Mw}, which is a purely imaginary (real) contribution to $M$, see also discussion after Eq.~\eqref{eq:ResponseFunction}. Thus, we can quantify whether the entanglement harvesting is genuine ($\phi_M = 1$) or communication-based ($\phi_M = 0$) via $\phi_M \equiv \mathrm{Re}[\me^{\mi \varphi_M}]$, where the argument $\varphi_M$ of $M$ is defined by $M/|M| \equiv \me^{\mi \varphi_M}$.

\subsection{Entanglement Witness}  \label{sec:EntanglementWitness}

Evaluating the entanglement measure of Eq.~\eqref{eq:Negativity} requires the complete knowledge of the underlying quantum state and, hence, the experimentally-challenging quantum state tomography of the unknown state. An entanglement witness $\hat{\mathcal{W}}$, on the other hand,is an observable of a bipartite system, whose expectation value is positive for all separable states, while there exists at least one entangled state with $\braket{\hat{W}} < 0 $. Thus, experimentally observing $\braket{\hat{W}} < 0 $ allows one to witness the non-separability of the underlying state. In Appendix \ref{app:EntanglementWintess}, we construct the following entanglement witness for the state of the two probe modes emerging from the crystal
\begin{align} \label{eq:WitnessLocal}
\hat{\mathcal{W}} =\frac{N_d}{4} (  G_{\Theta_{\varphi_M} , \Theta_{\varphi_M}} -  G_{ \Theta_{\varphi_M}^\prime , \Theta_{\varphi_M}^\prime }  )
+  \hat{a}_1^\dagger \hat{a}_1+  \hat{a}_2^\dagger \hat{a}_2 .
\end{align}
Here, $\Theta_{\varphi_M}$ and $\Theta_{\varphi_M}^\prime$ are defined via $P(\Theta_{\varphi_M}) = \me^{-\mi \varphi_M}$ and $P(\Theta_{\varphi_M}^\prime) = \mi \me^{-\mi \varphi_M}$, which can always be inverted to find $\Theta_{\varphi_M}$ and $\Theta_{\varphi_M}^\prime$. By construction, we find $\hat{\mathcal{W}}>0$ for all separable states, while for the state of the two probe modes $\hat{\rho}$ in Eq.~\eqref{eq:rhoDensitym}, we find $\braket{\hat{\mathcal{W}}} = L_{11} - |M|$. Thus, whenever we find that $\hat{\rho}$ is entangled according to the negativity, this is witnessed by $\hat{\mathcal{W}}$ in Eq.~\eqref{eq:WitnessLocal}.

The entanglement witness in Eq.~\eqref{eq:WitnessLocal} decomposes into four measurements, which can be individually performed: The first two terms in Eq.~\eqref{eq:WitnessLocal} correspond to two standard two-beam EOS correlation measurements [see Eq.~\eqref{eq:SignalG}] with $\theta_{1,2} = \Theta_{\varphi_M} $ and $ \theta_{1,2}= \Theta_{\varphi_M}^\prime$, respectively. Such measurements have been experimentally realized in Ref.~\cite{settembrini2022detection} for $\theta_{1,2}= \pi/2 $. The last two terms in Eq.~\eqref{eq:WitnessLocal} describe photon counting of $x$-polarized photons in the EOS basis modes $1$ and $2$, respectively. In principle, this could be achieved by using polarizing beam splitters to separate the coherent $y$-polarized laser pulses from the generated $x$-polarized probe photons and do a number resolved photodetection of the latter. In practice, however, the extinction ratio of the polarizer would have to be on the order of $L_{11}/N_{p}$ ($N_p$: total number of photons in each of the laser pulses). For the experimental setup used in Refs.~\cite{benea-chelmus_electric_2019,settembrini2022detection}, $L_{11}/N_{p}$ is on the order of $10^{-13}$. A more feasible alternative is offered by using single-beam EOS as performed in Refs.~\cite{riek_direct_2015,sulzer2020determination}. This allows one to access $\braket{\hat{a}_i^\dagger \hat{a}_i}$ on top of the shot noise of the laser pulses, see Appendix \ref{appsec:SingleStand}. In Appendix \ref{appsec:ShotNoiseRemoved}, we also introduce 
a shot-noise free detection scheme of $\hat{a}_i^\dagger \hat{a}_i$. Here, one of the laser pulses enters a beam splitter with reflectivity $R = \mi/\sqrt{2}$ and transmittivity $T = 1/\sqrt{2}$. The reflected and transmitted field enter two separate ellipsometry setups with respective wave-plates $\theta_R$ and $\theta_T$. The basic idea is that there is no shot-noise contribution in the correlation signal between the two ellipsometry setups as the noise in the two output ports of the beam splitter is uncorrelated. This might offer a valuable extension of general single-beam EOS correlation measurements considered, e.g., in Ref.~\cite{riek_direct_2015,sulzer2020determination,hubenschmid2022complete,onoe2022realizing}. The signal measured in this shot noise-removed single-beam EOS setup is derived in Appendix \ref{appsec:ShotNoiseRemoved} and it reads
\begin{multline} \label{eq:ShotNoiseFreeSignal}
    G^{(\mathrm{SB}i)}_{\theta_R \theta_T} = \frac{1}{N_d} ( P^\ast(\theta_R)P(\theta_T) \braket{\hat{a}_i^\dagger\hat{a}_i } \\
    - P(\theta_R)P(\theta_T) \braket{\hat{a}_i\hat{a}_i } + \mathrm{h.c.} ) .
\end{multline}
We find that one can access $\braket{\hat{a}_i^\dagger\hat{a}_i}$ without additional shot-noise contribution via two measurements 
\begin{align} \label{eq:aDagaShotNoiseFree}
    G^{(\mathrm{SB}i)}_{\frac{\pi}{2}\frac{\pi}{2}} +  G^{(\mathrm{SB}i)}_{\pi \pi} = \frac{4}{N_d} \braket{\hat{a}_i^\dagger\hat{a}_i }.
\end{align}
In total, we thus find
\begin{multline} \label{eq:WitnessMeas}
\braket{\hat{\mathcal{W} }} =\frac{N_d}{4} ( G_{\Theta_{\varphi_M} , \Theta_{\varphi_M}} -  G_{\Theta_{\varphi_M}^\prime ,  \Theta_{\varphi_M}^\prime}  
 \\
 + \left. \sum_{i =1,2} [ G^{(\mathrm{SB}i)}_{\frac{\pi}{2}\frac{\pi}{2}} +  G^{(\mathrm{SB}i)}_{\pi \pi} ] \right. ).
\end{multline}
To conclude, whenever entanglement has been harvested by the two probe modes according to the negativity entanglement measure, it can be witnessed by using a combination of standard two-beam correlation and (shot-noise removed) single-beam EOS measurements.

\subsection{Vacuum State}  \label{sec:EntanglementVacuum}

\begin{figure}
\includegraphics[width=1.\columnwidth]{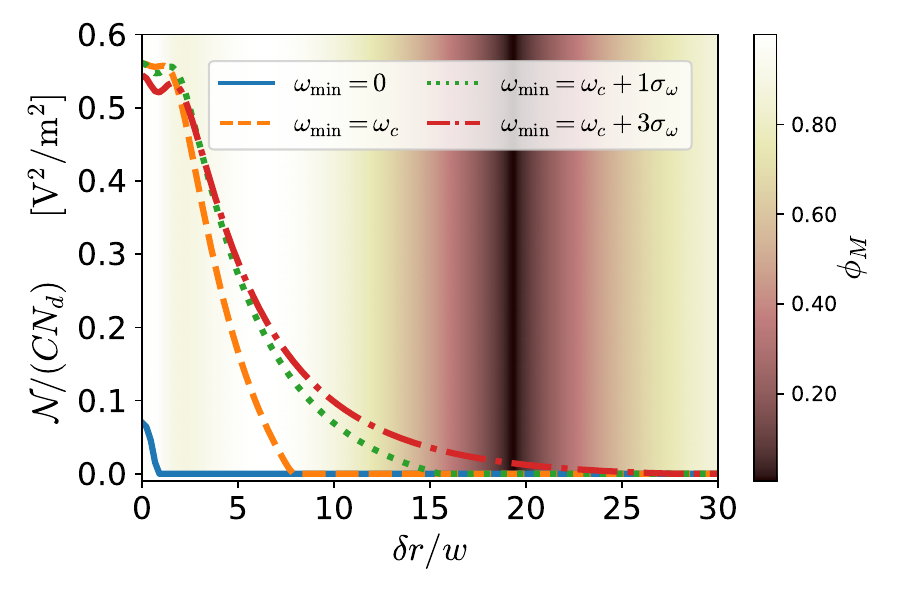}
\caption{\textit{Entanglement harvesting from the vacuum.} We show the negativity as a function of the beam separation $\delta r$ normalized by the beam width $w$ of the laser pulses in case the THz field is initially in its vacuum state. We vary the frequency of the high-pass filter $\omega_\mathrm{min}$, use $\delta t = 0$, and all other experimental parameters are fixed to the ones realized experimentally in Ref.~\cite{settembrini2022detection}, see caption of Fig.~\ref{fig:Thermal}. We distinguish genuine ($\phi_M = 1$) and communication-based ($\phi_M=0$) harvesting via the color of the background. $\phi_M$ is calculated for $\omega_\mathrm{min} = \omega_c + 3 \sigma_\omega$, but is qualitatively the same for the other values of $\omega_\mathrm{min}$ considered here.  }
\label{fig:Negativity}
\end{figure}

To analyze if entanglement can be harvested from the vacuum field in EOS experiments, we evaluate the negativity in Eq.~\eqref{eq:Negativity} with $E_1 = |M|-L_{11}$ for a vacuum THz field. This is done in Appendix \ref{appsec:EvaluateVacuum} by inserting the vacuum two-point correlation and response function into the expressions for $L_{11}$ and $M$ in Eqs.~\eqref{eq:L12w} and \eqref{eq:Mw}, respectively, and carrying out the integrals over the time and spatial coordinates. We use the same experimental parameters as in Ref.~\cite{settembrini2022detection}, see caption of Fig.~\ref{fig:Thermal}.  As a result, we show in Fig.~\ref{fig:Negativity} the negativity as a function of the beam separation for $\delta t=0$ and different high-pass frequency filterings $\omega_\mathrm{min}$ ($\omega_\mathrm{max} $ is set to infinity). The color in the background of Fig.~\ref{fig:Negativity} shows the parameter $\phi_M = \mathrm{Re}[M]/|M| $ as a function of the beam separation, which indicates whether the harvesting is genuine ($\phi_M = 1$), communication-based ($\phi_M = 0$), or communication assisted ($0<\phi_M <1$), see discussion at the end of Sec.~\ref{sec:EntanglementMeasure}.  

We find that in case all frequencies are detected ($\omega_\mathrm{min} \to 0$), entanglement is only harvested if the beams are overlapping ($\delta r < w$). By increasing the cut-off frequency of the high-pass filter $\omega_\mathrm{min}$, entanglement is harvested also for increasing values of $\delta r$. When choosing the lower frequency cut-off $\omega_\mathrm{min}$, there is thus a trade-off between finding entanglement for larger values of $\delta r$ and measuring enough photons to obtain a high enough signal to noise ratio.  Interestingly, for different beam separations the harvested entanglement can be genuine, communication-based or communication-assisted, such that all three processes can be accessed by changing $\delta r$. As was shown in Ref.~\cite{settembrini2022detection}, already for $\delta r \ge 5 w$ the majority of the accessed field correlations are ones between space-like separated regions, such that entanglement harvesting from space-like separated vacuum-field fluctuations is possible with state-of-the-art EOS setups. Note, however, that in this regime the laser pulses are still not entirely space-like separated, as an exchange of source radiation is still possible up to $\delta r \approx 28 w $ for the parameters under consideration, see Fig.~\ref{fig:Negativity}.

\begin{figure}
\includegraphics[width=1.\columnwidth]{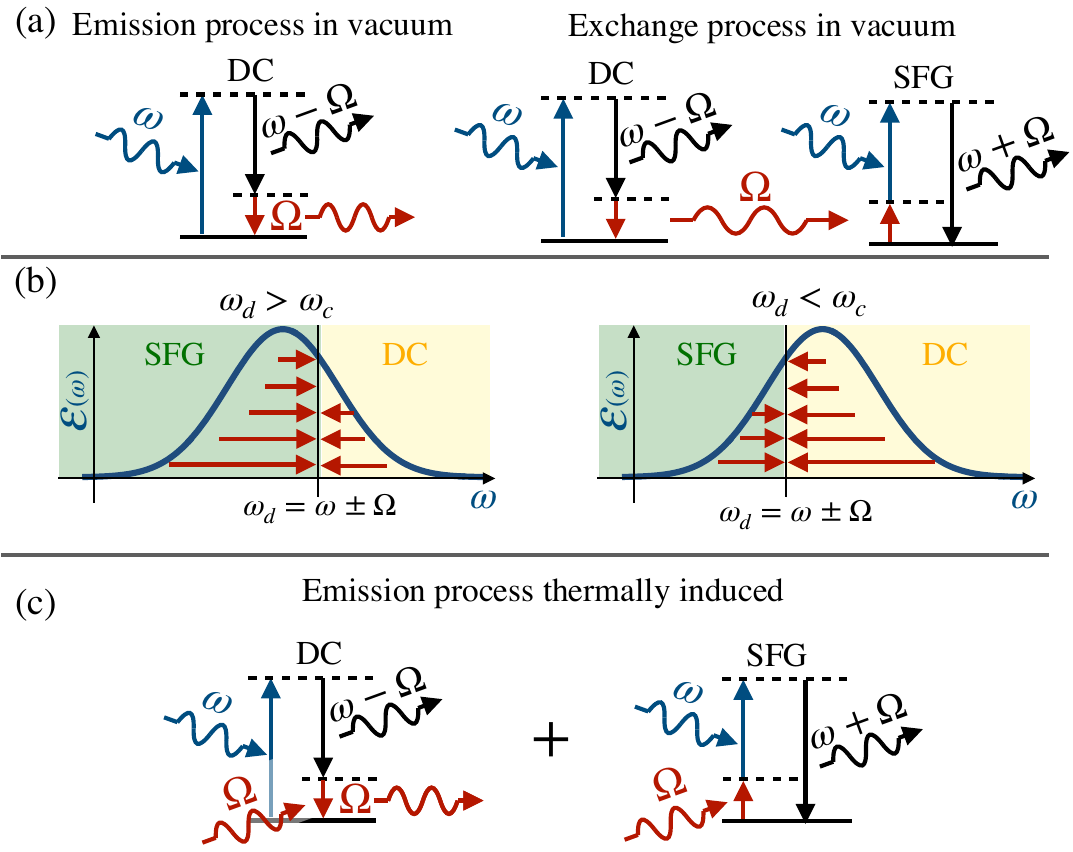}
\caption{\textit{Processes leading to $L_{ii}$ and $M$.} $L_{ii}$ corresponds to the local emission of a photon, whereas $M$ to an exchange process between the two probe modes. (a) Emission and exchange processes with a vacuum THz field. As no energy can be extracted from the vacuum, a photon can only be generated via a down-conversion (DC) process, in which a laser photon (blue) is down-converted to a THz (red) and a probe photon (black). The exchange term is a result of a sum-frequency generation (SFG) and a DC process. Note that not necessarily a real photon (source radiation) must be exchanged. (b) If the frequency of the detected probe photon $\omega_d$ is larger (smaller) than the central frequency of the laser pulses $\omega_c$, SFG (DC) processes dominate over DC (SFG) processes, as a larger part of the spectrum of the laser pulses $\mathcal{E}(\omega)$ is available for the process. (c) If thermal fluctuations are present, a photon can be generated by either an induced DC or a SFG process, in which energy is extracted from the thermal field. The thermally induced exchange processes (not shown) leading to $M$ still consist of a SFG and a DC process.}
\label{fig:Processes}
\end{figure}

To understand the dependence of the negativity on the cut-off frequency of the high-pass filter $\omega_\mathrm{min}$, we analyze the underlying nonlinear processes leading to $L_{11}$ and $|M|$. $L_{ii}$ describes the local emission of a photon in probe mode $i$ via a nonlinear process (sum-frequency generation (SFG) or down-conversion (DC) process), see left hand side Fig.~\ref{fig:Processes} (a). As no energy can be extracted from the vacuum, only the down-conversion process can lead to the emission of a photon for a vacuum THz field. Here, a photon from the laser pulse with frequency $\omega$ is down-converted to a THz photon with frequency $\Omega$ and a probe photon with frequency $\omega-\Omega$.  
$|M|$ is the result of an exchange process between the two probe modes, in which one laser pulse undergoes sum-frequency generation, while the other takes part in a down-conversion process, see right hand side of Fig.~\ref{fig:Processes} (a). No THz photon is emitted overall in this process, and energy is only conserved globally. The high-pass filter leads to a suppression of the down-conversion over the sum-frequency generation process, as a photon with a higher (lower) frequency than the detected photon frequency $\omega_d$ is needed to initiate the former (latter) process \cite{sulzer2020determination}, see left hand side of Fig.~\ref{fig:Processes} (b). The situation would be reversed for a low-pass filter as illustrated on right hand side of Fig.~\ref{fig:Processes} (b). As $L_{11}$ consists of only down-conversion processes, while $M$ on one down-conversion and one sum-frequency generation process, a high-pass filter suppresses $L_{11}$ more strongly than $|M|$, and thus leads to a wider parameter range for which $|M|>L_{11}$ such that $\mathcal{N}>0$.

\subsection{Thermal State}  \label{sec:EntanglementThermal}

\begin{figure}
\includegraphics[width=1.\columnwidth]{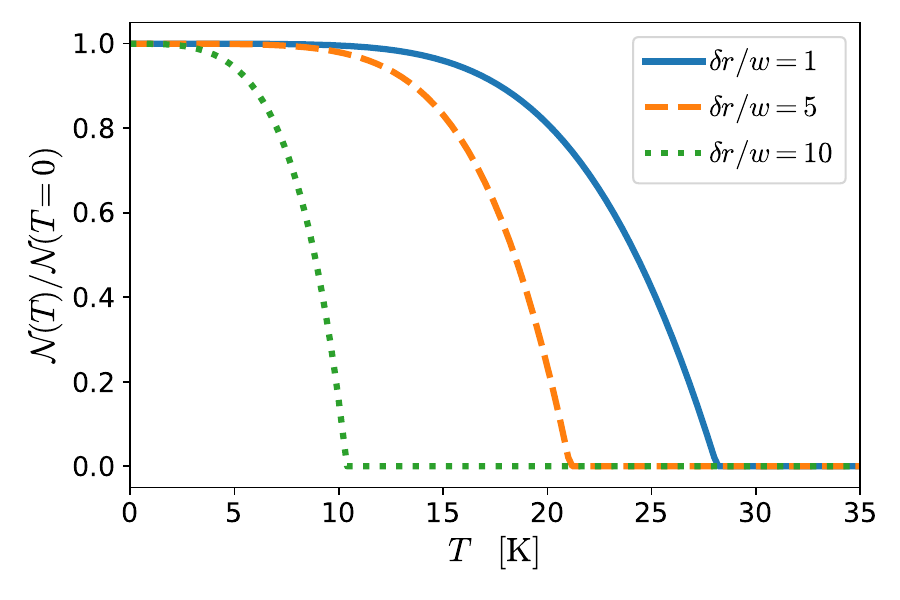}
\caption{\textit{Entanglement harvesting from thermal fluctuations.} We show the negativity as a function of the temperature $T$ for different beam separations $\delta r$ in case the THz field is initially in a thermal state. The cut-off frequency of the high-pass filter and time delay are set to $\omega_\mathrm{min} = \omega_c + 3 \sigma_\omega$ and $\delta t = 0$, respectively. All other experimental parameters are as in Ref.~\cite{settembrini2022detection}, see caption of Fig.~\ref{fig:Thermal}.  }
\label{fig:NegativityThermal}
\end{figure}

We quantify the entanglement harvested from thermal states, by evaluating the negativity in Eq.~\eqref{eq:Negativity} with $E_1 = |M| -L_{11}$ for different temperatures. In Appendix \ref{appsec:EvaluateThermal}, we simplify the expressions for $L_{11}$ and $|M|$ given by Eqs.~\eqref{eq:L12w} and \eqref{eq:Mw}, respectively, in case the THz field is initially in a thermal state \eqref{eq:ThermalState}. The resulting expressions are numerically integrated using the same experimental parameters as in the last section (see caption of Fig.~\ref{fig:Thermal}), to obtain the negativity $\mathcal{N}$ as a function of the temperature $T$ for different beam separations $\delta r$ and a low frequency cut-off $\omega_\mathrm{min} = \omega_c + 3 \sigma_\omega$ ($\omega_\mathrm{max} = \infty$). The result is shown in Fig.~\ref{fig:NegativityThermal}. We find that the negativity decreases monotonically with increasing temperature. With increasing beam separation this decrease occurs for smaller temperatures. That $\mathcal{N}(T)$ is a monotonically decreasing function can also be proven analytically. In Appendix \ref{app:ProofThermal} we follow Ref.~\cite{tjoa_harvesting_2020} and prove that $\mathcal{N}(T_1) \le \mathcal{N}(T_2)$ if $T_1 > T_2$. This also implies $\mathcal{N}(T) \le \mathcal{N}(T=0)$, where $\mathcal{N}(T=0)$ is the negativity in case the THz field is in its vacuum state. It follows that no entanglement can be harvested from thermal fluctuations. This was also previously shown for generic entanglement-harvesting protocols \cite{tjoa_harvesting_2020}, and signifies the classical nature of the thermal field.

The difference between entanglement harvesting from thermal and vacuum flcutuations can be understood on the basis of the nonlinear processes leading to $L_{11}$ and $M$.
In case thermal fluctuations are present, all the vacuum-induced nonlinear processes discussed in the last section and illustrated in Fig.~\ref{fig:Processes}(a) still occur, but they can now also be thermally induced. Additionally, in case real thermal photons are present, the emission process leading to $L_{11}$ can also stem from sum-frequency generation processes, in which energy is extracted from the thermal field, see Fig.~\ref{fig:NegativityThermal}. This leads to the crucial difference between vacuum- and thermally-induced processes: While for the former, $|M|/L_{11}$ can be increased via enhancing sum-frequency generation over down-conversion processes, for the latter this no longer applies as $L_{11}$ also has contributions from sum-frequency genration processes.

\section{Bell Nonlocality} \label{sec:Bell}

The vacuum field shows not only entanglement between different space-time regions but can even violate Bell inequalities \cite{summers_vacuum_1985,summers_bells_1987}, indicating that the vacuum field cannot be described by a local hidden variable (LHV) model in general experiments \cite{bell1964einstein}. We analyze whether Bell nonlocality can be harvested from the vacuum (and, thus, be observed) in EOS experiments. To do so, we show in Sec.~\ref{sec:BellViolation} that the state of the two probe modes $\hat{\rho}$ in Eq.~\eqref{eq:rhoDensitym} can indeed violate a Bell inequality. This Bell inequality was, however, shown to admit a loophole \cite{zukowski2016bell}. In Sec.~\ref{sec:BellDiscussion}, we discuss the loophole of the Bell inequality under consideration, connect our results to Bell-inequality violations found in generic entanglement-harvesting protocols \cite{reznik_violating_2005,matsumura_violation_2020}, and discuss open questions. For simplicity, we only consider the case of a vacuum THz field in this section.

\subsection{Violating the Bell Inequality} \label{sec:BellViolation}

In Ref.~\cite{reid1986violations}, a Bell inequality was formulated based on balanced homodyne measurements, which can be directly translated to the ellipsometry measurement of the two-beam EOS setup. To see this, we rewrite the two-beam EOS signal $G_{\Theta_1 \Theta_2}$ normalized to the total number of photons detected in both ellipsometry measurement schemes as (see Appendix \ref{app:EOSSetup} for details)
\begin{align} \label{eq:GinNumber}
G_{\Theta_1 \Theta_2} = \frac{\braket{(\hat{N}^{(1)}_{x^\prime }-\hat{N}^{(1)}_{y^\prime }) (\hat{N}^{(2)}_{x^\prime }-\hat{N}^{(2)}_{y^\prime })}}{\braket{[\hat{N}^{(1)}_{x^\prime }+\hat{N}^{(1)}_{y^\prime }][\hat{N}^{(2)}_{x^\prime }+\hat{N}^{(2)}_{y^\prime }]}}.
\end{align}
Here, $\hat{N}^{(i)}_{x^\prime }$ and $\hat{N}^{(i)}_{y^\prime}$ are the number operators of the photons detected by the two photodetectors of the ellipsometry scheme of mode $i$, see Fig.~\ref{fig:Setup}(b). Note that Eq.~\eqref{eq:GinNumber} reduces to Eq.~\eqref{eq:SignalG} in the limit of strong coherent laser pulses for which $\braket{[\hat{N}^{(1)}_{x^\prime }+\hat{N}^{(1)}_{y^\prime } ][\hat{N}^{(2)}_{x^\prime }+\hat{N}^{(2)}_{y^\prime }]} \approx N_d^2 $. Using Eq.~\eqref{eq:GinNumber}, the Bell inequality in Ref.~\cite{reid1986violations} reads
\begin{align} \label{eq:Bell}
\mathcal{B} = |G_{\Theta_1 \Theta_2} -G_{\Theta_1 \Theta_2^\prime} + G_{\Theta_1^\prime \Theta_2} + G_{\Theta_1^\prime \Theta_2^\prime}| \le 2.
\end{align}
We see that by performing four different two-beam EOS measurements with different angles $\Theta_i$, one can access $\mathcal{B}$ and thus potentially witness a violation of the Bell inequality $\mathcal{B}\le 2$ in Eq.~\eqref{eq:Bell}. 

The Bell inequality in Eq.~\eqref{eq:Bell} is only violated if the total number of detected photons in each of the two ellipsometry schemes and the correlation signal, i.e., the numerator and denominator of Eq.~\eqref{eq:GinNumber}, have the same order of magnitude. Thus, it is a prerequisite for a violation of Eq.~\eqref{eq:Bell} that we filter out most photons in the two $y$-polarized laser pulses emerging from the crystal before they act as local oscillators in the ellipsometry detecion scheme. The remaining number of photons in the pulses $N_{LO}$ should be on the same order of magnitude as the density matrix elements $|L_{ij}|$ and $ |M|$. The $x$-polarized probe fields should remain unchanged. This could be achieved via a polarizer \cite{hubenschmid2022complete} with an extinction ratio on the order of $|M|/N_p, |L_{ij}|/N_p  \thicksim  C $. This poses a technological challenge for the experimental setup in Ref.~\cite{settembrini2022detection}, where the detection efficiency is $C \thicksim 10^{-13}$ \cite{settembrini2022detection}. However, it might become feasible with improved efficiencies of emerging new EOS platforms \cite{benea2020electro,rajabali2023present}.

We evaluate the Bell inequality in Eq.~\eqref{eq:Bell} in the monochromatic case of narrow frequency filtering, i.e., the detected frequency interval is given by $[\omega_\mathrm{min}, \omega_\mathrm{max}] = [\omega_d - \Delta \omega/2,\omega_d +\Delta \omega/2]$ with $\Delta \omega \ll \omega_c, \sigma_\omega $. In this case $\hat{a}_i  \approx \hat{a}_i(\omega_d)$ and the density matrix in the corresponding EOS basis is still given by Eqs.~\eqref{eq:rhoDensitym} and \eqref{eq:RhoEntries} with 
\begin{multline} \label{eq:FiSmalldelw}
F_i(\vec{r}, t)  =  2 \chi^{(2)}   \sqrt{\frac{2\pi L \omega_c }{\hbar \eta} } \overline{\Ecli}(\vec{r}, t)  g_i (\vec{r}_\parallel) \\
\Delta \omega \overline{\mathcal{E}}(\omega_d) \me^{\mi (\omega_d-\omega_c)(t + \delta t_i -n_g z/c)}  .
\end{multline} 
We evaluate $G_{\Theta_1 \Theta_2}$ for the density matrix in Eq.~\eqref{eq:rhoDensitym} with $L_i =0$ (the THz field is assumed to be in its vacuum state). As the $y$-polarized field is now as weak as the density matrix elements $|L_{ij}|$ and $|M|$, which are in second order in $\chi^{(2)}$, the numerator and denominator of $G_{\Theta_1, \Theta_2}$ in Eq.~\eqref{eq:GinNumber} are in fourth order in $\chi^{(2)}$, such that we have to include the contribution from $X$. We find 
\begin{align} \label{eq:BellEv1}
G_{\Theta_1, \Theta_2}
 & =2 N_\mathrm{LO}\frac{ \mathrm{Re}[ P_1^\ast P_2 L_{12} + P_1 P_2 M  ]}{ N_\mathrm{LO}^2 + N_\mathrm{LO} L_{11} + X }.
\end{align}
Here, we used $L_{11} = L_{22}$ as before, and defined the shorthand notation $  P_i \equiv P(\Theta_i) $, see end of section \ref{sec:EOSSetup} for a definition of $P(\Theta_i)$. 
\begin{figure}
\includegraphics[width=1.\columnwidth]{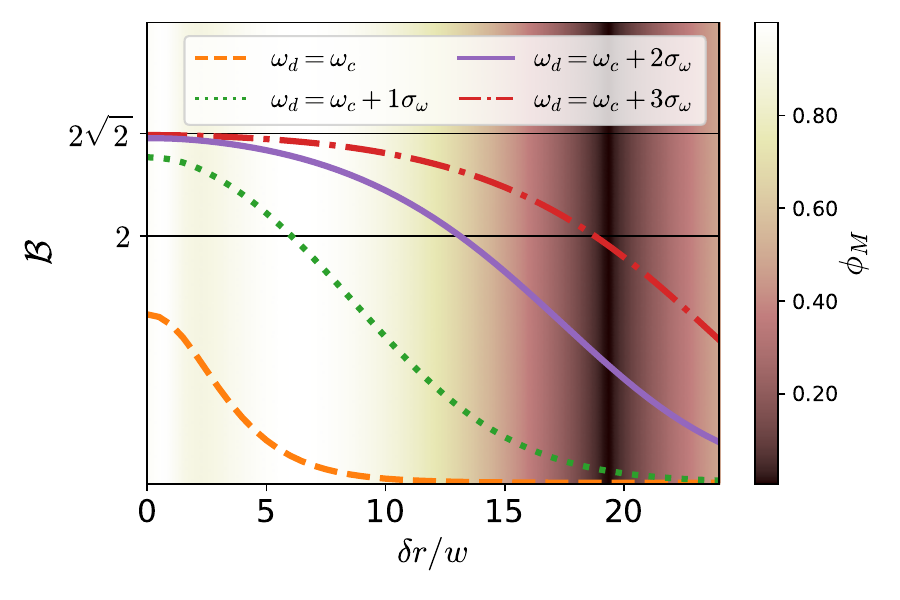}
\caption{\textit{Bell-inequality violation.} We compute $\mathcal{B}$ defined in Eq.~\eqref{eq:Bell} as a function of the beam separation $\delta r$ for different values of the detected frequency $\omega_d  $. For each value of $\delta r$, we optimize $\mathcal{B}$ over the angles $\Theta_i$, $\Theta_i^\prime$. The THz field is initially in its vacuum state and all other experimental parameters are as in Ref.~\cite{settembrini2022detection}, see caption of Fig.~\ref{fig:Thermal}. We find a violation of the Bell inequality if $\mathcal{B}> 2$, and a maximal violation for $\mathcal{B}= \sqrt{2}2$. }
\label{fig:Bell}
\end{figure}
We use the density matrix elements for a vacuum THz field obtained in Appendix \ref{appsec:EvaluateVacuum} to obtain $G_{\Theta_1 \Theta_2}$ in Eq.~\eqref{eq:BellEv1} for the same experimental parameters as in the laser sections (see caption of Fig.~\ref{fig:Thermal}) as a function of the distance between the pulses $\delta r$ for different values of the detected frequency $\omega_d$. We insert the resulting expression into the Bell inequality in Eq.~\eqref{eq:Bell}, set $N_\mathrm{LO} = |M|$ (this maximizes the $\mathcal{B}$ in the relevant regime considered here), and optimize for each value of $\delta r$ over the angles $\Theta_i$, $\Theta_i^\prime$. The result is shown in Fig.~\ref{fig:Bell}. We find that the Bell inequality can be violated in regimes for which the entanglement harvesting is genuine or communication based. As the beam separation $\delta r$ increases, the maximal value of $\mathcal{B}$ decreases. Conversely, as the detected frequency $\omega_d$ increases, the maximal value of the $\mathcal{B}$ also increases. It is even maximally violated ($\mathcal{B} \approx 2 \sqrt{2}$) for small beam separations and high frequency cut-offs, compare Fig.~\ref{fig:Bell}. In this regime, one finds that $L_{ij} \ll |M| $ due to a suppression of SFG compared to DC processes (see the discussion in Fig.~\ref{fig:Processes}), such that $X = |M|$, and $\phi_M = 1$. Using this in Eq.~\eqref{eq:rhoDensitym} we find that the density matrix reduces to $\hat{\rho} = \ket{\psi} \bra{\psi}$ with
\begin{align}
\ket{\psi} = \sqrt{1-M^2}    \ket{00} + M \ket{11}    ,
\end{align}
 resembling a two-mode squeezed vacuum. Furthermore, in this limit, Eq.~\eqref{eq:BellEv1} reads $G_{\Theta_1 \Theta_2}  =  \mathrm{Re}[ P_1 P_2   ]$. The resulting Bell inequality is given by
\begin{align} \label{eq:Bsimplified}
\mathcal{B } =  | \mathrm{Re}[ P_1 P_2  - P_1 P_2^\prime + P_1^\prime P_2 + P_1^\prime P_2^\prime  ] |\le 2.
\end{align}
Maximizing $\mathcal{B}$ in Eq.~\eqref{eq:Bsimplified} over the angles $\Theta_i$ and $\Theta_i^\prime$, we find that its maximal value is $2\sqrt{2}$, also known as the Tsirelson bound \cite{cirel1980quantum}.

\subsection{Discussion} \label{sec:BellDiscussion}

The Bell inequality of Eq.~\eqref{eq:Bell} tests whether the observed correlations can be described by a LHV model~\cite{reid1986violations}, i.e., whether we can find a model
\begin{equation}\label{eq:LHVmodel}
    \langle \hat{N}^{(1)}_{x^\prime } \hat{N}^{(2)}_{x^\prime }\rangle = \int \dif \lambda q(\lambda) N^{(1)}_{x^\prime }(\lambda,\Theta_1)N^{(2)}_{x^\prime }(\lambda,\Theta_2),
\end{equation}
with $0\leq q(\lambda)\leq 1$ and $\int \dif \lambda q(\lambda)=1$, and similarly for all other correlations and measurement settings. 

In Ref.~\cite{zukowski2016bell}, it was shown that, to demonstrate that Eq.~\eqref{eq:Bell} is fulfilled for such LHV models, one requires an additional assumption on the LHV model, i.e., that $N^{(i)}_{x^\prime }(\lambda,\Theta_i)+N^{(i)}_{y^\prime }(\lambda,\Theta_i)$ does not depend on the measurement setting $\Theta_i$ for $i=1,2$,. In other words, we must assume that for a given value $\lambda$ of the LHV, the total number of observed photons in each ellipsometry detection station does not depend on the local measurement setting. This assumption, though satisfied in quantum and classical electrodynamics, represents a loophole, i.e., it allows LHV models like Eq.~\eqref{eq:LHVmodel} to violate the Bell inequality of Eq.~\eqref{eq:Bell} if the model does not fulfill the additional assumption.

In Appendix~\ref{app:fairsampling}, we show that the additional assumption on the LHV model, that the total number of detected photons does not depend on the measurement settings, is closely connected to the fair sampling assumption~\cite{clauser_proposed_1969,berry_fair_2010,gebhart_extending_2022}. In the standard Bell scenario where a central source distributes two particles among two measurement parties, events for which only one of the parties detects their particle must usually be neglected to violate Bell inequalities. This postselection of data opens the detection loophole~\cite{pearle_hidden_1970} that can only be closed by using highly efficient detectors. If the loophole is not closed explicitly, which is still true for most of modern Bell experiments, one has to rely on the fair sampling assumption, i.e., one has to assume that the detection probability of each particle is independent on the local measurement setting. Furthermore, as we discuss in Appendix~\ref{app:fairsampling}, the fair sampling assumption must also be used in the traditional entanglement-harvesting scheme using two level systems as local probes~\cite{reznik_violating_2005,matsumura_violation_2020}. 

In summary, we can conclude that the discussed EOS experiment excludes a description of the measured correlations by a LHV model that fulfills the fair sampling assumption, an assumption that is widely used in Bell experiments. In Ref.~\cite{zukowski2016bell}, a refined Bell inequality is derived that closes the loophole. Whether this inequality can also be violated in EOS experiments is an interesting open question for further studies. As it is shown for the two-level-probe case that a local filtering operation (and thus the fair sampling assumption) is required to violate Bell inequalties~\cite{reznik_violating_2005,matsumura_violation_2020}, it, however, may well be that the assumption is required.  

\section{Discussion and outlook}

We have shown that genuine and communication-based entanglement harvesting from the vacuum field is possible with state-of-the-art EOS experimental setups. This allows to extend the experimental observation of two-point correlations in the vacuum field \cite{settembrini2022detection} to the quantum realm. The harvesting of genuine entanglement from space-like separated regions will prove that there is entanglement in the vacuum state \cite{de2023entanglement}. We have further shown that one cannot harvest entanglement from thermal fluctuations, indicating their classical nature, and that the same amount of entanglement can be harvested from all coherent states (including the vacuum state). From an applied perspective, our results enable the use of two-beam EOS for the characterization \cite{benea-chelmus_subcycle_2016,markmann2023electro} of quantum correlations present in arbitrary THz fields between different space-time regions, e.g., in broadband squeezed states \cite{riek2017subcycle} or two-photon states entangled in time/frequency relevant for quantum light spectroscopy \cite{dorfman_nonlinear_2016}. Significantly enhanced sensitivities via field confinement in antennas and a convenient chip-based implementation of the entanglement-harvesting protocols might be achievable using integrated photonics devices \cite{benea2020electro,rajabali2023present}.

The theoretical framework developed here establishes EOS as an experimental platform for entanglement harvesting, one of the basic work horses to study the interplay between quantum information theory and relativity \cite{fuentes2005alice,mann2012relativistic}. It will be interesting to extend the results of the current manuscript to reveal further findings from the field of relativistic quantum information, e.g.: The space-time geometry inside the nonlinear crystal can be effectively altered by applying an additional strong coherent laser pulse \cite{philbin_fiber-optical_2008,kizmann2019subcycle}, potentially allowing one to study entanglement harvesting in curved space-times \cite{fuentes2005alice,henderson2018harvesting}; tripartite entanglement harvesting \cite{mendez2022entanglement} and eavesdropping in entanglement-harvesting protocols \cite{sahu2022sabotaging} could be implemented via a third probe pulse in the EOS setup; questions of causal order in entanglement-harvesting protocols \cite{henderson2020quantum} could be potentially implemented in EOS experiments using quantum probe pulses that are initially in a coherent superposition state or entangled \cite{virally2021enhanced}.

\acknowledgements{F.L. is grateful to Andreas Buchleitner, Edoardo Carnio, Dominik Lentrodt and Andreas Woitzik for fruitful discussions, and acknowledges support from the Studienstiftung des deutschen Volkes. A.H. acknowledges financial support from Swiss National Science Foundation (SNSF) (grant 200020\_207795/1).}

\appendix 

\section{The Polaritonic Quantum Vacuum}  \label{appsec:Polaritonic}

We use macroscopic quantum electrodynamics \cite{scheel2009macroscopic,buhmann2013dispersion} to find the two-point correlation function of the electric field operator in general dispersive and absorbing environments described by the complex perimittivity $\epsilon(\Omega)$. We start with Fourier transforming the THz field
\begin{align} \label{eq:Fourier}
    \hat{\vec{E}}(\vec{r}, t) =  \int_{-\infty}^\infty \dif \Omega   \hat{\vec{E}}(\vec{r}, \Omega) \me^{-\mi \Omega t},
\end{align}
where $\hat{\vec{E}}(\vec{r}, \Omega) $ satisfies $\hat{\vec{E}}^\dagger(\vec{r}, \Omega) = \hat{\vec{E}}(\vec{r}, -\Omega)$ and is given by
\begin{multline} \label{eq:FieldExpansioninGreens}
\hat{\vec{E}}(\vec{r}, \Omega )  =  \mi \frac{\Omega^2}{c^2}  \int \dif^3 r^\prime \sqrt{\frac{\hbar}{\pi \epsilon_0} \mathrm{Im}[\epsilon(\vec{r}^\prime , \Omega)] } \\
\times\tens{D}(\vec{r}, \vec{r}^\prime, \Omega) \cdot \hat{\vec{f}}(\vec{r}^\prime, \Omega).
\end{multline}
Here, $\hat{\vec{f}}(\vec{r}, \Omega)$ and $\hat{\vec{f}}^\dagger(\vec{r}, \Omega)$ are polaritonic annihilation and creation operators \cite{scheel2009macroscopic} and we defined the Green tensor of the vector Helmoholtz equation via
\begin{align} \label{eq:GreensTensorDef}
\left(  \nabla \times \nabla \times - \frac{\Omega^2}{c^2}\epsilon(\Omega) \right) \tens{D}(\vec{r}, \vec{r}^\prime, \Omega) =    \boldsymbol{\delta}(\vec{r}-\vec{r}^\prime)      ,
\end{align}
and the boundary condition $\tens{D}(\vec{r}, \vec{r}^\prime, \Omega) \to 0 $ for \mbox{$|\vec{r}- \vec{r}^\prime |  \to \infty $}. In a bulk medium, the $xx$-component $\mathsf{D}_{xx} \equiv \mathsf{D}$ reads \cite{buhmann2013dispersion,lindel2023separately}
\begin{multline} \label{eq:GreensGeneral}
\mathsf{D}(\vec{r}, \vec{r}^\prime, \Omega) 
 =  \frac{\mi}{8\pi^2 } \int \dif^2 k_\parallel  \frac{\me^{\mi \vec{k}_\parallel \cdot (\vec{r}_\parallel - \vec{r}_\parallel)}}{k_z} \\
 \times  \left( 1- \frac{k_x^2}{k^2} \right) \me^{\mi k_z|z-z^\prime|} ,
\end{multline}
with $k_z = \sqrt{k^2 -k_\parallel^2}$, $k = n(\Omega) \Omega/c$, and $n(\Omega) = \sqrt{\epsilon(\Omega)}$. For the Green tensor in the near infrared, we apply the paraxial approximation, i.e., assume $k_\parallel \ll k$ such that the Green tensor reduces to 
\begin{align} \label{eq:GreensParaxial} 
\mathsf{D}(\vec{r}, \vec{r}^\prime, \omega) =  \frac{\mi}{2 k} \delta(\vec{r}_\parallel -\vec{r}_\parallel^\prime) \me^{\mi k (z-z^\prime)} .
\end{align}
Note that $\tens{D}$ is proportional to the Fourier transform of the response function of the electromagnetic field defined in Eq.~\eqref{eq:ResponseFunction} \cite{lindel2023separately}, i.e.,
\begin{align} \label{eq:ResponseGreen}
\mathcal{R}(\vec{r}, \vec{r}^\prime, \tau) 
 & =   \int \dif \omega \, \me^{-\mi \omega \tau }  \frac{\mu_0 \omega^2}{2\pi} \mathsf{D}( \vec{r}, \vec{r}^\prime \omega).
\end{align}
Assuming the field state to be given by the thermal state in Eq.~\eqref{eq:ThermalState}, one finds from the field expansion in Eq.~\eqref{eq:FieldExpansioninGreens} \cite{buhmann2013dispersion}
\begin{multline} \label{eq:EEdag}
    \braket{\hat{\vec{E}}(\vec{r}, \Omega) \hat{\vec{E}}^\dagger(\vec{r}, \Omega^\prime)} = \frac{\hbar \mu_0}{\pi} \Omega^2 \delta(\Omega-\Omega^\prime) \\
    \times [1 + 2 n_T(\Omega)] \mathrm{Im}[\tens{D}(\vec{r}, \vec{r}^\prime, \Omega)],
\end{multline}
and 
\begin{multline} \label{eq:EdagE}
    \braket{\hat{\vec{E}}^\dagger(\vec{r}, \Omega) \hat{\vec{E}}(\vec{r}, \Omega^\prime)} = \frac{2 \hbar \mu_0}{\pi} \Omega^2 \delta(\Omega-\Omega^\prime) \\ \times   n_T(\Omega)\mathrm{Im}[\tens{D}(\vec{r}, \vec{r}^\prime, \Omega)].
\end{multline}
Here, we assumed $\Omega, \Omega^\prime > 0 $, and $n_T(\Omega)$ is the thermal distribution defined in Eq.~\eqref{eq:ThermalDistr}. The correlation functions for the vacuum field are obtained from Eqs.~\eqref{eq:EEdag} and \eqref{eq:EdagE} in the limit $T \to 0$ in which case $n_T(\Omega)$ vanishes.

\section{Electro-optic Sampling in the Interaction Picture} 

We first recap some details of the EOS setup under consideration, before outlining a detailed derivation of the reduced density matrix of the two probe modes in Eq.~\eqref{eq:rhoDensitym}, and evaluating the latter in case the THz field is in the vacuum or a thermal state.

\subsection{Details of the EOS Setup} \label{app:EOSSetup}

\paragraph{Laser Pulses.}

The laser pulses are given by Eq.~\eqref{eq:laserPulse}, where we assumed that the they are focused to the center of the crystal, whose length $L$ is much smaller than the Rayleigh length $l_R = k w^2/2$ of the pulses \cite{moskalenko_paraxial_2015,lindel2021macroscopic}. We further assume that both laser pulses have the same Gaussian spectrum
\begin{align}
\mathcal{E}(\omega) = \left( \frac{\tau_\sigma^2}{2 \pi} \right)^{1/4} \me^{- \tau_\sigma^2 (|\omega| - \omega_c)^2/4}.
\end{align}
Assuming that the refractive index $ n(\omega)$ in the NIR is sufficiently flat in the spectral range of the laser pulses, we approximate the wave vector $k(\omega) = n(\omega) \omega/c$ of the laser pulses using a Tailor expansion around their central frequency $\pm\omega_c$, see Ref.~\cite{lindel2023separately}. Introducing the group refractive index $n_g = c \partial k_\omega / \partial \omega |_{\omega_c}$ and defining $n(\omega_c) \equiv n_\mathrm{c}$ we obtain the time-domain expression for the laser pulses
\begin{multline} \label{eq:laserPulseTimDomain}
\Ecli(\vec{r}, t) = \sqrt{ \frac{ \omega_c N_p \hbar L }{2 \epsilon_0 c n_\mathrm{c}} } \overline{\Eclo}(\vec{r}, t)  \\
\times 2 \mathrm{cos}\left[\omega_c \left( n_\mathrm{c} \frac{z}{c} - t - \delta t_i \right) \right],
\end{multline}
with the normalized pulse envelopes 
\begin{align}\label{eq:PulseEnvelopes}
\overline{\Eclo}^2(\vec{r}, t) & =  \frac{2^{3/2}\eta}{\pi^{3/2}\hbar \omega_c N_p \tau_\sigma w^2 L}  \me^{- 2\frac{(n_g\frac{z}{c}- t)^2}{\tau_\sigma^2} -2 \frac{ r_\parallel^2}{w^2}} .
\end{align}
Here, $N_p$ is the total number of photons in each of the two laser pulse
\begin{align}
    N_p  & = \eta \int \dif^2 r_\parallel \int_0^\infty \dif \omega \frac{1}{\hbar \omega} \mathcal{E}^{(i) \ast}(\vec{r}_\parallel, \omega) \mathcal{E}^{(i)}(\vec{r}_\parallel, \omega) \\ \label{eq:Np}
    & \approx \frac{\eta}{\hbar \omega_c} \int \dif^2 r_\parallel \int_0^\infty \dif \omega   \mathcal{E}^{(i) \ast}(\vec{r}_\parallel, \omega) \mathcal{E}^{(i)}(\vec{r}_\parallel, \omega).
\end{align}
Note, that we assume $N_p$ to be the same for both laser pulses. The number of photons $N_d$ within the frequency range $[\omega_\mathrm{min}, \omega_\mathrm{max}]$ is defined as in Eq.~\eqref{eq:Np} but with $\int_0^\infty \dif \omega \to \int_{\omega_\mathrm{min}}^{\omega_\mathrm{max}} \dif \omega$. For the second laser pulse we have $\overline{\Eclt}(\vec{r}, t) = \overline{\Eclo}(\vec{r} + \delta \vec{r}_\parallel, t+ \delta t)$. $\overline{\Ecli}(\vec{r}, t)$ is normalized such that $\int_{V_C}\dif ^3 r \int \dif t \overline{\Ecli}^2(\vec{r}, t) = 1$.

\paragraph{Detection Scheme.}

The ellipsometry measurement scheme is depcited in Fig.~\ref{fig:Setup} (b). Following Refs.~\cite{kizmann_quantum_2022}, we consider a waveplate whose fast axis is rotated by an angle $\alpha_i$ against the $y$-axis and which induces a phase shift $\theta_i$. It mixes the respectively $x$ and $y$ polarized fields $\hat{E}_x^{(i)} = \hat{E}^{(i)}$ and $\hat{E}_y^{(i)}$ emerging from the crystal via
\begin{align} \label{eq:WavePlate}
\left(  \begin{array}{c}
  \hat{E}_{x^\prime}^{(i)}   \\ \hat{E}_{y^\prime}^{(i)}
 \end{array} \right) & = M(\alpha_i) 
 \left( \begin{array}{cc}
    \me^{-\mi \frac{ \theta_i}{2}}  & 0 \\
    0  &  \me^{\mi \frac{ \theta_i}{2}}
 \end{array} \right) 
 M^T(\alpha_i) \left(  \begin{array}{c}
  \hat{E}^{(i)}   \\ \hat{E}_y^{(i)}
 \end{array} \right), \\
 M(\alpha_i)&  = \left( \begin{array}{cc}
    \mathrm{cos}(\alpha_i) & \mathrm{sin}(\alpha_i) \\
  -  \mathrm{sin}(\alpha_i)  &  \mathrm{cos}(\alpha_i)
 \end{array} \right) .
\end{align}
The Wollaston prism spatially separates $\hat{E}_{x^\prime}^{(i)}$ and $\hat{E}_{y^\prime}^{(i)}$, such that they can by individually detected by photodetectors. Eventually, the ellipsometry signal $\hat{S}^{(i)}_{\Theta_i}$ results by subtracting the currents of the two detectors and reads
\begin{align} \label{eq:SGeneral}
\hat{S}^{(i)}_{\Theta_i} = \hat{N}_{x^\prime }^{(i)}-\hat{N}_{y^\prime }^{(i)},
\end{align}
with \cite{raymer1995ultrafast,moskalenko_paraxial_2015}
\begin{align} \label{eq:NumberOp}
    \hat{N}_{j }^{(i)} =  \eta \int \dif^2 r_\parallel \int_{\omega_\mathrm{min}}^{\omega_\mathrm{max}} \dif \omega \frac{1}{\hbar \omega} \hat{E}_j^{(i)\dagger}(\vec{r}_\parallel, \omega) \hat{E}_j^{(i)}(\vec{r}_\parallel, \omega).
\end{align}
The total number of detected photons is given by $N^{(i)} =   \braket{\hat{N}_{x^\prime }^{(i)}+\hat{N}_{y^\prime }^{(i)}}$, such that, using Eq.~\eqref{eq:SGeneral} in Eq.~\eqref{eq:SignalG}, we find Eq.~\eqref{eq:GinNumber} of the main text. For balanced detection, we set $\alpha_i= \pm \mathrm{arccos}[-\mathrm{cot}^2(\theta_i/2)] $ for phase shifts $\theta_i \in [\pi/2, 3\pi/2]$ \cite{kizmann_quantum_2022}. Inserting Eqs.~\eqref{eq:WavePlate} and \eqref{eq:NumberOp} into Eq.~\eqref{eq:SGeneral}, we find 
\begin{multline} \label{eq:SignalFull}
  \hat{S}^{(i)}_{\Theta_i} = \eta \int \dif^2 r_\parallel \int_{\omega_\mathrm{min}}^{\omega_\mathrm{max}} \dif \omega \frac{1}{\hbar \omega} \\\times [P(\Theta_i) \mathcal{E}^{(i)\ast}(\vec{r}_\parallel, \omega)  \hat{E}^{(i)}(\vec{r}_\parallel, \omega) + \mathrm{h.c.}],
\end{multline}
and $N^{(i)} = \braket{\hat{N}_x^{(i)} + \hat{N}_y^{(i)}}$. Here, we also replaced the $y$-polarized field $\hat{E}_y^{(i)}$ by the coherent amplitude $\Ecli$. Inserting the paraxial mode expansion for the probe fields in Eq.~\eqref{eq:SignalFieldModeExp} into Eq.~\eqref{eq:SignalFull} and making use of Eq.~\eqref{eq:creation}, we find Eq.~\eqref{eq:SignalRed} of the main text.

\subsection{Reduced Density Matrix} \label{app:RedDenstiy}

Before the two laser pulses have entered the crystal, the THz and both NIR probe fields are in the ground state, such that the initial state reads $\hat{\rho}_\mathrm{vac} \otimes \hat{\rho}_\mathrm{vac}^{(\mathrm{THz})} $. The reduced density matrix of the two probe modes after the two laser pulses have emerged from the crystal $\hat{\rho}$ is given by Eq.~\eqref{eq:rhotGen} in the main text. To evaluate this expression, $\hat{U}$ is obtained in second order of $\chi^{(2)}$ by expanding the time-ordered exponential in a Dyson series: 
\begin{align} \label{eq:TimeEvoExpand}
\hat{U}  \approx \hat{U}^{(0)}+  \hat{U}^{(1)} +\hat{U}^{(2)},
\end{align}
with $\hat{U}^{(0)} = \hat{1}$, $\hat{U}^{(1)} \equiv - (\mi/\hbar) \int_{-\infty}^\infty \dif t \hat{H}_I (t)$, and we split the second order term into $\hat{U}^{(2)} = \hat{U}^{(2a)} + \hat{U}^{(2b)}$, with $\hat{U}^{(2a)} = (1/2) \hat{U}^{(1)} \hat{U}^{(1)}$, and 
\begin{align}
  \hat{U}^{(2b)}   = - \frac{1}{2\hbar^2} \int_{-\infty}^\infty \dif t^{\prime} \int_{-\infty}^{t^{\prime}} \dif t^{\prime\prime} [ \hat{H}_I(t^{\prime }) ,\hat{H}_I(t^{\prime\prime}) ].
\end{align}
Inserting Eq.~\eqref{eq:TimeEvoExpand} into Eq.~\eqref{eq:rhotGen} we obtain
\begin{align} \label{eq:rhoGeneral}
\hat{\rho} =\sum_{0  \le i+j \le 2 } \hat{\rho}^{(ij)},
\end{align}
where we defined 
\begin{align}
   \hat{\rho}^{(ij)} =\mathrm{tr}_\mathrm{THz} \{ \hat{U}^{(i)}\hat{\rho}_\mathrm{vac} \otimes \hat{\rho}_\mathrm{vac}^{(\mathrm{THz})}  \hat{U}^{(j)\dagger} \}. 
\end{align} 
Using the Hamiltonian in Eq.~\eqref{eq:HiEOS} we find 
\begin{multline} \label{eq:1rho}
 \hat{\rho}^{(10)}  = - \frac{2\mi \chi^{(2)}}{\hbar} \sum_{i =1,2} \int_{\vec{r}, t}  \braket{\hat{E}(\vec{r}, t)} \\ \times \Ecli(\vec{r}, t) \Ei(\vec{r}, t) \hat{\rho}_\mathrm{vac},
\end{multline}
with $\hat{\rho}_\mathrm{vac} = \ket{ 0,0 }\bra{ 0,0 }$. $ \hat{\rho}^{(01)}$ is the hermitian conjugate of $ \hat{\rho}^{(10)} $. Furthermore, we find 
\begin{multline} \label{eq:1rho1}
 \hat{\rho}^{(11)} = \frac{4\chi^{(2)2}}{\hbar^2} \sum_{i,j = 1,2}  \int_{\vec{r}, t,\vec{r}^\prime, t^\prime}  \hspace{-0.5cm} \braket{ \hat{E}(\vec{r}^\prime, t^\prime) \hat{E}(\vec{r}, t) }  \\ 
\times \Ecli (\vec{r}, t)  \Eclj (\vec{r}^\prime, t^\prime)  \Ei  (\vec{r}, t)    \hat{\rho}_\mathrm{vac}  \Ej (\vec{r}^\prime, t^\prime) ,
\end{multline}
\begin{multline}  \label{eq:2arho}
\hat{\rho}^{(2a 0)}  = -\frac{2\chi^{(2)2} }{\hbar^2}\sum_{i,j = 1,2} \int_{\vec{r}, t,\vec{r}^\prime, t^\prime}  \hspace{-0.5cm} \braket{ \hat{E}(\vec{r}, t) \hat{E}(\vec{r}^\prime, t^\prime) }  \\ 
 \times  \Ecli (\vec{r}, t)   \Eclj (\vec{r}^\prime, t^\prime)  \Ei  (\vec{r}, t)   \Ej (\vec{r}^\prime, t^\prime)  \hat{\rho}_\mathrm{vac},
\end{multline}
and 
\begin{multline}  \label{eq:2crho}
 \hat{\rho}^{(2b0)}   = \frac{2\chi^{(2)2}  \mi }{\hbar}\sum_{i,j = 1,2} \int_{\vec{r}, t,\vec{r}^\prime, t^\prime}  \hspace{-0.5cm}  \Ecli (\vec{r}, t) \Eclj (\vec{r}^\prime, t^\prime) \\
\times  \big\{ \delta_{ij} \mathcal{R}_{ii}(\vec{r}, \vec{r}^\prime, t-t^\prime)     \braket{ \hat{E}(\vec{r}, t) \hat{E}(\vec{r}^\prime, t^\prime) }   \\
 +   \mathcal{R}(\boldsymbol{\rho}, \tau) \hat{E}_{i}(\vec{r}, t) \hat{E}_{j} (\vec{r}^\prime, t^\prime)   \big\}  \hat{\rho}_\mathrm{vac}. 
\end{multline}
Here, $\mathcal{R}_{ii}$ is the response function of $\Ei$ and we used that $\Eo$ and $\hat{E}^{(2)}$ commute \cite{lindel2023separately}.

Expressing the reduced density matrix in Eqs.~\eqref{eq:rhoGeneral}--\eqref{eq:2crho} in the EOS basis defined via the creation and annihilation operators in Eq.~\eqref{eq:creation}, we find $\hat{\rho}$ given by Eqs.~\eqref{eq:rhoDensitym} and Eq.~\eqref{eq:RhoEntries} in the main text. To derive this result, we identified $F_i(\vec{r}, t)$ defined in Eq.~\eqref{eq:Fi} as
\begin{align}
 F_i(\vec{r}, t) =  \braket{0,0 | \hat{a}_i \underbrace{  \frac{2 \mi \chi^{(2)}}{\hbar} \Ecli(\vec{r}, t ) \Ei (\vec{r}, t )  }_{\Ei_\mathcal{E}(\vec{r}, t) }| 0,0 } ,
\end{align}
and made use of
\begin{align}
F_i(\vec{r}, t)  F_j(\vec{r}^\prime, t^\prime) &  = - \braket{1, 1 |\Ei_\mathcal{E}(\vec{r}, t)\Ej_\mathcal{E} (\vec{r}^\prime, t^\prime) | 0,0 }     ,\\
F_i^\ast(\vec{r}, t)  F_i(\vec{r}^\prime, t^\prime) & = \braket{0,0 |\Ei_\mathcal{E}(\vec{r}, t)\Ei_\mathcal{E} (\vec{r}^\prime, t^\prime) | 0,0 }     ,\\
 F_i(\vec{r}, t)  F_i(\vec{r}^\prime, t^\prime) & = -\braket{0, 0 |\frac{\hat{a}_i^2}{2} \Ei_\mathcal{E}(\vec{r}, t)\Ei_\mathcal{E} (\vec{r}^\prime, t^\prime) | 0,0 } .
\end{align} 
The fourth-order contribution $X$, is given by
\begin{align}
X &   = \braket{1,1| \mathrm{tr}_\mathrm{THz}\{ U^{(2)} \hat{\rho}_\mathrm{vac} \otimes \hat{\rho}_\mathrm{vac}^{(\mathrm{THz})}  U^{(2)\dagger} \}|1,1} .
\end{align}
For the THz field in its vacuum state, we can use Wicks theorem to find 
\begin{align} \label{eq:XDecompos}
X  & = L_{11} L_{22} + |L_{21}|^2+  |M|^2.
\end{align}

\section{Evaluation of the Reduced Density Matrix }

We explicitly evaluate $L_{ij}$ and $M$ in case the THz field is in its vacuum or in a thermal state in Sections \ref{appsec:EvaluateVacuum} and \ref{appsec:EvaluateThermal}, respectively. Note that $X$ reduces to combinations of $L_{ij}$ and $M$ for a vacuum THz field, compare Eq.~\eqref{eq:XDecompos}.    \\

\subsection{Vacuum State} \label{appsec:EvaluateVacuum}

Using the definition of the correlation and response function in Eqs.~\eqref{eq:CorrelationFunction} and \eqref{eq:ResponseFunction}, Fourier transforming the THz field to frequency space according to Eq.~\eqref{eq:Fourier}, using Eqs.~\eqref{eq:EEdag} and \eqref{eq:EdagE} with $n_T(\Omega) = 0$, as well as Eq.~\eqref{eq:ResponseGreen}, we find for a THz field in its vacuum state
\begin{multline} \label{eq:CRp}
\mathcal{C}(\boldsymbol{\rho}, \tau) - \mi \mathcal{R}^{\prime}(\boldsymbol{\rho}, \tau)   \\
= - \frac{\hbar \mu_0 \mi}{\pi} \int_0^\infty \dif \Omega \Omega^2 \mathsf{D}(\vec{r}, \vec{r}^\prime, \Omega) \cos[\Omega (t-t^\prime)],   
\end{multline}
and 
\begin{multline} \label{eq:CRpp}
\mathcal{C}(\boldsymbol{\rho}, \tau) - \mi \mathcal{R}^{\prime\prime}(\boldsymbol{\rho}, \tau)  \\
 = \frac{\hbar \mu_0}{\pi}  \int_0^\infty \dif \Omega \Omega^2 \mathrm{Im}[ \mathsf{D}(\vec{r}, \vec{r}^\prime, \Omega)] \me^{-\mi \Omega (t-t^\prime)} .
\end{multline}
Inserting Eqs.~\eqref{eq:CRp} and \eqref{eq:CRpp} together with the Green tensor in Eq.~\eqref{eq:GreensGeneral} into Eqs.~\eqref{eq:L12w} and Eq.~\eqref{eq:Mw}, we encounter Gaussian integrals for $t^{(\prime)}$ and $\vec{r}_\parallel^{(\prime)}$, which can be carried out analytically. We further carry out the $z$ and $z^\prime$ integrals using 
\begin{align}
\Pi(q_z, \Omega) &\equiv \int\limits_{-L/2}^{L/2} \frac{\dif z}{L}  \int \limits_{-L/2}^{L/2} \frac{\dif z^\prime}{L}  \me^{-\mi \Omega n_g \frac{z-z^\prime}{c}} \frac{\me^{ \mi q_z |z-z^\prime|}}{q_z}  \\ \nonumber
&= \frac{1}{L q_z}\big[ \frac{\mi}{  q_z + \Omega n_g/c }  + \frac{1- \me^{\mi L (  q_z + \Omega n_g/c )}}{L  (  q_z + \Omega n_g/c )^2}      \big]  \\
& \hspace{1cm}+ (n_g  \to -n_g ) ,
\end{align} 
where $+ (n_g  \to -n_g )$ indicates adding the preceding term subject to the replacement $n_g  \to -n_g$. In the resulting expression, we identify the normalized spectral auto-correlation function
\begin{align}\label{eq:fOmegaGen}
f(\Omega)   =   \frac{\int_{\omega_\mathrm{min}}^{\omega_\mathrm{max}} \dif \omega  \mathcal{E}(\omega) \mathcal{E}(\omega+ \Omega) }{ \int_{\omega_\mathrm{min}}^{\omega_\mathrm{max}} \dif \omega  \mathcal{E}^2(\omega)},
\end{align}
and define $ E_\mathrm{vac}^2 = \hbar n(\Omega)\Omega^3/(2\pi^2 \epsilon_0 c^3)$ to find 
\begin{align}  \label{eq:L11VacFinal}
L_{ii}^{(\mathrm{vac})}  =  \frac{  C N_d }{  4  } \int_0^\infty \dif \Omega \int \dif^2 q_\parallel   f^2(\Omega)      \mathrm{ Re}[R(\vec{q}_\parallel, \Omega)] , 
\end{align}
\begin{multline}  \label{eq:L12VacFinal}
L_{12}^{(\mathrm{vac})}  =  \frac{  C N_d }{  4  } \int_0^\infty \dif \Omega \int \dif^2 q_\parallel   f^2(\Omega)  \me^{\mi q_y \delta r }  \\
\times \me^{\mi \delta t \Omega}  \mathrm{ Re}[R(\vec{q}_\parallel, \Omega)] = L_{21}^\ast , 
\end{multline}
and
\begin{multline}\label{eq:MVacFinal} 
M^{(\mathrm{vac})}   =  \frac{ C N_d }{  4  } \int_0^\infty \dif \Omega \int \dif^2 q_\parallel f(\Omega)f(-\Omega)\me^{\mi q_y \delta r } \\ 
\times \cos \left[ \Omega \delta t \right]  R(\vec{q}_\parallel, \Omega) , 
\end{multline} 
with
\begin{align}
    R(\vec{q}_\parallel, \Omega) = \frac{E_\mathrm{vac}^2  \me^{-\frac{q_\parallel^2 \mathrm{w}^2}{4}} \left(1- \frac{q_x^2}{q^2} \right)   \Pi(q_z, \Omega)  }{4\pi\mathrm{Re}[q(\Omega)]}.
\end{align} 
For the monochromatic detection scheme, where $[\omega_\mathrm{min}, \omega_\mathrm{max}] = [\omega_d - \Delta \omega/2,\omega_d +\Delta \omega/2]$ with $\Delta \omega \ll \omega_c, \sigma_\omega $, we find 
\begin{align}
f(\Omega)  & =  \frac{\mathcal{E}(\omega_\mathrm{LO} + \Omega )}{\mathcal{E}(\omega_\mathrm{LO})}, \\
N_d &  = \frac{\eta \Delta \omega}{\hbar \omega_c} \mathcal{E}^2(\omega_\mathrm{LO}).
\end{align}

\subsection{Thermal State} \label{appsec:EvaluateThermal}

Using the definition of the correlation and response function in Eqs.~\eqref{eq:CorrelationFunction} and \eqref{eq:ResponseFunction}, Fourier transforming the THz field to frequency space according to Eq.~\eqref{eq:Fourier}, using Eqs.~\eqref{eq:EEdag} and \eqref{eq:EdagE}, as well as Eq.~\eqref{eq:ResponseGreen}, we find for a THz field in a thermal state
\begin{multline} \label{eq:CRpTemp}
\mathcal{C}(\boldsymbol{\rho}, \tau) - \mi \mathcal{R}^{\prime}(\boldsymbol{\rho}, \tau)   \\
 = - \frac{\hbar \mu_0 \mi}{\pi} \int_0^\infty \dif \Omega \Omega^2  \mathsf{D}(\vec{r}, \vec{r}^\prime, \Omega) \cos[\Omega (t-t^\prime)]\\
+  \frac{2\mu_0 \hbar}{\pi} \int_{0}^\infty \dif \Omega \Omega^2  n_T(\Omega)  \mathrm{Im}[\mathsf{D}(\vec{r}, \vec{r}^\prime, \omega)] \mathrm{cos}[\Omega \tau],
\end{multline}
and 
\begin{multline} \label{eq:CRppTemp}
\mathcal{C}(\boldsymbol{\rho}, \tau) - \mi \mathcal{R}^{\prime\prime}(\boldsymbol{\rho}, \tau)    \\
= \frac{\hbar \mu_0}{\pi} \int_0^\infty \dif \Omega \Omega^2 \mathrm{Im}[ \mathsf{D}(\vec{r}, \vec{r}^\prime, \Omega)] \me^{-\mi \Omega (t-t^\prime)}   \\ +  \frac{2\hbar \mu_0}{\pi} \int_0^\infty \dif \Omega \Omega^2 n_T(\Omega) \mathrm{Im}[ \mathsf{D}(\vec{r}, \vec{r}^\prime, \Omega)] \mathrm{cos}[\Omega (t-t^\prime)] .
\end{multline}
The first term on the right hand side of Eqs.~\eqref{eq:CRpTemp} and \eqref{eq:CRppTemp} are the same as in the zero temperature limit in Eqs.~\eqref{eq:CRp} and \eqref{eq:CRpp}, respectively. We thus find $M = M^{(\mathrm{vac})} + M^{(T)}$ and $L_{11}^{(\mathrm{vac})} + L_{11}^{(T)}$, where $M^{(T)}$ and $L_{11}^{(T)}$ arise due to the second terms on the right hand side of Eqs.~\eqref{eq:CRpTemp} and \eqref{eq:CRppTemp}, respectively. The evaluation of these additional contributions, stemming from thermal fluctuations, follows along similar lines as the calculation performed in the last section. We eventually find
\begin{multline} \label{eq:L11TFinal}
L_{ii}^{(T)}  =  \frac{  C N_d }{  4  } \int_0^\infty \dif \Omega  \int \dif^2 q_\parallel  \mathrm{Re}\left[ R(\vec{q}_\parallel, \Omega) \right] \\ \times   n_T(\Omega) [ f^2(\Omega)  +  f^2(-\Omega) ],
\end{multline} 
\begin{multline} \label{eq:L12TFinal}
L_{12}^{(T)}  =  \frac{  C N_d }{  4  } \int_0^\infty \dif \Omega  \int \dif^2 q_\parallel  \me^{\mi q_y \delta r } \mathrm{Re}\left[ R(\vec{q}_\parallel, \Omega) \right]    \\   n_T(\Omega)  [\me^{\mi \delta t \Omega} f^2(\Omega)  + \me^{-\mi \delta t \Omega}  f^2(-\Omega) ],
\end{multline} 
and 
\begin{multline}\label{eq:MTFinal} 
M^{(T)}   =  \frac{ C N_d}{ 2  } \int_0^\infty \dif \Omega \int \dif^2 q_\parallel   \me^{\mi q_y \delta r }\mathrm{Re}\left[ R(\vec{q}_\parallel, \Omega) \right] \\
\times n_T(\Omega)  f(\Omega) f(-\Omega)\mathrm{cos}(\Omega \delta t).
\end{multline}

\section{Constructing the Entanglement Witness in Eq.~\eqref{eq:WitnessLocal}}  \label{app:EntanglementWintess}

Having obtained the negativity in Section \ref{sec:EntanglementMeasure}, it is straightforward to construct an entanglement witness \cite{guhne2009entanglement}. We calculate the eigenvector $\ket{\psi^{(-)}}$ corresponding to the possibly negative eigenvalue $E_1$ of the partial transpose of the density matrix $\hat{\rho}$ in Eq.~\eqref{eq:rhoDensitym}. The following observable is then by construction an entanglement witness \cite{guhne2009entanglement}
\begin{align} \label{eq:EntanglementWitnessW} 
\hat{\mathcal{W}} & = \left( \ket{\psi^{(-)}}   \bra{\psi^{(-)}} \right)^{T_1}, \\ \label{eq:Eginevecotr1}
\ket{  \psi^{(-)}} & =  \frac{1 }{\sqrt{2}} \left( \ket{10} -  \me^{\mi \varphi_M} \ket{01} \right)
\end{align}
Remember that we defined $\varphi_M$ via $M = |M|  \me^{\mi \varphi_M} $. By construction, for any separable state $\hat{\rho}_\mathrm{sep}$ the expectation value of $\hat{\mathcal{W}}$ is positive \cite{guhne2009entanglement}, while for the reduced density matrix $\hat{\rho}$ in Eq.~\eqref{eq:rhoDensitym} we find $\braket{\hat{\mathcal{W}}} = L_{11} - |M| =  E_1$. 

We further account for the full infinite-dimensional Hilbert space of the two modes by defining a generalized witness in terms of the creation and annihilation operators in Eq.~\eqref{eq:creation}:
\begin{align} \label{eq:EntanglementWitnessW2}
\hat{\mathcal{W}}& = \sum_{n,m=0}^{\infty} \left( \ket{\psi^{(-)}(n,m)}   \bra{\psi^{(-)}(n,m)} \right)^{T_1},
\end{align}
with 
\begin{align}
    \ket{\psi^{(-)}(n,m)} = \frac{1}{n+m+2} (\hat{a}_1^\dagger-  \me^{\mi \varphi_M} \hat{a}^\dagger_2) \ket{n,m} .
\end{align}
Note that the witness in Eq.~\eqref{eq:EntanglementWitnessW2} reduces to the one in Eq.~\eqref{eq:EntanglementWitnessW} if one restricts the sum to the lowest excitation manifold $n=m=0$. Also, as $\hat{\mathcal{W}}$ in Eq.~\eqref{eq:EntanglementWitnessW2} is a sum of entanglement witnesses of the form \eqref{eq:EntanglementWitnessW}, it is itself an entanglement witness.

$\hat{\mathcal{W}}$ in Eq.~\eqref{eq:EntanglementWitnessW2} is an observable on the joint Hilbert space of both modes. To ease the experimental implementation, we decompose $\hat{\mathcal{W}}$ into local observables of the two subsystems, which leads to Eq.~\eqref{eq:WitnessLocal} of the main text.

\section{Single Beam EOS without Shot Noise} \label{app:SingleBeam}

\begin{figure}
\includegraphics[width=1.\columnwidth]{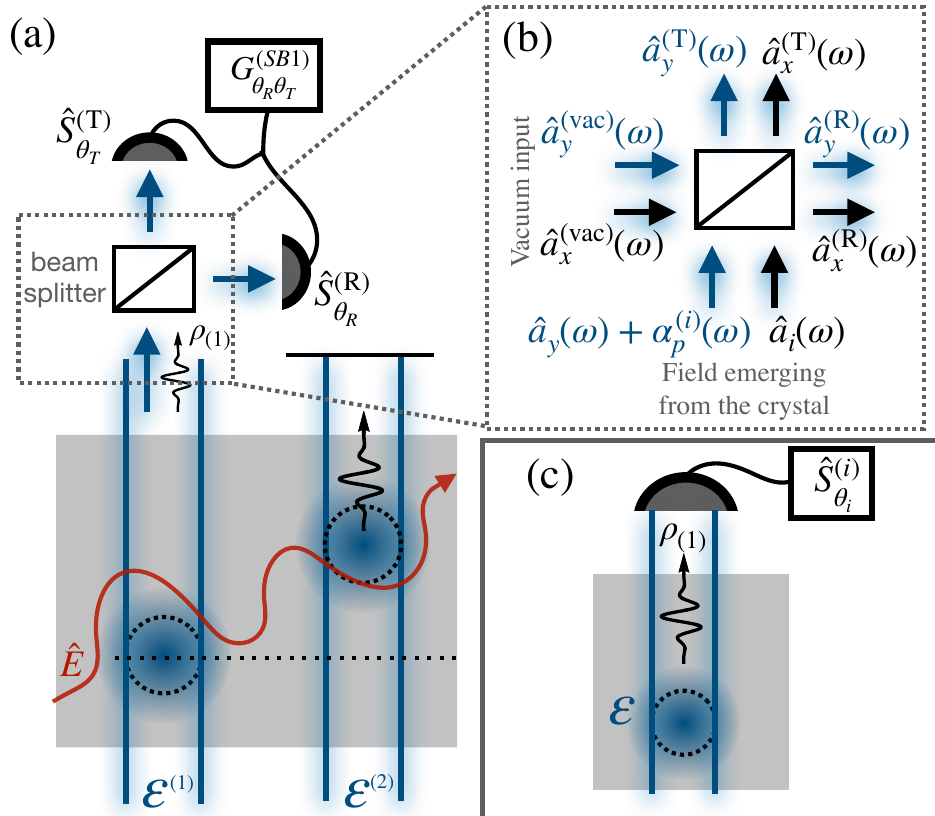}
\caption{\textit{Single-beam EOS setups.} (a) Shot-noise removed EOS setup. Laser pulse and probe mode $1$ enter a beam splitter after emerging from the crystal. The fields at the two output modes of the beam splitter are detected via two ellipsometry measurement schemes as in the two-beam EOS setup, see Fig.~\ref{fig:Setup} (a) and (b). (b) Different fields at the input and output ports of the beam splitter. (c) Standard single-beam EOS setup.}
\label{fig:SingleBeam}
\end{figure}

In subsection \ref{appsec:SingleStand}, we first recap standard single-beam EOS of field fluctuations [see Fig.~\ref{fig:SingleBeam} (c)] as experimentally realized in Ref.~\cite{riek_direct_2015,sulzer2020determination} and theoretically considered in, e.g., Refs.~\cite{moskalenko_paraxial_2015,onoe2022realizing,hubenschmid2022complete,gundougdu2023self}. We find that the measurement scheme is capable of detecting $L_{11}$, but is shot-noise limited. In subsection \ref{appsec:ShotNoiseRemoved}, we derive the signal of the shot-noise removed single-beam EOS setup [see Fig.~\ref{fig:SingleBeam} (a)] in Eq.~\eqref{eq:ShotNoiseFreeSignal}.

\subsection{Single-Beam EOS} \label{appsec:SingleStand}

The single-beam setup under consideration is illustrated in Fig.~\ref{fig:SingleBeam} (c). It is equivalent to the two-beam setup, only that the second laser pulse and its detection scheme is removed. The state $\hat{\rho}_{(i)}$ of the probe mode co-propagating with the single laser pulse can be obtained from the reduced density matrix obtained for the two-beam setup in Eq.~\eqref{eq:rhoDensitym} by tracing out one of the modes:
\begin{align} \label{eq:rhoDensityreducedSingleBeam}
\hat{\rho}_{(i)}(t) =   \left( \begin{array}{ccc}
1- L_{ii}- X& L_i^\ast & K_{i}^\ast \\
L_i & L_{ii} +  X& 0  \\
K_{i}  & 0 &0    \\
\end{array} \right).
\end{align}
The ellipsometry measurement accesses the operator $\hat{S}_{\theta_i}^{(i)}$ defined in Eq.~\eqref{eq:SGeneral}. We assume detection of a THz field that vanishes on average such that $L_i = 0$ and consequently $\langle  \hat{S}_{\theta_i}^{(i)} \rangle  =0$. The variance of $\hat{S}_{\theta_i}^{(i)}$, however, contains information about the fluctuations of the THz field inside the crystal, and reads $\Delta \hat{S}_{\theta_i}^{(i)} \equiv \sqrt{\braket{\hat{S}_{\theta_i}^{(i)2}} -\braket{\hat{S}_{\theta_i}^{(i)}}^2 } $. We find 
\begin{align} \label{eq:SingleBeamSignal}
\Delta \hat{S}_{\theta_i}^{(i)}  & =  N_d - N_d   [  P^2(\theta_i)  \hat{a}_i \hat{a}_i + \mathrm{h.c.} ] + 2 N_d \hat{a}_i^\dagger \hat{a}_i ,
\end{align}
where we used $ |P(\theta_i)|^2 = 1 $ and $[\hat{a}_i, \hat{a}_i^\dagger]  = 1$. By adding two measurements with $P(\pi/2) = \mi $ and $P(\pi) = 1$ we find
\begin{align} \label{eq:Try}
    \Delta \hat{S}_{\frac{\pi}{2}}^{(i)} + \Delta \hat{S}_{\pi}^{(i)}  &=N_d +  4 N_d \braket{ \hat{a}_i^\dagger \hat{a}_i} = N_d +  4 N_d L_{11}
\end{align}
To obtain the last equality sign, we used Eq.~\eqref{eq:rhoDensityreducedSingleBeam}. We thus find that single-beam EOS can be used to measure $\braket{\hat{a}_i^\dagger \hat{a}_i}$ on top of the shot-noise contribution given by the first term on the right hand side of Eqs.~\eqref{eq:SingleBeamSignal} and \eqref{eq:Try}. In case of no frequency filtering, we can use Eq.~\eqref{eq:FiNofilter} in Eq.~\eqref{eq:RhoEntries} to find that Eq.~\eqref{eq:SingleBeamSignal} reduces to
\begin{multline}
\Delta \hat{S}_{\theta_i}^{(i)}    =  N_d + \frac{1}{2}  C N_d^2 \int_{\vec{r}, t, \vec{r}^\prime, t^\prime} \hspace{-.7cm}    \overline{\Ecli}^2(\vec{r}, t)  \overline{\Ecli}^2(\vec{r}^\prime, t^\prime)  \\
\times  \{ \mathcal{C}(\boldsymbol{\rho}, \tau)      - \mathrm{Re} \left[ P^2(\theta_i) \left\{\mathcal{C}(\boldsymbol{\rho}, \tau) - \mi \hbar \mathcal{R}^\prime(\boldsymbol{\rho}, \tau)  \right\}   \right] \}.
\end{multline}
For a quarter wave plate ($\theta_i = \pi/2$) we get
\begin{align}
 \Delta \hat{S}_{\frac{\pi}{2}}^{(i)}  & =  N_d + C  N_d^2 \int_{\vec{r},\vec{r}^\prime, t, t^\prime} \hspace{-0.8cm}    \overline{\Ecli}^2(\vec{r}, t)  \overline{\Ecli}^2(\vec{r}^\prime, t^\prime)  \mathcal{C}(\boldsymbol{\rho}, \tau) .
\end{align} 
For a half wave plate ($\theta_i = 2\pi/3$) we find 
\begin{align}
 \Delta \hat{S}_{\frac{2\pi}{3}}^{(i)}   & =  N_d - \hbar C  N_d^2  \int_{\vec{r},\vec{r}^\prime, t, t^\prime} \hspace{-0.8cm}   \overline{\Ecli}^2(\vec{r}, t)  \overline{\Ecli}^2(\vec{r}^\prime, t^\prime)   \mathcal{R}^\prime(\boldsymbol{\rho}, \tau).
\end{align}
Thus, single-beam EOS can, similarly to the two-beam setup, be used to individually access the correlation function and the reactive part of the response function of the THz quantum field. However, these quantities are only averaged over the space-time volume of the single laser pulse and, thus, correlations and source radiation between different space-time regions can not be resolved, nor can the dissipative part of the response function be measured. Furthermore, these signals only arise on top of the shot noise of the probe beams. In the next section, we introduce a scheme, which overcomes this limitation.

\subsection{Shot-Noise Removed Single-Beam EOS}  \label{appsec:ShotNoiseRemoved}

 We consider the shot-noise removed single-beam setup depicted in Fig.~\ref{fig:SingleBeam} (a). Here, either mode $1$ or $2$ is disregarded, while the other enters one of the two input ports of a beam splitter after emerging from the crystal. The output modes of the beam splitter are expected to have uncorrelated shot noise, if the beam splitter is dissipationless \cite{loudon2000quantum}. This will lead to a shot-noise free detection, as we show below.

We start with deriving input-output relations for the creation and annihilation operator of the lowest-order Laguerre-Gauss modes entering and emerging from the beam splitter depicted in Fig.~\ref{fig:SingleBeam} (b). On the lower input port of the beam splitter in Fig.~\ref{fig:SingleBeam} (b), the $x$-polarized probe field is given by the annihilatoin operator of the probe mode $\hat{a}_i(\omega)$ ($i= 1,2$ depending on which beam is analyzed) while the $y$-polarized field co-propagating with it is given in the vacuum picture \cite{allen1983concepts} as $\hat{a}_y(\omega) + \alpha^{(i)}_p(\omega)$ ($\hat{a}_y(\omega)$: bosonic annihilation operator) with the coherent amplitude of the laser pulse $\alpha_p^{(i)}(\omega) = \mathcal{E}^{(i)}(\omega) \sqrt{4\pi \epsilon_0 n c/\hbar |\omega|}/\mathrm{sgn}[\omega]$ (to match the definitions of the paraxial field and the laser pulses in Eqs.~\eqref{eq:SignalFieldModeExp} and \eqref{eq:laserPulse}, respectively). On the other input port, the field is in its vacuum state with $x$ and $y$-polarized creation and annihilation operators $\hat{a}_x^{(\mathrm{vac})}(\omega)$ and $\hat{a}_y^{(\mathrm{vac})}(\omega)$. The input-output relations, relating the output modes $\hat{a}_{x,y}^{(\mathrm{T})}(\omega)$ and $\hat{a}_{x,y}^{(\mathrm{R})}(\omega)$ to the input modes read \cite{loudon2000quantum}
\begin{align} \nonumber
\hat{a}_y^{(\mathrm{T})}(\omega) & = T [\alpha_p^{(i)}(\omega) + \hat{a}_y(\omega)] + R \hat{a}_y^\mathrm{(vac)}(\omega) \\
&\approx T \alpha_p^{(i)}(\omega) \\ \nonumber
\hat{a}_y^{(\mathrm{R})}(\omega)& = R [\alpha_p^{(i)}(\omega) + \hat{a}_y(\omega)] + T \hat{a}_y^\mathrm{(vac)} (\omega) \\
&\approx R \alpha_p^{(i)}(\omega), \\ \label{eq:InputOutputx1}
\hat{a}_x^{(\mathrm{T})}(\omega) & = T \hat{a}_i(\omega) + R \hat{a}_x^\mathrm{(vac)} (\omega) \\ \label{eq:InputOutputx2}
\hat{a}_x^{(\mathrm{R})}(\omega)& = R \hat{a}_i (\omega)+ T \hat{a}_x^\mathrm{(vac)}(\omega) .
\end{align}
Here, we defined the transmission ($T$) and reflection ($R$) coefficients of the dissipationless beam splitter, which satisfy $T R^\ast + T^\ast R = 0$ and $|T|^2 + |R|^2 = 1$. Also, we used that the laser pulse is sufficiently strong, such that we can replace all $y$ polarized fields by their coherent amplitude. We find a situation very similar to the two-beam EOS setup: We have two modes labelled by $T$ and $R$ in both of which the $y$-polarized field is assumed to be a strong, coherent field, acting as local oscillator in the ellipsometry scheme, while the $x$ polarized field carries the signal. The two elispommetry measurements on mode $T$ and $R$  and its two-point correlation reads in analogy to Eq.~\eqref{eq:SignalG}:
\begin{subequations}\label{eq:SignalOut}.
\begin{align} \label{eq:SignalOuta}
 G^{(\mathrm{SB}i)}_{\theta_T \theta_R} &=  \frac{1}{|R|^2 |T|^2 N_d^2} \left( \braket{\hat{S}^{(\mathrm{T})}_{\theta_T} \hat{S}^{(\mathrm{R})}_{\theta_R}  }  - \braket{ \hat{S}^{(\mathrm{T})}_{\theta_T}} \braket{\hat{S}^{(\mathrm{R})}_{\theta_R}  } \right) \\
\hat{S}^{(\mathrm{J})}_{\theta_J} &= \eta \int_{\omega_\mathrm{min}}^{\omega_\mathrm{max}} \dif \omega \frac{\mathcal{E}(\omega)}{\sqrt{\hbar \omega}} \left[ \mi J^\ast P(\theta_J)   \hat{a}_x^{(\mathrm{J})}(\omega) + \mathrm{h.c.}  \right],
\end{align}
\end{subequations}
with $J=R,T$. We insert Eqs.~\eqref{eq:InputOutputx1} and \eqref{eq:InputOutputx2} into Eq.~\eqref{eq:SignalOut} and trace out the $x$-polarized vacuum input using $\braket{\hat{a}^{(\mathrm{vac})\dagger}_x(\omega) \hat{a}^{(\mathrm{vac})}_x(\omega) } = 0 $ and $[ \hat{a}_x^{(\mathrm{vac})}(\omega) , \hat{a}_x^{(\mathrm{vac})\dagger}(\omega^\prime)] =   \delta (\omega- \omega^\prime )$. Note that $\hat{a}_x^{(\mathrm{R})}(\omega)$ and $\hat{a}_x^{(\mathrm{T})}(\omega)$ are uncorrelated, i.e., they commute, since $T R^\ast + T^\ast R = 0$. We find
\begin{multline} \label{eq:SRSTCorr}
 \braket{\hat{S}^{(\mathrm{T})}_{\theta_1} \hat{S}^{(\mathrm{R})}_{\theta_2}  }  
 =   \frac{N_d}{2}\left(  \mathrm{Re}[P^\ast(\theta_T)P(\theta_R)] \braket{\hat{a}_i^\dagger  \hat{a}_i } \right. \\
 \left. - \sqrt{2}\mathrm{Re}[P(\theta_T)P(\theta_R) \braket{ \hat{a}_i   \hat{a}_i }  ]   \right)  ,
\end{multline} 
where $\hat{a}_i$ is defined in Eq.~\eqref{eq:creation}. Here, we also assumed $T = 1/\sqrt{2}$ and $R = \mi/\sqrt{2}$. If the THz field vanishes on average, such that $L_i \propto \braket{\hat{\vec{E}}(\vec{r}, t)} =  0$, we find Eq.~\eqref{eq:ShotNoiseFreeSignal} of the main text by inserting Eq.~\eqref{eq:SRSTCorr} into Eq.~\eqref{eq:SignalOuta}. 

For completeness, we also give the correlation signal  $G^{(\mathrm{SB}i)}_{\theta_T \theta_R}$ in terms of the components of the reduced density matrix in Eq.~\eqref{eq:rhoDensitym}. Also accounting for non-vanishing $L_{i}$, we find 
\begin{multline}
 G^{(\mathrm{SB}i)}_{\theta_R \theta_T}  =     \frac{2}{N_d}\left( \mathrm{Re}[P^\ast(\theta_R)P(\theta_T)] L_{ii}    \right. \\ \left.  - \sqrt{2}\mathrm{Re}[P(\theta_R)P(\theta_T) K_i   ] \right. \\\left. -    2 \mathrm{Im}\left[ P(\theta_R) L_{i}  \right] \mathrm{Im}\left[ P(\theta_T) L_{i}  \right] \right).
\end{multline}
We see that using different wave-plates in the detection setup, one can use the shot-noise removed single-beam setup to access $\overline{L}_{ii}$ and $ K_i $ without additional shot-noise contribution.

\section{Monotony of the Temperature-Dependent Negativity} \label{app:ProofThermal}

We consider the difference between the negativity for two different temperatures $T_1 $ and $T_2$ with $T_1> T_2$. As $M =M^{\mathrm{(vac)}} + M^{(T)}$ and $L =L^{\mathrm{(vac)}} + L^{(T)}$, we find
\begin{align} \label{eq:ThermalProof1}
\mathcal{N}(T_1)- \mathcal{N}(T_2) = | M^{(T_1)}| -| M^{(T_2)}| - L_{11}^{(T_1)} + L_{11}^{(T_2)} .
\end{align}
Using Eq.~\eqref{eq:MTFinal} we estimate
\begin{multline} \label{eq:ThermalProof2}
 | M^{(T_1)}| -| M^{(T_2)}|\le  \frac{ C }{ 2  N} \int_0^\infty \dif \Omega \int \dif^2 q_\parallel \\
 \times |R(\vec{q}_\parallel, \Omega) | [ n_{T_1}(\Omega) - n_{T_2}(\Omega)]   f(\Omega)f(-\Omega) .
\end{multline}
Since $L_{ii}>0$ ($L_{ii}$ is a diagonal entry of the positive definite reduced density matrix) we find $L_{11}^{(T_2)} - L_{11}^{(T_1)}  =  |L_{11}^{(T_2)} |-  | L_{11}^{(T_1)} |$  and then from Eq.~\eqref{eq:L11TFinal} 
\begin{multline} \label{eq:ThermalProof3}
  |L_{11}^{(T_2)} |-  | L_{11}^{(T_1)} | \le \frac{  C }{  4  N} \int_0^\infty \dif \Omega \int \dif^2 q_\parallel |R(\vec{q}_\parallel, \Omega) | \\
  \times \left[ n_{T_2}(\Omega)- n_{T_1}(\Omega) \right] [ f^2(\Omega)   +  f^2(-\Omega) ].
\end{multline}
Combining Eqs.~\eqref{eq:ThermalProof1}, \eqref{eq:ThermalProof2} and \eqref{eq:ThermalProof3}, we find
\begin{multline}
\mathcal{N}^{(T_1)}- \mathcal{N}^{(T_2)} \le   \frac{ C }{ 2  N} \int_0^\infty \dif \Omega \int \dif^2 q_\parallel|R(\vec{q}_\parallel, \Omega)| \\
\times [ n_{T_2}(\Omega) - n_{T_1}(\Omega)] \left[   f(\Omega) - f(-\Omega) \right]^2 \le 0 ,
\end{multline}
such that 
\begin{align}
\mathcal{N}^{(T_1)} \le  \mathcal{N}^{(T_2)} , \quad \text{for   } T_1> T_2.
\end{align}
Here, we used that $ n_{T_2}(\Omega) < n_{T_1}(\Omega)$ for $T_1 > T_2$.

\section{Fair Sampling Assumption in the EOS Experiment} \label{app:fairsampling}




Before we discuss the additional assumption required for the derivation of the Bell inequality of Eq.~\eqref{eq:Bell}~\cite{zukowski2016bell}, we argue why a similar assumption, the fair sampling assumption, is also necessary to violate Bell inequalities in the traditional entanglement-harvesting protocols using two-level probe systems~\cite{reznik_violating_2005,matsumura_violation_2020}. Here, the state of the two probes after their interaction with the vacuum field 
is entangled but does not lead to a violation of Bell inequalities~\cite{werner_quantum_1989}. Nonlocality has to be `distilled' by means of local filtering operations~\cite{gisin_hidden_1996}: instead of directly measuring each probe, a filtering operation is applied locally before the local measurements. Therefore, each measurement party has a finite probability to not obtain a measurement outcome. To test the Bell inequality, one then only uses the experimental runs for which both local probes produce a measurement outcome~\cite{reznik_violating_2005,matsumura_violation_2020}. Crucially, even though the local filtering operations are independent, the postselection of the measured correlations is collective and not local, i.e., it cannot be decided by each party on their own, and thus requires communication between the parties. This enables LHV models to describe apparent violations of Bell inequalities with postselected correlations if the local detection efficiency depends on the measurement setting, an effect known as the detection loophole~\cite{pearle_hidden_1970}. We note that in the quantum mechanical description, the filtering operation is performed before the final measurement that depends on the setting~\cite{reznik_violating_2005,matsumura_violation_2020}, and thus the detection efficiency is independent of the measurement setting. However, in Bell experiments, one wants to exclude all possible LHV models independently of the quantum description of the setup. As the local filtering operation is not spacelike separated from the final measurement (they together constitute a local measurement station), one requires the fair sampling assumption (i.e., the assumption that, given the LHV $\lambda$, the detection probability is independent of the local measurement setting) on the LHV model to demonstrate Bell inequaltities~\cite{clauser_proposed_1969,berry_fair_2010,gebhart_extending_2022}.

\begin{figure}
\includegraphics[width=1.\columnwidth]{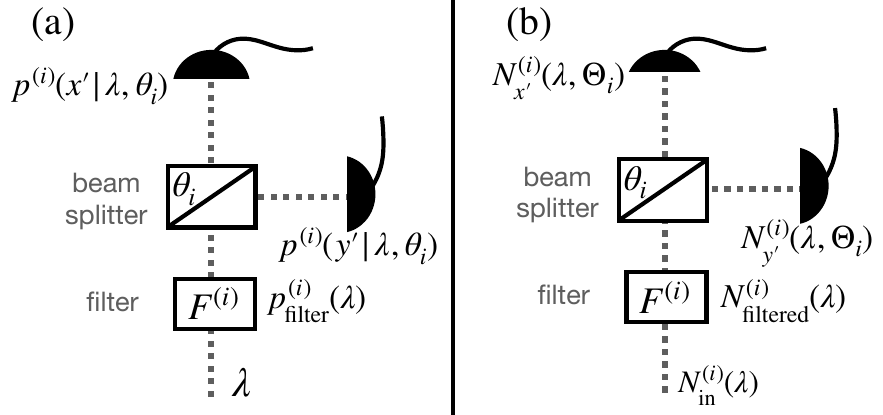}
\caption{\textit{Fair sampling assumption.} (a) LHV model of a single-photon measurement station in standard Bell experiments with the fair sampling assumption. The incoming state labeled by the LHV $\lambda$ (depicted as a dashed line) passes a local filter $F^{(i)}$ with filter probability $p_\mathrm{filter}^{(i)}(\lambda)$, before entering a detector with setting $\theta_i$ that, given the incoming state $\lambda$, gives the measurement outcomes $x'$ and $y'$ with probability $p^{(i)}(x'|\lambda,\theta_i)$ and $p^{(i)}(y'|\lambda,\theta_i)=1-p^{(i)}(x|\lambda,\theta_i)$, respectively.  (b) LHV model of the EOS experiment considered in this work with the fair sampling assumption. The incoming state $N_\textrm{in}^{(i)}(\lambda)$ is transformed to the state $N_\textrm{filtered}^{(i)}(\lambda)$ by a local filter $F^{(i)}$. This state is then split by the detector with measurement setting $\Theta_i$ to yield the final measurement currents $N^{(i)}_{x^\prime }(\lambda,\Theta_i)$ and $N^{(i)}_{y^\prime }(\lambda,\Theta_i)= N_\textrm{filtered}^{(i)}(\lambda)-N^{(i)}_{x^\prime }(\lambda,\Theta_i)$.}
\label{fig:app_fair}
\end{figure}

In the EOS experiment, instead of two-level probe systems, we measure the local probe fields with an ellipsometry detection scheme, see Fig.~\ref{fig:Setup}(b). We thus first outline the fair sampling assumption in typical photonic Bell experiments where a single photon enters a polarizing beam splitter that can be tuned with a local measurement setting $\theta_i$. Each of the outgoing arms then enters a single-photon detector. In the corresponding LHV description, the incoming LHV state labeled by the LHV $\lambda$ is split at the beam splitter and is measured at the $x'$-detector ($y'$-detector) with a probability of $p^{(i)}(x'|\lambda,\theta_i)$ ($p^{(i)}(y'|\lambda,\theta_i)$), see Fig.~\ref{fig:app_fair}(a). The detection efficiency is defined as 
\begin{equation}\label{eq:app:efficiency}
    \eta^{(i)}_\mathrm{det}(\lambda,\theta_i)=p^{(i)}(x|\lambda,\theta_i)+p^{(i)}(y|\lambda,\theta_i).
\end{equation}
If every incoming photon is measured, $\eta_\mathrm{det}^{(i)}(\lambda,\theta_i)=1$, no postselection is needed and the detection loophole is closed. If $\eta^{(i)}_\mathrm{det}(\lambda,\theta_i)<1$ and we postselect the statistics as described above, we must assume fair sampling, i.e., that the detection efficiency $\eta_\mathrm{det}^{(i)}(\lambda,\theta)$ is independent of $\theta_i$. In Fig.~\ref{fig:app_fair}(a), this can be included by assuming $p^{(i)}(x'|\lambda,\theta_i)+p^{(i)}(y'|\lambda,\theta_i)=1$ and by adding a local filter $F^{(i)}$ in front of the beam splitter, that results in a null detection with probability $p_\mathrm{filter}^{(i)}(\lambda)$. In this way, the total detection probability of the measurement station for any $\lambda$ is independent of $\theta_i$. 

In a LHV of the EOS experiment, Eq.~\eqref{eq:LHVmodel}, the incoming state corresponding to the LHV $\lambda$ is given by $N_\textrm{in}^{(i)}(\lambda)$ ($i=1,2$), see Fig.~\ref{fig:app_fair}(b). Assuming fair sampling in this case can be included by adding a local filtering operation $F^{(i)}$ such that the filtered state is given by $N_\textrm{filtered}^{(i)}(\lambda)$. Note that we do not need to assume that $N_\textrm{filtered}^{(i)}(\lambda)=[1-p_\mathrm{filter}^{(i)}(\lambda)]N_\textrm{in}^{(i)}(\lambda)$, which would correspond more directly to the fair sampling assumption for the single photon case, Fig.~\ref{fig:app_fair}(a). Finally, a $\Theta_i$-dependent beam splitter separates the filtered state to result in the measured currents $N^{(i)}_{x^\prime }(\lambda,\Theta_i)$ and $N^{(i)}_{y^\prime }(\lambda,\Theta_i)$, respectively. We assume no losses at this step, $N^{(i)}_{x^\prime }(\lambda,\Theta_i)+N^{(i)}_{y^\prime }(\lambda,\Theta_i)= N_\textrm{filtered}^{(i)}(\lambda)$, such that the assumption needed to prove the Bell inequality of Eq.~\eqref{eq:Bell}, see Ref.~\cite{zukowski2016bell}, is satisfied.

We want to emphasize that, contrary to the two-level-probe entanglement-harvesting protocols where it is necessary to physically add local filters to the experimental setup~\cite{reznik_violating_2005,matsumura_violation_2020}, in the EOS experiment, no additional filters need to be included. Instead, due to the loophole discussed in Ref.~\cite{zukowski2016bell}, a concept similar to these local filters has to be assumed in the LHV descriptions if one wants to use the Bell inequality of Eq.~\eqref{eq:Bell} to exclude LHV models. However, as we discussed in the first paragraph of the Appendix, such an assumption on the LHV models is also needed to show nonlocality in two-level-probe entanglement-harvesting protocols.


\begin{thebibliography}{84}%
\makeatletter
\providecommand \@ifxundefined [1]{%
 \@ifx{#1\undefined}
}%
\providecommand \@ifnum [1]{%
 \ifnum #1\expandafter \@firstoftwo
 \else \expandafter \@secondoftwo
 \fi
}%
\providecommand \@ifx [1]{%
 \ifx #1\expandafter \@firstoftwo
 \else \expandafter \@secondoftwo
 \fi
}%
\providecommand \natexlab [1]{#1}%
\providecommand \enquote  [1]{``#1''}%
\providecommand \bibnamefont  [1]{#1}%
\providecommand \bibfnamefont [1]{#1}%
\providecommand \citenamefont [1]{#1}%
\providecommand \href@noop [0]{\@secondoftwo}%
\providecommand \href [0]{\begingroup \@sanitize@url \@href}%
\providecommand \@href[1]{\@@startlink{#1}\@@href}%
\providecommand \@@href[1]{\endgroup#1\@@endlink}%
\providecommand \@sanitize@url [0]{\catcode `\\12\catcode `\$12\catcode
  `\&12\catcode `\#12\catcode `\^12\catcode `\_12\catcode `\%12\relax}%
\providecommand \@@startlink[1]{}%
\providecommand \@@endlink[0]{}%
\providecommand \url  [0]{\begingroup\@sanitize@url \@url }%
\providecommand \@url [1]{\endgroup\@href {#1}{\urlprefix }}%
\providecommand \urlprefix  [0]{URL }%
\providecommand \Eprint [0]{\href }%
\providecommand \doibase [0]{https://doi.org/}%
\providecommand \selectlanguage [0]{\@gobble}%
\providecommand \bibinfo  [0]{\@secondoftwo}%
\providecommand \bibfield  [0]{\@secondoftwo}%
\providecommand \translation [1]{[#1]}%
\providecommand \BibitemOpen [0]{}%
\providecommand \bibitemStop [0]{}%
\providecommand \bibitemNoStop [0]{.\EOS\space}%
\providecommand \EOS [0]{\spacefactor3000\relax}%
\providecommand \BibitemShut  [1]{\csname bibitem#1\endcsname}%
\let\auto@bib@innerbib\@empty
\bibitem [{\citenamefont {Einstein}\ \emph {et~al.}(1935)\citenamefont
  {Einstein}, \citenamefont {Podolsky},\ and\ \citenamefont
  {Rosen}}]{einstein1935can}%
  \BibitemOpen
  \bibfield  {author} {\bibinfo {author} {\bibfnamefont {A.}~\bibnamefont
  {Einstein}}, \bibinfo {author} {\bibfnamefont {B.}~\bibnamefont {Podolsky}},\
  and\ \bibinfo {author} {\bibfnamefont {N.}~\bibnamefont {Rosen}},\ }\bibfield
   {title} {\bibinfo {title} {Can quantum-mechanical description of physical
  reality be considered complete?},\ }\href@noop {} {\bibfield  {journal}
  {\bibinfo  {journal} {Phys. Rev.}\ }\textbf {\bibinfo {volume} {47}},\
  \bibinfo {pages} {777} (\bibinfo {year} {1935})}\BibitemShut {NoStop}%
\bibitem [{\citenamefont {Bell}(1964)}]{bell1964einstein}%
  \BibitemOpen
  \bibfield  {author} {\bibinfo {author} {\bibfnamefont {J.~S.}\ \bibnamefont
  {Bell}},\ }\bibfield  {title} {\bibinfo {title} {On the einstein podolsky
  rosen paradox},\ }\href@noop {} {\bibfield  {journal} {\bibinfo  {journal}
  {Phys. Phys. Fiz.}\ }\textbf {\bibinfo {volume} {1}},\ \bibinfo {pages} {195}
  (\bibinfo {year} {1964})}\BibitemShut {NoStop}%
\bibitem [{\citenamefont {Horodecki}\ \emph {et~al.}(2009)\citenamefont
  {Horodecki}, \citenamefont {Horodecki}, \citenamefont {Horodecki},\ and\
  \citenamefont {Horodecki}}]{horodecki2009quantum}%
  \BibitemOpen
  \bibfield  {author} {\bibinfo {author} {\bibfnamefont {R.}~\bibnamefont
  {Horodecki}}, \bibinfo {author} {\bibfnamefont {P.}~\bibnamefont
  {Horodecki}}, \bibinfo {author} {\bibfnamefont {M.}~\bibnamefont
  {Horodecki}},\ and\ \bibinfo {author} {\bibfnamefont {K.}~\bibnamefont
  {Horodecki}},\ }\bibfield  {title} {\bibinfo {title} {Quantum entanglement},\
  }\href {https://journals.aps.org/rmp/abstract/10.1103/RevModPhys.81.865} {\bibfield  {journal} {\bibinfo  {journal} {Rev. Mod. Phys.}\
  }\textbf {\bibinfo {volume} {81}},\ \bibinfo {pages} {865} (\bibinfo {year}
  {2009})}\BibitemShut {NoStop}%
\bibitem [{\citenamefont {Horodecki}\ and\ \citenamefont
  {Oppenheim}(2013)}]{horodecki2013quantumness}%
  \BibitemOpen
  \bibfield  {author} {\bibinfo {author} {\bibfnamefont {M.}~\bibnamefont
  {Horodecki}}\ and\ \bibinfo {author} {\bibfnamefont {J.}~\bibnamefont
  {Oppenheim}},\ }\bibfield  {title} {\bibinfo {title} {(Quantumness in the
  context of) resource theories},\ }\href {https://www.worldscientific.com/doi/abs/10.1142/S0217979213450197} {\bibfield  {journal}
  {\bibinfo  {journal} {Int. J. Mod. Phys. B}\ }\textbf {\bibinfo {volume}
  {27}},\ \bibinfo {pages} {1345019} (\bibinfo {year} {2013})}\BibitemShut
  {NoStop}%
\bibitem [{\citenamefont {Chitambar}\ and\ \citenamefont
  {Gour}(2019)}]{chitambar2019quantum}%
  \BibitemOpen
  \bibfield  {author} {\bibinfo {author} {\bibfnamefont {E.}~\bibnamefont
  {Chitambar}}\ and\ \bibinfo {author} {\bibfnamefont {G.}~\bibnamefont
  {Gour}},\ }\bibfield  {title} {\bibinfo {title} {Quantum resource theories},\
  }\href {https://journals.aps.org/rmp/abstract/10.1103/RevModPhys.91.025001} {\bibfield  {journal} {\bibinfo  {journal} {Rev. Mod. Phys.}\
  }\textbf {\bibinfo {volume} {91}},\ \bibinfo {pages} {025001} (\bibinfo
  {year} {2019})}\BibitemShut {NoStop}%
\bibitem [{\citenamefont {Nielsen}\ and\ \citenamefont
  {Chuang}(2010)}]{nielsen2010quantum}%
  \BibitemOpen
  \bibfield  {author} {\bibinfo {author} {\bibfnamefont {M.~A.}\ \bibnamefont
  {Nielsen}}\ and\ \bibinfo {author} {\bibfnamefont {I.~L.}\ \bibnamefont
  {Chuang}},\ }\href@noop {} {\emph {\bibinfo {title} {Quantum computation and
  quantum information}}}\ (\bibinfo  {publisher} {Cambridge University Press,
  Cambridge},\ \bibinfo {year} {2010})\BibitemShut {NoStop}%
\bibitem [{\citenamefont {Gisin}\ \emph {et~al.}(2002)\citenamefont {Gisin},
  \citenamefont {Ribordy}, \citenamefont {Tittel},\ and\ \citenamefont
  {Zbinden}}]{gisin2002quantum}%
  \BibitemOpen
  \bibfield  {author} {\bibinfo {author} {\bibfnamefont {N.}~\bibnamefont
  {Gisin}}, \bibinfo {author} {\bibfnamefont {G.}~\bibnamefont {Ribordy}},
  \bibinfo {author} {\bibfnamefont {W.}~\bibnamefont {Tittel}},\ and\ \bibinfo
  {author} {\bibfnamefont {H.}~\bibnamefont {Zbinden}},\ }\bibfield  {title}
  {\bibinfo {title} {Quantum cryptography},\ }\href {https://journals.aps.org/rmp/abstract/10.1103/RevModPhys.74.145} {\bibfield
  {journal} {\bibinfo  {journal} {Rev. Mod. Phys.}\ }\textbf {\bibinfo {volume}
  {74}},\ \bibinfo {pages} {145} (\bibinfo {year} {2002})}\BibitemShut
  {NoStop}%
\bibitem [{\citenamefont {Summers}\ and\ \citenamefont
  {Werner}(1985)}]{summers_vacuum_1985}%
  \BibitemOpen
  \bibfield  {author} {\bibinfo {author} {\bibfnamefont {S.~J.}\ \bibnamefont
  {Summers}}\ and\ \bibinfo {author} {\bibfnamefont {R.}~\bibnamefont
  {Werner}},\ }\bibfield  {title} {\bibinfo {title} {The vacuum violates
  {Bell}'s inequalities},\ }\href
  {https://doi.org/10.1016/0375-9601(85)90093-3} {\bibfield  {journal}
  {\bibinfo  {journal} {Phys. Lett. A}\ }\textbf {\bibinfo {volume} {110}},\
  \bibinfo {pages} {257} (\bibinfo {year} {1985})}\BibitemShut {NoStop}%
\bibitem [{\citenamefont {Summers}\ and\ \citenamefont
  {Werner}(1987)}]{summers_bells_1987}%
  \BibitemOpen
  \bibfield  {author} {\bibinfo {author} {\bibfnamefont {S.~J.}\ \bibnamefont
  {Summers}}\ and\ \bibinfo {author} {\bibfnamefont {R.}~\bibnamefont
  {Werner}},\ }\bibfield  {title} {\bibinfo {title} {Bell's inequalities and
  quantum field theory. {I}. {General} setting},\ }\href
  {https://doi.org/10.1063/1.527733} {\bibfield  {journal} {\bibinfo  {journal}
  {J. Math. Phys.}\ }\textbf {\bibinfo {volume} {28}},\ \bibinfo {pages} {2440}
  (\bibinfo {year} {1987})}\BibitemShut {NoStop}%
\bibitem [{\citenamefont {Calabrese}\ and\ \citenamefont
  {Cardy}(2004)}]{calabrese2004entanglement}%
  \BibitemOpen
  \bibfield  {author} {\bibinfo {author} {\bibfnamefont {P.}~\bibnamefont
  {Calabrese}}\ and\ \bibinfo {author} {\bibfnamefont {J.}~\bibnamefont
  {Cardy}},\ }\bibfield  {title} {\bibinfo {title} {Entanglement entropy and
  quantum field theory},\ }\href {https://iopscience.iop.org/article/10.1088/1742-5468/2004/06/P06002/meta} {\bibfield  {journal} {\bibinfo
  {journal} {J. Stat. Mech. Theory Exp.}\ }\textbf {\bibinfo {volume} {2004}},\
  \bibinfo {pages} {P06002} (\bibinfo {year} {2004})}\BibitemShut {NoStop}%
\bibitem [{\citenamefont {Witten}(2018)}]{witten2018aps}%
  \BibitemOpen
  \bibfield  {author} {\bibinfo {author} {\bibfnamefont {E.}~\bibnamefont
  {Witten}},\ }\bibfield  {title} {\bibinfo {title} {Aps medal for exceptional
  achievement in research: Invited article on entanglement properties of
  quantum field theory},\ }\href {https://journals.aps.org/rmp/abstract/10.1103/RevModPhys.90.045003} {\bibfield  {journal} {\bibinfo
  {journal} {Rev. Mod. Phys.}\ }\textbf {\bibinfo {volume} {90}},\ \bibinfo
  {pages} {045003} (\bibinfo {year} {2018})}\BibitemShut {NoStop}%
\bibitem [{\citenamefont {Nishioka}(2018)}]{nishioka2018entanglement}%
  \BibitemOpen
  \bibfield  {author} {\bibinfo {author} {\bibfnamefont {T.}~\bibnamefont
  {Nishioka}},\ }\bibfield  {title} {\bibinfo {title} {Entanglement entropy:
  holography and renormalization group},\ }\href {https://journals.aps.org/rmp/abstract/10.1103/RevModPhys.90.035007} {\bibfield  {journal}
  {\bibinfo  {journal} {Rev. Mod. Phys.}\ }\textbf {\bibinfo {volume} {90}},\
  \bibinfo {pages} {035007} (\bibinfo {year} {2018})}\BibitemShut {NoStop}%
\bibitem [{\citenamefont {Preskill}(1992)}]{preskill1992black}%
  \BibitemOpen
  \bibfield  {author} {\bibinfo {author} {\bibfnamefont {J.}~\bibnamefont
  {Preskill}},\ }\bibfield  {title} {\bibinfo {title} {Do black holes destroy
  information},\ }in\ \href {https://www.worldscientific.com/doi/abs/10.1142/9789814536752#page=37} {\emph {\bibinfo {booktitle} {Proceedings
  of the International symposium on black holes, membranes, wormholes and
  superstrings.}}},\ \bibinfo {organization} {World Scientific}\ (\bibinfo
  {publisher} {World Scientific, Singapore},\ \bibinfo {year} {1992})\ pp.\
  \bibinfo {pages} {22--39}\BibitemShut {NoStop}%
\bibitem [{\citenamefont {Hawking}(2005)}]{hawking2005information}%
  \BibitemOpen
  \bibfield  {author} {\bibinfo {author} {\bibfnamefont {S.~W.}\ \bibnamefont
  {Hawking}},\ }\bibfield  {title} {\bibinfo {title} {Information loss in black
  holes},\ }\href {https://journals.aps.org/prd/abstract/10.1103/PhysRevD.48.3743} {\bibfield  {journal} {\bibinfo  {journal} {Phys.
  Rev. D}\ }\textbf {\bibinfo {volume} {72}},\ \bibinfo {pages} {084013}
  (\bibinfo {year} {2005})}\BibitemShut {NoStop}%
\bibitem [{\citenamefont {Susskind}\ \emph {et~al.}(1993)\citenamefont
  {Susskind}, \citenamefont {Thorlacius},\ and\ \citenamefont
  {Uglum}}]{susskind1993stretched}%
  \BibitemOpen
  \bibfield  {author} {\bibinfo {author} {\bibfnamefont {L.}~\bibnamefont
  {Susskind}}, \bibinfo {author} {\bibfnamefont {L.}~\bibnamefont
  {Thorlacius}},\ and\ \bibinfo {author} {\bibfnamefont {J.}~\bibnamefont
  {Uglum}},\ }\bibfield  {title} {\bibinfo {title} {The stretched horizon and
  black hole complementarity},\ }\href@noop {} {\bibfield  {journal} {\bibinfo
  {journal} {Phys. Rev. D}\ }\textbf {\bibinfo {volume} {48}},\ \bibinfo
  {pages} {3743} (\bibinfo {year} {1993})}\BibitemShut {NoStop}%
\bibitem [{\citenamefont {Van~Raamsdonk}(2010)}]{van2010building}%
  \BibitemOpen
  \bibfield  {author} {\bibinfo {author} {\bibfnamefont {M.}~\bibnamefont
  {Van~Raamsdonk}},\ }\bibfield  {title} {\bibinfo {title} {Building up
  space--time with quantum entanglement},\ }\href {https://www.worldscientific.com/doi/abs/10.1142/S0218271810018529} {\bibfield  {journal}
  {\bibinfo  {journal} {Int. J. Mod. Phys. D}\ }\textbf {\bibinfo {volume}
  {19}},\ \bibinfo {pages} {2429} (\bibinfo {year} {2010})}\BibitemShut
  {NoStop}%
\bibitem [{\citenamefont {Lashkari}\ \emph {et~al.}(2014)\citenamefont
  {Lashkari}, \citenamefont {McDermott},\ and\ \citenamefont
  {Van~Raamsdonk}}]{lashkari2014gravitational}%
  \BibitemOpen
  \bibfield  {author} {\bibinfo {author} {\bibfnamefont {N.}~\bibnamefont
  {Lashkari}}, \bibinfo {author} {\bibfnamefont {M.~B.}\ \bibnamefont
  {McDermott}},\ and\ \bibinfo {author} {\bibfnamefont {M.}~\bibnamefont
  {Van~Raamsdonk}},\ }\bibfield  {title} {\bibinfo {title} {Gravitational
  dynamics from entanglement “thermodynamics”},\ }\href {https://link.springer.com/article/10.1007/JHEP04(2014)195} {\bibfield
  {journal} {\bibinfo  {journal} {J. High Energy Phys.}\ }\textbf {\bibinfo
  {volume} {2014}}\bibinfo  {number} { (4)},\ \bibinfo {pages} {1}}\BibitemShut
  {NoStop}%
\bibitem [{\citenamefont {Cao}\ and\ \citenamefont
  {Carroll}(2018)}]{cao2018bulk}%
  \BibitemOpen
\bibfield  {number} {  }\bibfield  {author} {\bibinfo {author} {\bibfnamefont
  {C.~J.}\ \bibnamefont {Cao}}\ and\ \bibinfo {author} {\bibfnamefont {S.~M.}\
  \bibnamefont {Carroll}},\ }\bibfield  {title} {\bibinfo {title} {Bulk
  entanglement gravity without a boundary: Towards finding einstein’s
  equation in hilbert space},\ }\href {https://journals.aps.org/prd/abstract/10.1103/PhysRevD.95.024031} {\bibfield  {journal} {\bibinfo
  {journal} {Phys. Rev. D}\ }\textbf {\bibinfo {volume} {97}},\ \bibinfo
  {pages} {086003} (\bibinfo {year} {2018})}\BibitemShut {NoStop}%
\bibitem [{\citenamefont {Valentini}(1991)}]{valentini_non-local_1991}%
  \BibitemOpen
  \bibfield  {author} {\bibinfo {author} {\bibfnamefont {A.}~\bibnamefont
  {Valentini}},\ }\bibfield  {title} {\bibinfo {title} {Non-local correlations
  in quantum electrodynamics},\ }\href
  {https://doi.org/10.1016/0375-9601(91)90952-5} {\bibfield  {journal}
  {\bibinfo  {journal} {Phys. Lett. A}\ }\textbf {\bibinfo {volume} {153}},\
  \bibinfo {pages} {321} (\bibinfo {year} {1991})}\BibitemShut {NoStop}%
\bibitem [{\citenamefont {Reznik}\ \emph {et~al.}(2005)\citenamefont {Reznik},
  \citenamefont {Retzker},\ and\ \citenamefont
  {Silman}}]{reznik_violating_2005}%
  \BibitemOpen
  \bibfield  {author} {\bibinfo {author} {\bibfnamefont {B.}~\bibnamefont
  {Reznik}}, \bibinfo {author} {\bibfnamefont {A.}~\bibnamefont {Retzker}},\
  and\ \bibinfo {author} {\bibfnamefont {J.}~\bibnamefont {Silman}},\
  }\bibfield  {title} {\bibinfo {title} {Violating {Bell}'s inequalities in
  vacuum},\ }\href {https://doi.org/10.1103/PhysRevA.71.042104} {\bibfield
  {journal} {\bibinfo  {journal} {Phys. Rev. A}\ }\textbf {\bibinfo {volume}
  {71}},\ \bibinfo {pages} {042104} (\bibinfo {year} {2005})}\BibitemShut
  {NoStop}%
\bibitem [{\citenamefont {Franson}(2008)}]{franson_generation_2008}%
  \BibitemOpen
  \bibfield  {author} {\bibinfo {author} {\bibfnamefont {J.~D.}\ \bibnamefont
  {Franson}},\ }\bibfield  {title} {\bibinfo {title} {Generation of
  {Entanglement} {Outside} of the {Light} {Cone}},\ }\href  {https://www.tandfonline.com/doi/full/10.1080/09500340801983129?casa_token=SEe9WCrR-8gAAAAA%3AyrzGKzT6pg6Pp_0Yvf9uMEAewDEn7IVv_q9oO_FoprIc5BJASnhSlhVSYYMGHshvaQCPK6csFRWHOx4} {\bibfield
  {journal} {\bibinfo  {journal} {J. Mod. Opt.} \textbf {\bibinfo {volume} {55}} ,\ \bibinfo {pages} {2117}}
  (\bibinfo {year} {2008})}\BibitemShut {NoStop}%
\bibitem [{\citenamefont {Salton}\ \emph {et~al.}(2015)\citenamefont {Salton},
  \citenamefont {Mann},\ and\ \citenamefont
  {Menicucci}}]{salton2015acceleration}%
  \BibitemOpen
  \bibfield  {author} {\bibinfo {author} {\bibfnamefont {G.}~\bibnamefont
  {Salton}}, \bibinfo {author} {\bibfnamefont {R.~B.}\ \bibnamefont {Mann}},\
  and\ \bibinfo {author} {\bibfnamefont {N.~C.}\ \bibnamefont {Menicucci}},\
  }\bibfield  {title} {\bibinfo {title} {Acceleration-assisted entanglement
  harvesting and rangefinding},\ }\href {https://iopscience.iop.org/article/10.1088/1367-2630/17/3/035001/meta} {\bibfield  {journal} {\bibinfo
   {journal} {New J. Phys.}\ }\textbf {\bibinfo {volume} {17}},\ \bibinfo
  {pages} {035001} (\bibinfo {year} {2015})}\BibitemShut {NoStop}%
\bibitem [{\citenamefont {Pozas-Kerstjens}\ and\ \citenamefont
  {Mart\'{i}n-Mart\'{i}nez}(2015)}]{pozas-kerstjens_harvesting_2015}%
  \BibitemOpen
  \bibfield  {author} {\bibinfo {author} {\bibfnamefont {A.}~\bibnamefont
  {Pozas-Kerstjens}}\ and\ \bibinfo {author} {\bibfnamefont {E.}~\bibnamefont
  {Mart\'{i}n-Mart\'{i}nez}},\ }\bibfield  {title} {\bibinfo {title}
  {Harvesting correlations from the quantum vacuum},\ }\href
  {https://doi.org/10.1103/PhysRevD.92.064042} {\bibfield  {journal} {\bibinfo
  {journal} {Phys. Rev. D}\ }\textbf {\bibinfo {volume} {92}},\ \bibinfo
  {pages} {064042} (\bibinfo {year} {2015})}\BibitemShut {NoStop}%
\bibitem [{\citenamefont {Simidzija}\ and\ \citenamefont
  {Mart\'{i}n-Mart\'{i}nez}(2018)}]{simidzija_harvesting_2018}%
  \BibitemOpen
  \bibfield  {author} {\bibinfo {author} {\bibfnamefont {P.}~\bibnamefont
  {Simidzija}}\ and\ \bibinfo {author} {\bibfnamefont {E.}~\bibnamefont
  {Mart\'{i}n-Mart\'{i}nez}},\ }\bibfield  {title} {\bibinfo {title}
  {Harvesting correlations from thermal and squeezed coherent states},\ }\href
  {https://doi.org/10.1103/PhysRevD.98.085007} {\bibfield  {journal} {\bibinfo
  {journal} {Phys. Rev. D}\ }\textbf {\bibinfo {volume} {98}},\ \bibinfo
  {pages} {085007} (\bibinfo {year} {2018})}\BibitemShut {NoStop}%
\bibitem [{\citenamefont {Henderson}\ \emph
  {et~al.}(2018{\natexlab{a}})\citenamefont {Henderson}, \citenamefont
  {Hennigar}, \citenamefont {Mann}, \citenamefont {Smith},\ and\ \citenamefont
  {Zhang}}]{henderson_harvesting_2018}%
  \BibitemOpen
  \bibfield  {author} {\bibinfo {author} {\bibfnamefont {L.~J.}\ \bibnamefont
  {Henderson}}, \bibinfo {author} {\bibfnamefont {R.~A.}\ \bibnamefont
  {Hennigar}}, \bibinfo {author} {\bibfnamefont {R.~B.}\ \bibnamefont {Mann}},
  \bibinfo {author} {\bibfnamefont {A.~R.~H.}\ \bibnamefont {Smith}},\ and\
  \bibinfo {author} {\bibfnamefont {J.}~\bibnamefont {Zhang}},\ }\bibfield
  {title} {\bibinfo {title} {Harvesting entanglement from the black hole
  vacuum},\ }\href {https://doi.org/10.1088/1361-6382/aae27e} {\bibfield
  {journal} {\bibinfo  {journal} {Class. Quantum Gravity}\ }\textbf {\bibinfo
  {volume} {35}},\ \bibinfo {pages} {21LT02} (\bibinfo {year}
  {2018}{\natexlab{a}})}\BibitemShut {NoStop}%
\bibitem [{\citenamefont {Tjoa}\ and\ \citenamefont
  {Mart\'in-Mart\'inez}(2021)}]{tjoa_when_2021}%
  \BibitemOpen
  \bibfield  {author} {\bibinfo {author} {\bibfnamefont {E.}~\bibnamefont
  {Tjoa}}\ and\ \bibinfo {author} {\bibfnamefont {E.}~\bibnamefont
  {Mart\'in-Mart\'inez}},\ }\bibfield  {title} {\bibinfo {title} {When
  entanglement harvesting is not really harvesting},\ }\href
  {https://doi.org/10.1103/PhysRevD.104.125005} {\bibfield  {journal} {\bibinfo
   {journal} {Phys. Rev. D}\ }\textbf {\bibinfo {volume} {104}},\ \bibinfo
  {pages} {125005} (\bibinfo {year} {2021})}\BibitemShut {NoStop}%
\bibitem [{\citenamefont {Gooding}\ \emph {et~al.}(2023)\citenamefont
  {Gooding}, \citenamefont {Sachs}, \citenamefont {Mann},\ and\ \citenamefont
  {Weinfurtner}}]{gooding2023vacuum}%
  \BibitemOpen
  \bibfield  {author} {\bibinfo {author} {\bibfnamefont {C.}~\bibnamefont
  {Gooding}}, \bibinfo {author} {\bibfnamefont {A.}~\bibnamefont {Sachs}},
  \bibinfo {author} {\bibfnamefont {R.~B.}\ \bibnamefont {Mann}},\ and\
  \bibinfo {author} {\bibfnamefont {S.}~\bibnamefont {Weinfurtner}},\
  }\bibfield  {title} {\bibinfo {title} {Vacuum entanglement probes for
  ultra-cold atom systems},\ }\href {https://arxiv.org/abs/2308.07892} {\bibfield  {journal} {\bibinfo
  {journal} {arXiv preprint arXiv:2308.07892}\ } (\bibinfo {year}
  {2023})}\BibitemShut {NoStop}%
\bibitem [{\citenamefont {Matsumura}\ and\ \citenamefont
  {Nambu}(2020)}]{matsumura_violation_2020}%
  \BibitemOpen
  \bibfield  {author} {\bibinfo {author} {\bibfnamefont {A.}~\bibnamefont
  {Matsumura}}\ and\ \bibinfo {author} {\bibfnamefont {Y.}~\bibnamefont
  {Nambu}},\ }\bibfield  {title} {\bibinfo {title} {Violation of {Bell}-{CHSH}
  {Inequalities} through {Optimal} {Local} {Filters} in the {Vacuum}},\ }\href
  {https://doi.org/10.3390/quantum2040038} {\bibfield  {journal} {\bibinfo
  {journal} {Quantum Rep.}\ }\textbf {\bibinfo {volume} {2}},\ \bibinfo {pages}
  {542} (\bibinfo {year} {2020})}\BibitemShut {NoStop}%
\bibitem [{\citenamefont {Lindel}\ \emph {et~al.}(2023)\citenamefont {Lindel},
  \citenamefont {Herter}, \citenamefont {Faist},\ and\ \citenamefont
  {Buhmann}}]{lindel2023separately}%
  \BibitemOpen
  \bibfield  {author} {\bibinfo {author} {\bibfnamefont {F.}~\bibnamefont
  {Lindel}}, \bibinfo {author} {\bibfnamefont {A.}~\bibnamefont {Herter}},
  \bibinfo {author} {\bibfnamefont {J.}~\bibnamefont {Faist}},\ and\ \bibinfo
  {author} {\bibfnamefont {S.~Y.}\ \bibnamefont {Buhmann}},\ }\bibfield
  {title} {\bibinfo {title} {How to separately probe vacuum field fluctuations
  and source radiation in space and time},\ }\href {https://arxiv.org/abs/2305.06387} {\bibfield
  {journal} {\bibinfo  {journal} {arXiv preprint arXiv:2305.06387}\ } (\bibinfo
  {year} {2023})}\BibitemShut {NoStop}%
\bibitem [{\citenamefont {de~S.~L.~Torres}\ \emph {et~al.}(2023)\citenamefont
  {de~S.~L.~Torres}, \citenamefont {Wurtz}, \citenamefont {Polo-G{\'o}mez},\
  and\ \citenamefont {Mart{\'\i}n-Mart{\'\i}nez}}]{de2023entanglement}%
  \BibitemOpen
  \bibfield  {author} {\bibinfo {author} {\bibfnamefont {B.}~\bibnamefont
  {de~S.~L.~Torres}}, \bibinfo {author} {\bibfnamefont {K.}~\bibnamefont
  {Wurtz}}, \bibinfo {author} {\bibfnamefont {J.}~\bibnamefont
  {Polo-G{\'o}mez}},\ and\ \bibinfo {author} {\bibfnamefont {E.}~\bibnamefont
  {Mart{\'\i}n-Mart{\'\i}nez}},\ }\bibfield  {title} {\bibinfo {title}
  {Entanglement structure of quantum fields through local probes},\ }\href
  {https://link.springer.com/article/10.1007/jhep05(2023)058} {\bibfield  {journal} {\bibinfo  {journal} {J. High Energy Phys.}\
  }\textbf {\bibinfo {volume} {2023}}\bibinfo  {number} { (5)},\ \bibinfo
  {pages} {1}}\BibitemShut {NoStop}%
\bibitem [{\citenamefont {Mart\'{i}n-Mart\'{i}nez}\ \emph
  {et~al.}(2013)\citenamefont {Mart\'{i}n-Mart\'{i}nez}, \citenamefont {Brown},
  \citenamefont {Donnelly},\ and\ \citenamefont
  {Kempf}}]{martin-martinez_sustainable_2013}%
  \BibitemOpen
\bibfield  {number} {  }\bibfield  {author} {\bibinfo {author} {\bibfnamefont
  {E.}~\bibnamefont {Mart\'{i}n-Mart\'{i}nez}}, \bibinfo {author}
  {\bibfnamefont {E.~G.}\ \bibnamefont {Brown}}, \bibinfo {author}
  {\bibfnamefont {W.}~\bibnamefont {Donnelly}},\ and\ \bibinfo {author}
  {\bibfnamefont {A.}~\bibnamefont {Kempf}},\ }\bibfield  {title} {\bibinfo
  {title} {Sustainable entanglement production from a quantum field},\ }\href
  {https://doi.org/10.1103/PhysRevA.88.052310} {\bibfield  {journal} {\bibinfo
  {journal} {Phys. Rev. A}\ }\textbf {\bibinfo {volume} {88}},\ \bibinfo
  {pages} {052310} (\bibinfo {year} {2013})}\BibitemShut {NoStop}%
\bibitem [{\citenamefont {Mann}\ and\ \citenamefont
  {Ralph}(2012)}]{mann2012relativistic}%
  \BibitemOpen
  \bibfield  {author} {\bibinfo {author} {\bibfnamefont {R.~B.}\ \bibnamefont
  {Mann}}\ and\ \bibinfo {author} {\bibfnamefont {T.~C.}\ \bibnamefont
  {Ralph}},\ }\bibfield  {title} {\bibinfo {title} {Relativistic quantum
  information},\ }\href {https://iopscience.iop.org/article/10.1088/0264-9381/29/22/220301/meta} {\bibfield  {journal} {\bibinfo  {journal}
  {Class. Quantum Gravity}\ }\textbf {\bibinfo {volume} {29}},\ \bibinfo
  {pages} {220301} (\bibinfo {year} {2012})}\BibitemShut {NoStop}%
\bibitem [{\citenamefont {Wu}\ and\ \citenamefont {Zhang}(1995)}]{wu1995free}%
  \BibitemOpen
  \bibfield  {author} {\bibinfo {author} {\bibfnamefont {Q.}~\bibnamefont
  {Wu}}\ and\ \bibinfo {author} {\bibfnamefont {X.-C.}\ \bibnamefont {Zhang}},\
  }\bibfield  {title} {\bibinfo {title} {Free-space electro-optic sampling of
  terahertz beams},\ }\href {https://pubs.aip.org/aip/apl/article/67/24/3523/521992/Free-space-electro-optic-sampling-of-terahertz} {\bibfield  {journal} {\bibinfo  {journal}
  {Appl. Phys. Lett.}\ }\textbf {\bibinfo {volume} {67}},\ \bibinfo {pages}
  {3523} (\bibinfo {year} {1995})}\BibitemShut {NoStop}%
\bibitem [{\citenamefont {Wu}\ and\ \citenamefont
  {Zhang}(1996)}]{wu1996ultrafast}%
  \BibitemOpen
  \bibfield  {author} {\bibinfo {author} {\bibfnamefont {Q.}~\bibnamefont
  {Wu}}\ and\ \bibinfo {author} {\bibfnamefont {X.-C.}\ \bibnamefont {Zhang}},\
  }\bibfield  {title} {\bibinfo {title} {Ultrafast electro-optic field
  sensors},\ }\href {https://pubs.aip.org/aip/apl/article/68/12/1604/65288/Ultrafast-electro-optic-field-sensors} {\bibfield  {journal} {\bibinfo  {journal} {Appl.
  Phys. Lett.}\ }\textbf {\bibinfo {volume} {68}},\ \bibinfo {pages} {1604}
  (\bibinfo {year} {1996})}\BibitemShut {NoStop}%
\bibitem [{\citenamefont {Unruh}(1976)}]{unruh1976notes}%
  \BibitemOpen
  \bibfield  {author} {\bibinfo {author} {\bibfnamefont {W.~G.}\ \bibnamefont
  {Unruh}},\ }\bibfield  {title} {\bibinfo {title} {Notes on black-hole
  evaporation},\ }\href {https://journals.aps.org/prd/abstract/10.1103/PhysRevD.14.870} {\bibfield  {journal} {\bibinfo  {journal}
  {Phys. Rev. D}\ }\textbf {\bibinfo {volume} {14}},\ \bibinfo {pages} {870}
  (\bibinfo {year} {1976})}\BibitemShut {NoStop}%
\bibitem [{\citenamefont {Unruh}\ and\ \citenamefont
  {Wald}(1984)}]{unruh1984happens}%
  \BibitemOpen
  \bibfield  {author} {\bibinfo {author} {\bibfnamefont {W.~G.}\ \bibnamefont
  {Unruh}}\ and\ \bibinfo {author} {\bibfnamefont {R.~M.}\ \bibnamefont
  {Wald}},\ }\bibfield  {title} {\bibinfo {title} {What happens when an
  accelerating observer detects a rindler particle},\ }\href {https://journals.aps.org/prd/abstract/10.1103/PhysRevD.29.1047} {\bibfield
   {journal} {\bibinfo  {journal} {Phys. Rev. D}\ }\textbf {\bibinfo {volume}
  {29}},\ \bibinfo {pages} {1047} (\bibinfo {year} {1984})}\BibitemShut
  {NoStop}%
\bibitem [{\citenamefont {Onoe}\ \emph {et~al.}(2022)\citenamefont {Onoe},
  \citenamefont {Guedes}, \citenamefont {Moskalenko}, \citenamefont
  {Leitenstorfer}, \citenamefont {Burkard},\ and\ \citenamefont
  {Ralph}}]{onoe2022realizing}%
  \BibitemOpen
  \bibfield  {author} {\bibinfo {author} {\bibfnamefont {S.}~\bibnamefont
  {Onoe}}, \bibinfo {author} {\bibfnamefont {T.~L.~M.}\ \bibnamefont {Guedes}},
  \bibinfo {author} {\bibfnamefont {A.~S.}\ \bibnamefont {Moskalenko}},
  \bibinfo {author} {\bibfnamefont {A.}~\bibnamefont {Leitenstorfer}}, \bibinfo
  {author} {\bibfnamefont {G.}~\bibnamefont {Burkard}},\ and\ \bibinfo {author}
  {\bibfnamefont {T.~C.}\ \bibnamefont {Ralph}},\ }\bibfield  {title} {\bibinfo
  {title} {Realizing a rapidly switched unruh-dewitt detector through
  electro-optic sampling of the electromagnetic vacuum},\ }\href {https://journals.aps.org/prd/abstract/10.1103/PhysRevD.105.056023}
  {\bibfield  {journal} {\bibinfo  {journal} {Phys. Rev. D}\ }\textbf {\bibinfo
  {volume} {105}},\ \bibinfo {pages} {056023} (\bibinfo {year}
  {2022})}\BibitemShut {NoStop}%
\bibitem [{\citenamefont {Leitenstorfer}\ \emph {et~al.}(1999)\citenamefont
  {Leitenstorfer}, \citenamefont {Hunsche}, \citenamefont {Shah}, \citenamefont
  {Nuss},\ and\ \citenamefont {Knox}}]{leitenstorfer1999detectors}%
  \BibitemOpen
  \bibfield  {author} {\bibinfo {author} {\bibfnamefont {A.}~\bibnamefont
  {Leitenstorfer}}, \bibinfo {author} {\bibfnamefont {S.}~\bibnamefont
  {Hunsche}}, \bibinfo {author} {\bibfnamefont {J.}~\bibnamefont {Shah}},
  \bibinfo {author} {\bibfnamefont {M.}~\bibnamefont {Nuss}},\ and\ \bibinfo
  {author} {\bibfnamefont {W.}~\bibnamefont {Knox}},\ }\bibfield  {title}
  {\bibinfo {title} {Detectors and sources for ultrabroadband electro-optic
  sampling: Experiment and theory},\ }\href {https://pubs.aip.org/aip/apl/article/74/11/1516/69791/Detectors-and-sources-for-ultrabroadband-electro} {\bibfield  {journal}
  {\bibinfo  {journal} {Appl. Phys. Lett.}\ }\textbf {\bibinfo {volume} {74}},\
  \bibinfo {pages} {1516} (\bibinfo {year} {1999})}\BibitemShut {NoStop}%
\bibitem [{\citenamefont {Riek}\ \emph {et~al.}(2015)\citenamefont {Riek},
  \citenamefont {Seletskiy}, \citenamefont {Moskalenko}, \citenamefont
  {Schmidt}, \citenamefont {Krauspe}, \citenamefont {Eckart}, \citenamefont
  {Eggert}, \citenamefont {Burkard},\ and\ \citenamefont
  {Leitenstorfer}}]{riek_direct_2015}%
  \BibitemOpen
  \bibfield  {author} {\bibinfo {author} {\bibfnamefont {C.}~\bibnamefont
  {Riek}}, \bibinfo {author} {\bibfnamefont {D.~V.}\ \bibnamefont {Seletskiy}},
  \bibinfo {author} {\bibfnamefont {A.~S.}\ \bibnamefont {Moskalenko}},
  \bibinfo {author} {\bibfnamefont {J.~F.}\ \bibnamefont {Schmidt}}, \bibinfo
  {author} {\bibfnamefont {P.}~\bibnamefont {Krauspe}}, \bibinfo {author}
  {\bibfnamefont {S.}~\bibnamefont {Eckart}}, \bibinfo {author} {\bibfnamefont
  {S.}~\bibnamefont {Eggert}}, \bibinfo {author} {\bibfnamefont
  {G.}~\bibnamefont {Burkard}},\ and\ \bibinfo {author} {\bibfnamefont
  {A.}~\bibnamefont {Leitenstorfer}},\ }\bibfield  {title} {\bibinfo {title}
  {Direct sampling of electric-field vacuum fluctuations},\ }\href
  {https://doi.org/10.1126/science.aac9788} {\bibfield  {journal} {\bibinfo
  {journal} {Science}\ }\textbf {\bibinfo {volume} {350}},\ \bibinfo {pages}
  {420} (\bibinfo {year} {2015})}\BibitemShut {NoStop}%
\bibitem [{\citenamefont {Riek}\ \emph {et~al.}(2017)\citenamefont {Riek},
  \citenamefont {Sulzer}, \citenamefont {Seeger}, \citenamefont {Moskalenko},
  \citenamefont {Burkard}, \citenamefont {Seletskiy},\ and\ \citenamefont
  {Leitenstorfer}}]{riek2017subcycle}%
  \BibitemOpen
  \bibfield  {author} {\bibinfo {author} {\bibfnamefont {C.}~\bibnamefont
  {Riek}}, \bibinfo {author} {\bibfnamefont {P.}~\bibnamefont {Sulzer}},
  \bibinfo {author} {\bibfnamefont {M.}~\bibnamefont {Seeger}}, \bibinfo
  {author} {\bibfnamefont {A.~S.}\ \bibnamefont {Moskalenko}}, \bibinfo
  {author} {\bibfnamefont {G.}~\bibnamefont {Burkard}}, \bibinfo {author}
  {\bibfnamefont {D.~V.}\ \bibnamefont {Seletskiy}},\ and\ \bibinfo {author}
  {\bibfnamefont {A.}~\bibnamefont {Leitenstorfer}},\ }\bibfield  {title}
  {\bibinfo {title} {Subcycle quantum electrodynamics},\ }\href {https://www.nature.com/articles/nature21024}
  {\bibfield  {journal} {\bibinfo  {journal} {Nature}\ }\textbf {\bibinfo
  {volume} {541}},\ \bibinfo {pages} {376} (\bibinfo {year}
  {2017})}\BibitemShut {NoStop}%
\bibitem [{\citenamefont {Benea-Chelmus}\ \emph {et~al.}(2019)\citenamefont
  {Benea-Chelmus}, \citenamefont {Settembrini}, \citenamefont {Scalari},\ and\
  \citenamefont {Faist}}]{benea-chelmus_electric_2019}%
  \BibitemOpen
  \bibfield  {author} {\bibinfo {author} {\bibfnamefont {I.-C.}\ \bibnamefont
  {Benea-Chelmus}}, \bibinfo {author} {\bibfnamefont {F.~F.}\ \bibnamefont
  {Settembrini}}, \bibinfo {author} {\bibfnamefont {G.}~\bibnamefont
  {Scalari}},\ and\ \bibinfo {author} {\bibfnamefont {J.}~\bibnamefont
  {Faist}},\ }\bibfield  {title} {\bibinfo {title} {Electric field correlation
  measurements on the electromagnetic vacuum state},\ }\href
  {https://doi.org/10.1038/s41586-019-1083-9} {\bibfield  {journal} {\bibinfo
  {journal} {Nature}\ }\textbf {\bibinfo {volume} {568}},\ \bibinfo {pages}
  {202} (\bibinfo {year} {2019})}\BibitemShut {NoStop}%
\bibitem [{\citenamefont {Settembrini}\ \emph {et~al.}(2022)\citenamefont
  {Settembrini}, \citenamefont {Lindel}, \citenamefont {Herter}, \citenamefont
  {Buhmann},\ and\ \citenamefont {Faist}}]{settembrini2022detection}%
  \BibitemOpen
  \bibfield  {author} {\bibinfo {author} {\bibfnamefont {F.~F.}\ \bibnamefont
  {Settembrini}}, \bibinfo {author} {\bibfnamefont {F.}~\bibnamefont {Lindel}},
  \bibinfo {author} {\bibfnamefont {A.~M.}\ \bibnamefont {Herter}}, \bibinfo
  {author} {\bibfnamefont {S.~Y.}\ \bibnamefont {Buhmann}},\ and\ \bibinfo
  {author} {\bibfnamefont {J.}~\bibnamefont {Faist}},\ }\bibfield  {title}
  {\bibinfo {title} {Detection of quantum-vacuum field correlations outside the
  light cone},\ }\href {https://www.nature.com/articles/s41467-022-31081-1} {\bibfield  {journal} {\bibinfo  {journal} {Nat.
  Comm.}\ }\textbf {\bibinfo {volume} {13}},\ \bibinfo {pages} {3383} (\bibinfo
  {year} {2022})}\BibitemShut {NoStop}%
\bibitem [{\citenamefont {Benea-Chelmus}\ \emph {et~al.}(2020)\citenamefont
  {Benea-Chelmus}, \citenamefont {Salamin}, \citenamefont {Settembrini},
  \citenamefont {Fedoryshyn}, \citenamefont {Heni}, \citenamefont {Elder},
  \citenamefont {Dalton}, \citenamefont {Leuthold},\ and\ \citenamefont
  {Faist}}]{benea2020electro}%
  \BibitemOpen
  \bibfield  {author} {\bibinfo {author} {\bibfnamefont {I.-C.}\ \bibnamefont
  {Benea-Chelmus}}, \bibinfo {author} {\bibfnamefont {Y.}~\bibnamefont
  {Salamin}}, \bibinfo {author} {\bibfnamefont {F.~F.}\ \bibnamefont
  {Settembrini}}, \bibinfo {author} {\bibfnamefont {Y.}~\bibnamefont
  {Fedoryshyn}}, \bibinfo {author} {\bibfnamefont {W.}~\bibnamefont {Heni}},
  \bibinfo {author} {\bibfnamefont {D.~L.}\ \bibnamefont {Elder}}, \bibinfo
  {author} {\bibfnamefont {L.~R.}\ \bibnamefont {Dalton}}, \bibinfo {author}
  {\bibfnamefont {J.}~\bibnamefont {Leuthold}},\ and\ \bibinfo {author}
  {\bibfnamefont {J.}~\bibnamefont {Faist}},\ }\bibfield  {title} {\bibinfo
  {title} {Electro-optic interface for ultrasensitive intracavity electric
  field measurements at microwave and terahertz frequencies},\ }\href {https://opg.optica.org/optica/fulltext.cfm?uri=optica-7-5-498&id=431641}
  {\bibfield  {journal} {\bibinfo  {journal} {Optica}\ }\textbf {\bibinfo
  {volume} {7}},\ \bibinfo {pages} {498} (\bibinfo {year} {2020})}\BibitemShut
  {NoStop}%
\bibitem [{\citenamefont {Salamin}\ \emph {et~al.}(2019)\citenamefont
  {Salamin}, \citenamefont {Benea-Chelmus}, \citenamefont {Fedoryshyn},
  \citenamefont {Heni}, \citenamefont {Elder}, \citenamefont {Dalton},
  \citenamefont {Faist},\ and\ \citenamefont {Leuthold}}]{salamin2019compact}%
  \BibitemOpen
  \bibfield  {author} {\bibinfo {author} {\bibfnamefont {Y.}~\bibnamefont
  {Salamin}}, \bibinfo {author} {\bibfnamefont {I.-C.}\ \bibnamefont
  {Benea-Chelmus}}, \bibinfo {author} {\bibfnamefont {Y.}~\bibnamefont
  {Fedoryshyn}}, \bibinfo {author} {\bibfnamefont {W.}~\bibnamefont {Heni}},
  \bibinfo {author} {\bibfnamefont {D.~L.}\ \bibnamefont {Elder}}, \bibinfo
  {author} {\bibfnamefont {L.~R.}\ \bibnamefont {Dalton}}, \bibinfo {author}
  {\bibfnamefont {J.}~\bibnamefont {Faist}},\ and\ \bibinfo {author}
  {\bibfnamefont {J.}~\bibnamefont {Leuthold}},\ }\bibfield  {title} {\bibinfo
  {title} {Compact and ultra-efficient broadband plasmonic terahertz field
  detector},\ }\href {https://www.nature.com/articles/s41467-019-13490-x} {\bibfield  {journal} {\bibinfo  {journal} {Nat.
  Comm.}\ }\textbf {\bibinfo {volume} {10}},\ \bibinfo {pages} {5550} (\bibinfo
  {year} {2019})}\BibitemShut {NoStop}%
\bibitem [{\citenamefont {Kizmann}\ \emph {et~al.}(2022)\citenamefont
  {Kizmann}, \citenamefont {Moskalenko}, \citenamefont {Leitenstorfer},
  \citenamefont {Burkard},\ and\ \citenamefont
  {Mukamel}}]{kizmann_quantum_2022}%
  \BibitemOpen
  \bibfield  {author} {\bibinfo {author} {\bibfnamefont {M.}~\bibnamefont
  {Kizmann}}, \bibinfo {author} {\bibfnamefont {A.~S.}\ \bibnamefont
  {Moskalenko}}, \bibinfo {author} {\bibfnamefont {A.}~\bibnamefont
  {Leitenstorfer}}, \bibinfo {author} {\bibfnamefont {G.}~\bibnamefont
  {Burkard}},\ and\ \bibinfo {author} {\bibfnamefont {S.}~\bibnamefont
  {Mukamel}},\ }\bibfield  {title} {\bibinfo {title} {Quantum
  {Susceptibilities} in {Time} {Domain} {Sampling} of {Electric} {Field}
  fluctuations},\ }\href {https://doi.org/10.1002/lpor.202100423} {\bibfield
  {journal} {\bibinfo  {journal} {Laser Photonics Rev.}\ }\textbf {\bibinfo
  {volume} {16}},\ \bibinfo {pages} {2100423} (\bibinfo {year}
  {2022})}\BibitemShut {NoStop}%
\bibitem [{\citenamefont {Hubenschmid}\ \emph {et~al.}(2022)\citenamefont
  {Hubenschmid}, \citenamefont {Guedes},\ and\ \citenamefont
  {Burkard}}]{hubenschmid2022complete}%
  \BibitemOpen
  \bibfield  {author} {\bibinfo {author} {\bibfnamefont {E.}~\bibnamefont
  {Hubenschmid}}, \bibinfo {author} {\bibfnamefont {T.~L.~M.}\ \bibnamefont
  {Guedes}},\ and\ \bibinfo {author} {\bibfnamefont {G.}~\bibnamefont
  {Burkard}},\ }\bibfield  {title} {\bibinfo {title} {Complete positive
  operator-valued measure description of multichannel quantum electro-optic
  sampling with monochromatic field modes},\ }\href {https://journals.aps.org/pra/abstract/10.1103/PhysRevA.106.043713} {\bibfield
  {journal} {\bibinfo  {journal} {Phys. Rev. A}\ }\textbf {\bibinfo {volume}
  {106}},\ \bibinfo {pages} {043713} (\bibinfo {year} {2022})}\BibitemShut
  {NoStop}%
\bibitem [{\citenamefont {Hubenschmid}\ \emph {et~al.}(2023)\citenamefont
  {Hubenschmid}, \citenamefont {Guedes},\ and\ \citenamefont
  {Burkard}}]{hubenschmid2023optical}%
  \BibitemOpen
  \bibfield  {author} {\bibinfo {author} {\bibfnamefont {E.}~\bibnamefont
  {Hubenschmid}}, \bibinfo {author} {\bibfnamefont {T.~L.}\ \bibnamefont
  {Guedes}},\ and\ \bibinfo {author} {\bibfnamefont {G.}~\bibnamefont
  {Burkard}},\ }\bibfield  {title} {\bibinfo {title} {Optical time-domain
  quantum state tomography on a subcycle scale},\ }\href {https://arxiv.org/abs/2307.13090} {\bibfield
  {journal} {\bibinfo  {journal} {arXiv preprint arXiv:2307.13090}\ } (\bibinfo
  {year} {2023})}\BibitemShut {NoStop}%
\bibitem [{\citenamefont {Onoe}\ \emph {et~al.}(2023)\citenamefont {Onoe},
  \citenamefont {Virally},\ and\ \citenamefont {Seletskiy}}]{onoe2023direct}%
  \BibitemOpen
  \bibfield  {author} {\bibinfo {author} {\bibfnamefont {S.}~\bibnamefont
  {Onoe}}, \bibinfo {author} {\bibfnamefont {S.}~\bibnamefont {Virally}},\ and\
  \bibinfo {author} {\bibfnamefont {D.~V.}\ \bibnamefont {Seletskiy}},\
  }\bibfield  {title} {\bibinfo {title} {Direct measurement of the Husimi-Q
  function of the electric-field in the time-domain},\ }\href {https://arxiv.org/abs/2307.13088}
  {\bibfield  {journal} {\bibinfo  {journal} {arXiv preprint arXiv:2307.13088}\
  } (\bibinfo {year} {2023})}\BibitemShut {NoStop}%
\bibitem [{\citenamefont {Yang}\ \emph {et~al.}(2023)\citenamefont {Yang},
  \citenamefont {Virally},\ and\ \citenamefont {Seletskiy}}]{yang2023}%
  \BibitemOpen
  \bibfield  {author} {\bibinfo {author} {\bibfnamefont {G.}~\bibnamefont
  {Yang}}, \bibinfo {author} {\bibfnamefont {M.}~\bibnamefont {Kizmann}}, \bibinfo {author} {\bibfnamefont {A.}~\bibnamefont {Leitenstorfer}},\ and\
  \bibinfo {author} {\bibfnamefont {A.~S.}\ \bibnamefont {Moskalenko}},\
  }\bibfield  {title} {\bibinfo {title} {Subcycle tomography of quantum light},\ }\href {https://arxiv.org/abs/2307.12812}
  {\bibfield  {journal} {\bibinfo  {journal} {arXiv preprint arXiv:2307.12812}\
  } (\bibinfo {year} {2023})}\BibitemShut {NoStop}%
\bibitem [{\citenamefont {Moskalenko}\ \emph {et~al.}(2015)\citenamefont
  {Moskalenko}, \citenamefont {Riek}, \citenamefont {Seletskiy}, \citenamefont
  {Burkard},\ and\ \citenamefont {Leitenstorfer}}]{moskalenko_paraxial_2015}%
  \BibitemOpen
  \bibfield  {author} {\bibinfo {author} {\bibfnamefont {A.~S.}\ \bibnamefont
  {Moskalenko}}, \bibinfo {author} {\bibfnamefont {C.}~\bibnamefont {Riek}},
  \bibinfo {author} {\bibfnamefont {D.~V.}~\bibnamefont {Seletskiy}}, \bibinfo
  {author} {\bibfnamefont {G.}~\bibnamefont {Burkard}},\ and\ \bibinfo {author}
  {\bibfnamefont {A.}~\bibnamefont {Leitenstorfer}},\ }\bibfield  {title}
  {\bibinfo {title} {Paraxial {Theory} of {Direct} {Electro}-optic {Sampling}
  of the {Quantum} {Vacuum}},\ }\href
  {https://doi.org/10.1103/PhysRevLett.115.263601} {\bibfield  {journal}
  {\bibinfo  {journal} {Phys. Rev. Lett.}\ }\textbf {\bibinfo {volume} {115}},\
  \bibinfo {pages} {263601} (\bibinfo {year} {2015})}\BibitemShut {NoStop}%
\bibitem [{\citenamefont {Lindel}\ \emph {et~al.}(2020)\citenamefont {Lindel},
  \citenamefont {Bennett},\ and\ \citenamefont {Buhmann}}]{lindel2020theory}%
  \BibitemOpen
  \bibfield  {author} {\bibinfo {author} {\bibfnamefont {F.}~\bibnamefont
  {Lindel}}, \bibinfo {author} {\bibfnamefont {R.}~\bibnamefont {Bennett}},\
  and\ \bibinfo {author} {\bibfnamefont {S.~Y.}\ \bibnamefont {Buhmann}},\
  }\bibfield  {title} {\bibinfo {title} {Theory of polaritonic quantum-vacuum
  detection},\ }\href {https://journals.aps.org/pra/abstract/10.1103/PhysRevA.102.041701} {\bibfield  {journal} {\bibinfo  {journal} {Phys.
  Rev. A}\ }\textbf {\bibinfo {volume} {102}},\ \bibinfo {pages} {041701(R)}
  (\bibinfo {year} {2020})}\BibitemShut {NoStop}%
\bibitem [{\citenamefont {Lindel}\ \emph {et~al.}(2021)\citenamefont {Lindel},
  \citenamefont {Bennett},\ and\ \citenamefont
  {Buhmann}}]{lindel2021macroscopic}%
  \BibitemOpen
  \bibfield  {author} {\bibinfo {author} {\bibfnamefont {F.}~\bibnamefont
  {Lindel}}, \bibinfo {author} {\bibfnamefont {R.}~\bibnamefont {Bennett}},\
  and\ \bibinfo {author} {\bibfnamefont {S.~Y.}\ \bibnamefont {Buhmann}},\
  }\bibfield  {title} {\bibinfo {title} {Macroscopic quantum electrodynamics
  approach to nonlinear optics and application to polaritonic quantum-vacuum
  detection},\ }\href {https://journals.aps.org/pra/abstract/10.1103/PhysRevA.103.033705} {\bibfield  {journal} {\bibinfo  {journal} {Phys.
  Rev. A}\ }\textbf {\bibinfo {volume} {103}},\ \bibinfo {pages} {033705}
  (\bibinfo {year} {2021})}\BibitemShut {NoStop}%
\bibitem [{\citenamefont {Lindel}\ \emph {et~al.}(2022)\citenamefont {Lindel},
  \citenamefont {Settembrini}, \citenamefont {Bennett},\ and\ \citenamefont
  {Buhmann}}]{lindel2022probing}%
  \BibitemOpen
  \bibfield  {author} {\bibinfo {author} {\bibfnamefont {F.}~\bibnamefont
  {Lindel}}, \bibinfo {author} {\bibfnamefont {F.~F.}\ \bibnamefont
  {Settembrini}}, \bibinfo {author} {\bibfnamefont {R.}~\bibnamefont
  {Bennett}},\ and\ \bibinfo {author} {\bibfnamefont {S.~Y.}\ \bibnamefont
  {Buhmann}},\ }\bibfield  {title} {\bibinfo {title} {Probing the purcell
  effect without radiative decay: lessons in the frequency and time domains},\
  }\href {https://iopscience.iop.org/article/10.1088/1367-2630/ac434e/meta} {\bibfield  {journal} {\bibinfo  {journal} {New J. Phys.}\
  }\textbf {\bibinfo {volume} {24}},\ \bibinfo {pages} {013006} (\bibinfo
  {year} {2022})}\BibitemShut {NoStop}%
\bibitem [{\citenamefont {Pozas-Kerstjens}\ and\ \citenamefont
  {Mart\'{i}n-Mart\'{i}nez}(2016)}]{pozas-kerstjens_entanglement_2016}%
  \BibitemOpen
  \bibfield  {author} {\bibinfo {author} {\bibfnamefont {A.}~\bibnamefont
  {Pozas-Kerstjens}}\ and\ \bibinfo {author} {\bibfnamefont {E.}~\bibnamefont
  {Mart\'{i}n-Mart\'{i}nez}},\ }\bibfield  {title} {\bibinfo {title}
  {Entanglement harvesting from the electromagnetic vacuum with hydrogenlike
  atoms},\ }\href {https://doi.org/10.1103/PhysRevD.94.064074} {\bibfield
  {journal} {\bibinfo  {journal} {Phys. Rev. D}\ }\textbf {\bibinfo {volume}
  {94}},\ \bibinfo {pages} {064074} (\bibinfo {year} {2016})}\BibitemShut
  {NoStop}%
\bibitem [{\citenamefont {Scheel}\ and\ \citenamefont
  {Buhmann}(2008)}]{scheel2009macroscopic}%
  \BibitemOpen
  \bibfield  {author} {\bibinfo {author} {\bibfnamefont {S.}~\bibnamefont
  {Scheel}}\ and\ \bibinfo {author} {\bibfnamefont {S.~Y.}\ \bibnamefont
  {Buhmann}},\ }\bibfield  {title} {\bibinfo {title} {Macroscopic qed-concepts
  and applications},\ }\href@noop {} {\bibfield  {journal} {\bibinfo  {journal}
  {Acta Phys. Slovaca}\ }\textbf {\bibinfo {volume} {58}},\ \bibinfo {pages}
  {675} (\bibinfo {year} {2008})}\BibitemShut {NoStop}%
\bibitem [{\citenamefont {Buhmann}(2012)}]{buhmann2013dispersion}%
  \BibitemOpen
  \bibfield  {author} {\bibinfo {author} {\bibfnamefont {S.~Y.}\ \bibnamefont
  {Buhmann}},\ }\href@noop {} {\emph {\bibinfo {title} {Dispersion Forces I}}}\
  (\bibinfo  {publisher} {Springer, Heidelberg},\ \bibinfo {year}
  {2012})\BibitemShut {NoStop}%
\bibitem [{\citenamefont {Loudon}(2000)}]{loudon2000quantum}%
  \BibitemOpen
  \bibfield  {author} {\bibinfo {author} {\bibfnamefont {R.}~\bibnamefont
  {Loudon}},\ }\href@noop {} {\emph {\bibinfo {title} {The quantum theory of
  light}}}\ (\bibinfo  {publisher} {Oxford University Press, New York},\
  \bibinfo {year} {2000})\BibitemShut {NoStop}%
\bibitem [{Note1()}]{Note1}%
  \BibitemOpen
  \bibinfo {note} {Note that we do not normalize the signal by the detection
  efficiency as in Refs.~\cite
  {benea-chelmus_electric_2019,lindel2021macroscopic,settembrini2022detection,lindel2023separately}.
  Also, we included the term $\propto -\langle \protect \hat {S}^{(1)}(\theta
  _1) \rangle \langle \protect \hat {S}^{(2)}(\theta _2) \rangle $ as in
  Ref.~\cite {lindel2023separately}, which vanishes in second order in $\chi
  ^{(2)}$ in case the THz field is in its vacuum or thermal state, as
  considered in Refs.~\cite
  {benea-chelmus_electric_2019,lindel2021macroscopic,settembrini2022detection}.}\BibitemShut
  {Stop}%
\bibitem [{\citenamefont {Tjoa}\ and\ \citenamefont
  {Mann}(2020)}]{tjoa_harvesting_2020}%
  \BibitemOpen
  \bibfield  {author} {\bibinfo {author} {\bibfnamefont {E.}~\bibnamefont
  {Tjoa}}\ and\ \bibinfo {author} {\bibfnamefont {R.~B.}\ \bibnamefont
  {Mann}},\ }\bibfield  {title} {\bibinfo {title} {Harvesting correlations in
  {Schwarzschild} and collapsing shell spacetimes},\ }\href
  {https://doi.org/10.1007/JHEP08(2020)155} {\bibfield  {journal} {\bibinfo
  {journal} {J. High Energy Phys.}\ }\textbf {\bibinfo {volume} {2020}}\bibinfo
   {number} { (8)},\ \bibinfo {pages} {155}}\BibitemShut {NoStop}%
\bibitem [{\citenamefont {Sulzer}\ \emph {et~al.}(2020)\citenamefont {Sulzer},
  \citenamefont {Oguchi}, \citenamefont {Huster}, \citenamefont {Kizmann},
  \citenamefont {Guedes}, \citenamefont {Liehl}, \citenamefont {Beckh},
  \citenamefont {Moskalenko}, \citenamefont {Burkard}, \citenamefont
  {Seletskiy} \emph {et~al.}}]{sulzer2020determination}%
  \BibitemOpen
\bibfield  {number} {  }\bibfield  {author} {\bibinfo {author} {\bibfnamefont
  {P.}~\bibnamefont {Sulzer}}, \bibinfo {author} {\bibfnamefont
  {K.}~\bibnamefont {Oguchi}}, \bibinfo {author} {\bibfnamefont
  {J.}~\bibnamefont {Huster}}, \bibinfo {author} {\bibfnamefont
  {M.}~\bibnamefont {Kizmann}}, \bibinfo {author} {\bibfnamefont {T.~L.}\
  \bibnamefont {Guedes}}, \bibinfo {author} {\bibfnamefont {A.}~\bibnamefont
  {Liehl}}, \bibinfo {author} {\bibfnamefont {C.}~\bibnamefont {Beckh}},
  \bibinfo {author} {\bibfnamefont {A.~S.}\ \bibnamefont {Moskalenko}},
  \bibinfo {author} {\bibfnamefont {G.}~\bibnamefont {Burkard}}, \bibinfo
  {author} {\bibfnamefont {D.~V.}\ \bibnamefont {Seletskiy}}, \bibinfo
  {author} {\bibfnamefont {A.}\ \bibnamefont {Leitenstorfer}},\
  }\bibfield  {title} {\bibinfo {title} {Determination of the electric field
  and its hilbert transform in femtosecond electro-optic sampling},\
  }\href {https://journals.aps.org/pra/abstract/10.1103/PhysRevA.101.033821} {\bibfield  {journal} {\bibinfo  {journal} {Phys. Rev. A}\
  }\textbf {\bibinfo {volume} {101}},\ \bibinfo {pages} {033821} (\bibinfo
  {year} {2020})}\BibitemShut {NoStop}%
\bibitem [{\citenamefont {{\.Z}ukowski}\ \emph {et~al.}(2016)\citenamefont
  {{\.Z}ukowski}, \citenamefont {Wie{\'s}niak},\ and\ \citenamefont
  {Laskowski}}]{zukowski2016bell}%
  \BibitemOpen
  \bibfield  {author} {\bibinfo {author} {\bibfnamefont {M.}~\bibnamefont
  {{\.Z}ukowski}}, \bibinfo {author} {\bibfnamefont {M.}~\bibnamefont
  {Wie{\'s}niak}},\ and\ \bibinfo {author} {\bibfnamefont {W.}~\bibnamefont
  {Laskowski}},\ }\bibfield  {title} {\bibinfo {title} {Bell inequalities for
  quantum optical fields},\ }\href {https://journals.aps.org/pra/abstract/10.1103/PhysRevA.94.020102} {\bibfield  {journal} {\bibinfo
  {journal} {Phys. Rev. A}\ }\textbf {\bibinfo {volume} {94}},\ \bibinfo
  {pages} {020102(R)} (\bibinfo {year} {2016})}\BibitemShut {NoStop}%
\bibitem [{\citenamefont {Reid}\ and\ \citenamefont
  {Walls}(1986)}]{reid1986violations}%
  \BibitemOpen
  \bibfield  {author} {\bibinfo {author} {\bibfnamefont {M.~D.}\ \bibnamefont
  {Reid}}\ and\ \bibinfo {author} {\bibfnamefont {D.~F.}\ \bibnamefont
  {Walls}},\ }\bibfield  {title} {\bibinfo {title} {Violations of classical
  inequalities in quantum optics},\ }\href {https://journals.aps.org/pra/abstract/10.1103/PhysRevA.34.1260} {\bibfield  {journal}
  {\bibinfo  {journal} {Phys. Rev. A}\ }\textbf {\bibinfo {volume} {34}},\
  \bibinfo {pages} {1260} (\bibinfo {year} {1986})}\BibitemShut {NoStop}%
\bibitem [{\citenamefont {Rajabali}\ and\ \citenamefont
  {Benea-Chelmus}(2023)}]{rajabali2023present}%
  \BibitemOpen
  \bibfield  {author} {\bibinfo {author} {\bibfnamefont {S.}~\bibnamefont
  {Rajabali}}\ and\ \bibinfo {author} {\bibfnamefont {I.-C.}\ \bibnamefont
  {Benea-Chelmus}},\ }\bibfield  {title} {\bibinfo {title} {Present and future
  of terahertz integrated photonic devices},\ }\href {https://pubs.aip.org/aip/app/article/8/8/080901/2905252} {\bibfield
  {journal} {\bibinfo  {journal} {APL Photonics}\ }\textbf {\bibinfo {volume}
  {8}} (\bibinfo {year} {2023})}\BibitemShut {NoStop}%
\bibitem [{\citenamefont {Tsirelsoon}(1980)}]{cirel1980quantum}%
  \BibitemOpen
  \bibfield  {author} {\bibinfo {author} {\bibfnamefont {B.~S.}\ \bibnamefont
  {Tsirelsoon}},\ }\bibfield  {title} {\bibinfo {title} {Quantum
  generalizations of bell's inequality},\ }\href {https://link.springer.com/article/10.1007/bf00417500} {\bibfield  {journal}
  {\bibinfo  {journal} {Lett. Math. Phys.}\ }\textbf {\bibinfo {volume} {4}},\
  \bibinfo {pages} {93} (\bibinfo {year} {1980})}\BibitemShut {NoStop}%
\bibitem [{\citenamefont {Clauser}\ \emph {et~al.}(1969)\citenamefont
  {Clauser}, \citenamefont {Horne}, \citenamefont {Shimony},\ and\
  \citenamefont {Holt}}]{clauser_proposed_1969}%
  \BibitemOpen
  \bibfield  {author} {\bibinfo {author} {\bibfnamefont {J.~F.}\ \bibnamefont
  {Clauser}}, \bibinfo {author} {\bibfnamefont {M.~A.}\ \bibnamefont {Horne}},
  \bibinfo {author} {\bibfnamefont {A.}~\bibnamefont {Shimony}},\ and\ \bibinfo
  {author} {\bibfnamefont {R.~A.}\ \bibnamefont {Holt}},\ }\bibfield  {title}
  {\bibinfo {title} {Proposed experiment to test local hidden-variable
  theories},\ }\href {https://doi.org/10.1103/PhysRevLett.23.880} {\bibfield
  {journal} {\bibinfo  {journal} {Phys. Rev. Lett.}\ }\textbf {\bibinfo
  {volume} {23}},\ \bibinfo {pages} {880} (\bibinfo {year} {1969})}\BibitemShut
  {NoStop}%
\bibitem [{\citenamefont {Berry}\ \emph {et~al.}(2010)\citenamefont {Berry},
  \citenamefont {Jeong}, \citenamefont {Stobi\ifmmode~\acute{n}\else
  \'{n}\fi{}ska},\ and\ \citenamefont {Ralph}}]{berry_fair_2010}%
  \BibitemOpen
  \bibfield  {author} {\bibinfo {author} {\bibfnamefont {D.~W.}\ \bibnamefont
  {Berry}}, \bibinfo {author} {\bibfnamefont {H.}~\bibnamefont {Jeong}},
  \bibinfo {author} {\bibfnamefont {M.}~\bibnamefont
  {Stobi\ifmmode~\acute{n}\else \'{n}\fi{}ska}},\ and\ \bibinfo {author}
  {\bibfnamefont {T.~C.}\ \bibnamefont {Ralph}},\ }\bibfield  {title} {\bibinfo
  {title} {Fair-sampling assumption is not necessary for testing local
  realism},\ }\href {https://doi.org/10.1103/PhysRevA.81.012109} {\bibfield
  {journal} {\bibinfo  {journal} {Phys. Rev. A}\ }\textbf {\bibinfo {volume}
  {81}},\ \bibinfo {pages} {012109} (\bibinfo {year} {2010})}\BibitemShut
  {NoStop}%
\bibitem [{\citenamefont {Gebhart}\ and\ \citenamefont
  {Smerzi}(2023)}]{gebhart_extending_2022}%
  \BibitemOpen
  \bibfield  {author} {\bibinfo {author} {\bibfnamefont {V.}~\bibnamefont
  {Gebhart}}\ and\ \bibinfo {author} {\bibfnamefont {A.}~\bibnamefont
  {Smerzi}},\ }\bibfield  {title} {\bibinfo {title} {Extending the fair
  sampling assumption using causal diagrams},\ }\href
  {https://doi.org/10.22331/q-2023-01-13-897} {\bibfield  {journal} {\bibinfo
  {journal} {{Quantum}}\ }\textbf {\bibinfo {volume} {7}},\ \bibinfo {pages}
  {897} (\bibinfo {year} {2023})}\BibitemShut {NoStop}%
\bibitem [{\citenamefont {Pearle}(1970)}]{pearle_hidden_1970}%
  \BibitemOpen
  \bibfield  {author} {\bibinfo {author} {\bibfnamefont {P.~M.}\ \bibnamefont
  {Pearle}},\ }\bibfield  {title} {\bibinfo {title} {Hidden-variable example
  based upon data rejection},\ }\href {https://doi.org/10.1103/PhysRevD.2.1418}
  {\bibfield  {journal} {\bibinfo  {journal} {Phys. Rev. D}\ }\textbf {\bibinfo
  {volume} {2}},\ \bibinfo {pages} {1418} (\bibinfo {year} {1970})}\BibitemShut
  {NoStop}%
\bibitem [{\citenamefont {Benea-Chelmus}\ \emph {et~al.}(2016)\citenamefont
  {Benea-Chelmus}, \citenamefont {Bonzon}, \citenamefont {Maissen},
  \citenamefont {Scalari}, \citenamefont {Beck},\ and\ \citenamefont
  {Faist}}]{benea-chelmus_subcycle_2016}%
  \BibitemOpen
  \bibfield  {author} {\bibinfo {author} {\bibfnamefont {I.-C.}\ \bibnamefont
  {Benea-Chelmus}}, \bibinfo {author} {\bibfnamefont {C.}~\bibnamefont
  {Bonzon}}, \bibinfo {author} {\bibfnamefont {C.}~\bibnamefont {Maissen}},
  \bibinfo {author} {\bibfnamefont {G.}~\bibnamefont {Scalari}}, \bibinfo
  {author} {\bibfnamefont {M.}~\bibnamefont {Beck}},\ and\ \bibinfo {author}
  {\bibfnamefont {J.}~\bibnamefont {Faist}},\ }\bibfield  {title} {\bibinfo
  {title} {Subcycle measurement of intensity correlations in the terahertz
  frequency range},\ }\href {https://doi.org/10.1103/PhysRevA.93.043812}
  {\bibfield  {journal} {\bibinfo  {journal} {Phys. Rev. A}\ }\textbf {\bibinfo
  {volume} {93}},\ \bibinfo {pages} {043812} (\bibinfo {year}
  {2016})}\BibitemShut {NoStop}%
\bibitem [{\citenamefont {Markmann}\ \emph {et~al.}(2023)\citenamefont
  {Markmann}, \citenamefont {Stark}, \citenamefont {Singleton}, \citenamefont
  {Beck}, \citenamefont {Faist},\ and\ \citenamefont
  {Scalari}}]{markmann2023electro}%
  \BibitemOpen
  \bibfield  {author} {\bibinfo {author} {\bibfnamefont {S.}~\bibnamefont
  {Markmann}}, \bibinfo {author} {\bibfnamefont {D.}~\bibnamefont {Stark}},
  \bibinfo {author} {\bibfnamefont {M.}~\bibnamefont {Singleton}}, \bibinfo
  {author} {\bibfnamefont {M.}~\bibnamefont {Beck}}, \bibinfo {author}
  {\bibfnamefont {J.}~\bibnamefont {Faist}},\ and\ \bibinfo {author}
  {\bibfnamefont {G.}~\bibnamefont {Scalari}},\ }\bibfield  {title} {\bibinfo
  {title} {Electro-optic sampling of a free-running terahertz
  quantum-cascade-laser frequency comb},\ }\href {https://journals.aps.org/prapplied/abstract/10.1103/PhysRevApplied.19.064063} {\bibfield  {journal}
  {\bibinfo  {journal} {Phys. Rev. Appl.}\ }\textbf {\bibinfo {volume} {19}},\
  \bibinfo {pages} {064063} (\bibinfo {year} {2023})}\BibitemShut {NoStop}%
\bibitem [{\citenamefont {Dorfman}\ \emph {et~al.}(2016)\citenamefont
  {Dorfman}, \citenamefont {Schlawin},\ and\ \citenamefont
  {Mukamel}}]{dorfman_nonlinear_2016}%
  \BibitemOpen
  \bibfield  {author} {\bibinfo {author} {\bibfnamefont {K.~E.}\ \bibnamefont
  {Dorfman}}, \bibinfo {author} {\bibfnamefont {F.}~\bibnamefont {Schlawin}},\
  and\ \bibinfo {author} {\bibfnamefont {S.}~\bibnamefont {Mukamel}},\
  }\bibfield  {title} {\bibinfo {title} {Nonlinear optical signals and
  spectroscopy with quantum light},\ }\href
  {https://doi.org/10.1103/RevModPhys.88.045008} {\bibfield  {journal}
  {\bibinfo  {journal} {Rev. Mod. Phys.}\ }\textbf {\bibinfo {volume} {88}},\
  \bibinfo {pages} {045008} (\bibinfo {year} {2016})}\BibitemShut {NoStop}%
\bibitem [{\citenamefont {Fuentes-Schuller}\ and\ \citenamefont
  {Mann}(2005)}]{fuentes2005alice}%
  \BibitemOpen
  \bibfield  {author} {\bibinfo {author} {\bibfnamefont {I.}~\bibnamefont
  {Fuentes-Schuller}}\ and\ \bibinfo {author} {\bibfnamefont {R.~B.}\
  \bibnamefont {Mann}},\ }\bibfield  {title} {\bibinfo {title} {Alice falls
  into a black hole: entanglement in noninertial frames},\ }\href {https://journals.aps.org/prl/abstract/10.1103/PhysRevLett.95.120404}
  {\bibfield  {journal} {\bibinfo  {journal} {Phys. Rev. Lett.}\ }\textbf
  {\bibinfo {volume} {95}},\ \bibinfo {pages} {120404} (\bibinfo {year}
  {2005})}\BibitemShut {NoStop}%
\bibitem [{\citenamefont {Philbin}\ \emph {et~al.}(2008)\citenamefont
  {Philbin}, \citenamefont {Kuklewicz}, \citenamefont {Robertson},
  \citenamefont {Hill}, \citenamefont {K\"{o}nig},\ and\ \citenamefont
  {Leonhardt}}]{philbin_fiber-optical_2008}%
  \BibitemOpen
  \bibfield  {author} {\bibinfo {author} {\bibfnamefont {T.~G.}\ \bibnamefont
  {Philbin}}, \bibinfo {author} {\bibfnamefont {C.}~\bibnamefont {Kuklewicz}},
  \bibinfo {author} {\bibfnamefont {S.}~\bibnamefont {Robertson}}, \bibinfo
  {author} {\bibfnamefont {S.}~\bibnamefont {Hill}}, \bibinfo {author}
  {\bibfnamefont {F.}~\bibnamefont {K\"{o}nig}},\ and\ \bibinfo {author}
  {\bibfnamefont {U.}~\bibnamefont {Leonhardt}},\ }\bibfield  {title} {\bibinfo
  {title} {Fiber-{Optical} {Analog} of the {Event} {Horizon}},\ }\href
  {https://doi.org/10.1126/science.1153625} {\bibfield  {journal} {\bibinfo
  {journal} {Science}\ }\textbf {\bibinfo {volume} {319}},\ \bibinfo {pages}
  {1367} (\bibinfo {year} {2008})}\BibitemShut {NoStop}%
\bibitem [{\citenamefont {Kizmann}\ \emph {et~al.}(2019)\citenamefont
  {Kizmann}, \citenamefont {Guedes}, \citenamefont {Seletskiy}, \citenamefont
  {Moskalenko}, \citenamefont {Leitenstorfer},\ and\ \citenamefont
  {Burkard}}]{kizmann2019subcycle}%
  \BibitemOpen
  \bibfield  {author} {\bibinfo {author} {\bibfnamefont {M.}~\bibnamefont
  {Kizmann}}, \bibinfo {author} {\bibfnamefont {T.~L.~M.}\ \bibnamefont
  {Guedes}}, \bibinfo {author} {\bibfnamefont {D.~V.}\ \bibnamefont
  {Seletskiy}}, \bibinfo {author} {\bibfnamefont {A.~S.}\ \bibnamefont
  {Moskalenko}}, \bibinfo {author} {\bibfnamefont {A.}~\bibnamefont
  {Leitenstorfer}},\ and\ \bibinfo {author} {\bibfnamefont {G.}~\bibnamefont
  {Burkard}},\ }\bibfield  {title} {\bibinfo {title} {Subcycle squeezing of
  light from a time flow perspective},\ }\href {https://www.nature.com/articles/s41567-019-0560-2} {\bibfield  {journal}
  {\bibinfo  {journal} {Nat. Phys.}\ }\textbf {\bibinfo {volume} {15}},\
  \bibinfo {pages} {960} (\bibinfo {year} {2019})}\BibitemShut {NoStop}%
\bibitem [{\citenamefont {Henderson}\ \emph
  {et~al.}(2018{\natexlab{b}})\citenamefont {Henderson}, \citenamefont
  {Hennigar}, \citenamefont {Mann}, \citenamefont {Smith},\ and\ \citenamefont
  {Zhang}}]{henderson2018harvesting}%
  \BibitemOpen
  \bibfield  {author} {\bibinfo {author} {\bibfnamefont {L.~J.}\ \bibnamefont
  {Henderson}}, \bibinfo {author} {\bibfnamefont {R.~A.}\ \bibnamefont
  {Hennigar}}, \bibinfo {author} {\bibfnamefont {R.~B.}\ \bibnamefont {Mann}},
  \bibinfo {author} {\bibfnamefont {A.~R.}\ \bibnamefont {Smith}},\ and\
  \bibinfo {author} {\bibfnamefont {J.}~\bibnamefont {Zhang}},\ }\bibfield
  {title} {\bibinfo {title} {Harvesting entanglement from the black hole
  vacuum},\ }\href {https://iopscience.iop.org/article/10.1088/1361-6382/aae27e/meta} {\bibfield  {journal} {\bibinfo  {journal} {Class.
  Quantum Gravity}\ }\textbf {\bibinfo {volume} {35}},\ \bibinfo {pages}
  {21LT02} (\bibinfo {year} {2018}{\natexlab{b}})}\BibitemShut {NoStop}%
\bibitem [{\citenamefont {Mendez-Avalos}\ \emph {et~al.}(2022)\citenamefont
  {Mendez-Avalos}, \citenamefont {Henderson}, \citenamefont
  {Gallock-Yoshimura},\ and\ \citenamefont {Mann}}]{mendez2022entanglement}%
  \BibitemOpen
  \bibfield  {author} {\bibinfo {author} {\bibfnamefont {D.}~\bibnamefont
  {Mendez-Avalos}}, \bibinfo {author} {\bibfnamefont {L.~J.}\ \bibnamefont
  {Henderson}}, \bibinfo {author} {\bibfnamefont {K.}~\bibnamefont
  {Gallock-Yoshimura}},\ and\ \bibinfo {author} {\bibfnamefont {R.~B.}\
  \bibnamefont {Mann}},\ }\bibfield  {title} {\bibinfo {title} {Entanglement
  harvesting of three unruh-dewitt detectors},\ }\href {https://link.springer.com/article/10.1007/s10714-022-02956-x} {\bibfield
  {journal} {\bibinfo  {journal} {Gen. Relativ. Gravit.}\ }\textbf {\bibinfo
  {volume} {54}},\ \bibinfo {pages} {87} (\bibinfo {year} {2022})}\BibitemShut
  {NoStop}%
\bibitem [{\citenamefont {Sahu}\ \emph {et~al.}(2022)\citenamefont {Sahu},
  \citenamefont {Melgarejo-Lermas},\ and\ \citenamefont
  {Mart{\'{\i}}n-Mart{\'{i}}nez}}]{sahu2022sabotaging}%
  \BibitemOpen
  \bibfield  {author} {\bibinfo {author} {\bibfnamefont {A.}~\bibnamefont
  {Sahu}}, \bibinfo {author} {\bibfnamefont {I.}~\bibnamefont
  {Melgarejo-Lermas}},\ and\ \bibinfo {author} {\bibfnamefont {E.}~\bibnamefont
  {Mart{\'{\i}}n-Mart{\'{i}}nez}},\ }\bibfield  {title} {\bibinfo {title}
  {Sabotaging the harvesting of correlations from quantum fields},\ }\href
  {https://journals.aps.org/prd/abstract/10.1103/PhysRevD.105.065011} {\bibfield  {journal} {\bibinfo  {journal} {Phys. Rev. D}\ }\textbf
  {\bibinfo {volume} {105}},\ \bibinfo {pages} {065011} (\bibinfo {year}
  {2022})}\BibitemShut {NoStop}%
\bibitem [{\citenamefont {Henderson}\ \emph {et~al.}(2020)\citenamefont
  {Henderson}, \citenamefont {Belenchia}, \citenamefont {Castro-Ruiz},
  \citenamefont {Budroni}, \citenamefont {Zych}, \citenamefont {Brukner},\ and\
  \citenamefont {Mann}}]{henderson2020quantum}%
  \BibitemOpen
  \bibfield  {author} {\bibinfo {author} {\bibfnamefont {L.~J.}\ \bibnamefont
  {Henderson}}, \bibinfo {author} {\bibfnamefont {A.}~\bibnamefont
  {Belenchia}}, \bibinfo {author} {\bibfnamefont {E.}~\bibnamefont
  {Castro-Ruiz}}, \bibinfo {author} {\bibfnamefont {C.}~\bibnamefont
  {Budroni}}, \bibinfo {author} {\bibfnamefont {M.}~\bibnamefont {Zych}},
  \bibinfo {author} {\bibfnamefont {{\v{C}}.}~\bibnamefont {Brukner}},\ and\
  \bibinfo {author} {\bibfnamefont {R.~B.}\ \bibnamefont {Mann}},\ }\bibfield
  {title} {\bibinfo {title} {Quantum temporal superposition: the case of
  quantum field theory},\ }\href {https://journals.aps.org/prl/abstract/10.1103/PhysRevLett.125.131602} {\bibfield  {journal} {\bibinfo
  {journal} {Phys. Rev. Lett.}\ }\textbf {\bibinfo {volume} {125}},\ \bibinfo
  {pages} {131602} (\bibinfo {year} {2020})}\BibitemShut {NoStop}%
\bibitem [{\citenamefont {Virally}\ \emph {et~al.}(2021)\citenamefont
  {Virally}, \citenamefont {Cusson},\ and\ \citenamefont
  {Seletskiy}}]{virally2021enhanced}%
  \BibitemOpen
  \bibfield  {author} {\bibinfo {author} {\bibfnamefont {S.}~\bibnamefont
  {Virally}}, \bibinfo {author} {\bibfnamefont {P.}~\bibnamefont {Cusson}},\
  and\ \bibinfo {author} {\bibfnamefont {D.~V.}\ \bibnamefont {Seletskiy}},\
  }\bibfield  {title} {\bibinfo {title} {Enhanced electro-optic sampling with
  quantum probes},\ }\href {https://journals.aps.org/prl/abstract/10.1103/PhysRevLett.127.270504} {\bibfield  {journal} {\bibinfo  {journal}
  {Phys. Rev. Lett.}\ }\textbf {\bibinfo {volume} {127}},\ \bibinfo {pages}
  {270504} (\bibinfo {year} {2021})}\BibitemShut {NoStop}%
\bibitem [{\citenamefont {Raymer}\ \emph {et~al.}(1995)\citenamefont {Raymer},
  \citenamefont {Cooper}, \citenamefont {Carmichael}, \citenamefont {Beck},\
  and\ \citenamefont {Smithey}}]{raymer1995ultrafast}%
  \BibitemOpen
  \bibfield  {author} {\bibinfo {author} {\bibfnamefont {M.}~\bibnamefont
  {Raymer}}, \bibinfo {author} {\bibfnamefont {J.}~\bibnamefont {Cooper}},
  \bibinfo {author} {\bibfnamefont {H.}~\bibnamefont {Carmichael}}, \bibinfo
  {author} {\bibfnamefont {M.}~\bibnamefont {Beck}},\ and\ \bibinfo {author}
  {\bibfnamefont {D.}~\bibnamefont {Smithey}},\ }\bibfield  {title} {\bibinfo
  {title} {Ultrafast measurement of optical-field statistics by dc-balanced
  homodyne detection},\ }\href {https://opg.optica.org/josab/fulltext.cfm?uri=josab-12-10-1801&id=33583} {\bibfield  {journal} {\bibinfo
  {journal} {JOSA B}\ }\textbf {\bibinfo {volume} {12}},\ \bibinfo {pages}
  {1801} (\bibinfo {year} {1995})}\BibitemShut {NoStop}%
\bibitem [{\citenamefont {G{\"u}hne}\ and\ \citenamefont
  {T{\'o}th}(2009)}]{guhne2009entanglement}%
  \BibitemOpen
  \bibfield  {author} {\bibinfo {author} {\bibfnamefont {O.}~\bibnamefont
  {G{\"u}hne}}\ and\ \bibinfo {author} {\bibfnamefont {G.}~\bibnamefont
  {T{\'o}th}},\ }\bibfield  {title} {\bibinfo {title} {Entanglement
  detection},\ }\href {https://www.sciencedirect.com/science/article/pii/S0370157309000623} {\bibfield  {journal} {\bibinfo  {journal} {Phys.
  Rep.}\ }\textbf {\bibinfo {volume} {474}},\ \bibinfo {pages} {1} (\bibinfo
  {year} {2009})}\BibitemShut {NoStop}%
\bibitem [{\citenamefont {G{\"u}ndo{\u{g}}du}\ \emph
  {et~al.}(2023)\citenamefont {G{\"u}ndo{\u{g}}du}, \citenamefont {Virally},
  \citenamefont {Scaglia}, \citenamefont {Seletskiy},\ and\ \citenamefont
  {Moskalenko}}]{gundougdu2023self}%
  \BibitemOpen
  \bibfield  {author} {\bibinfo {author} {\bibfnamefont {S.}~\bibnamefont
  {G{\"u}ndo{\u{g}}du}}, \bibinfo {author} {\bibfnamefont {S.}~\bibnamefont
  {Virally}}, \bibinfo {author} {\bibfnamefont {M.}~\bibnamefont {Scaglia}},
  \bibinfo {author} {\bibfnamefont {D.~V.}\ \bibnamefont {Seletskiy}},\ and\
  \bibinfo {author} {\bibfnamefont {A.~S.}\ \bibnamefont {Moskalenko}},\
  }\bibfield  {title} {\bibinfo {title} {Self-referenced subcycle metrology of
  quantum fields},\ }\href {https://onlinelibrary.wiley.com/doi/abs/10.1002/lpor.202200706} {\bibfield  {journal} {\bibinfo  {journal}
  {Laser Photonics Rev.}\ }\textbf {\bibinfo {volume} {17}},\ \bibinfo {pages}
  {2200706} (\bibinfo {year} {2023})}\BibitemShut {NoStop}%
\bibitem [{\citenamefont {Allen}\ and\ \citenamefont
  {Knight}(1983)}]{allen1983concepts}%
  \BibitemOpen
  \bibfield  {author} {\bibinfo {author} {\bibfnamefont {L.}~\bibnamefont
  {Allen}}\ and\ \bibinfo {author} {\bibfnamefont {P.}~\bibnamefont {Knight}},\
  }\href {} {\emph {\bibinfo {title} {Concepts of Quantum optics}}}\
  (\bibinfo  {publisher} {Pergamon Press, Oxford},\ \bibinfo {year}
  {1983})\BibitemShut {NoStop}%
\bibitem [{\citenamefont {Werner}(1989)}]{werner_quantum_1989}%
  \BibitemOpen
  \bibfield  {author} {\bibinfo {author} {\bibfnamefont {R.~F.}\ \bibnamefont
  {Werner}},\ }\bibfield  {title} {\bibinfo {title} {Quantum states with
  einstein-podolsky-rosen correlations admitting a hidden-variable model},\
  }\href {https://doi.org/10.1103/PhysRevA.40.4277} {\bibfield  {journal}
  {\bibinfo  {journal} {Phys. Rev. A}\ }\textbf {\bibinfo {volume} {40}},\
  \bibinfo {pages} {4277} (\bibinfo {year} {1989})}\BibitemShut {NoStop}%
\bibitem [{\citenamefont {Gisin}(1996)}]{gisin_hidden_1996}%
  \BibitemOpen
  \bibfield  {author} {\bibinfo {author} {\bibfnamefont {N.}~\bibnamefont
  {Gisin}},\ }\bibfield  {title} {\bibinfo {title} {Hidden quantum nonlocality
  revealed by local filters},\ }\href
  {https://doi.org/https://doi.org/10.1016/S0375-9601(96)80001-6} {\bibfield
  {journal} {\bibinfo  {journal} {Phys. Lett. A}\ }\textbf {\bibinfo {volume}
  {210}},\ \bibinfo {pages} {151} (\bibinfo {year} {1996})}\BibitemShut
  {NoStop}%
\end{thebibliography}

%

\end{document}